\def\LeftLineBreak{\noindent\rule{20.5pc}{0.1 pt}\rule{0.1pt}{5 pt}}
\def\RightLineBreak{\hspace{22pc}\rule{0.1pt}{5 pt}\rule[5 pt]{20.5pc}{0.1 pt}}
\documentstyle[aps,prc,eqsecnum,multicol,epsf]{revtex}
\begin{document}
\title{A Consistent Meson-Field-Theoretical Description of PP-Bremsstrahlung\footnote{
Also to appear in Phys.~Rev.~{\bf C53}, 1102 (1996)}}
\author{J. A. Eden and M. F. Gari}
\address{ Institut f\"ur Theoretische Physik\\
	  Ruhr Universit\"at Bochum, D-44780 Bochum, Germany\\
	  {\rm email: jamie@deuteron.tp2.ruhr-uni-bochum.de}}
\date{31 March 1995, Revised 28 September 1995}
\maketitle
\tighten
\begin{abstract}
A parameter-free and relativistic extension of the RuhrPot meson-baryon model 
is used to define the dominant isoscalar meson-exchange currents. We compute
pp-bremsstrahlung observables below the $\pi-$production threshold using 
a relativistic hadronic current density that includes impulse, wave function 
re-orthonormalization, meson-recoil, $\bar{{\rm N}}$N creation and 
annihilation, 
$\rho\pi\gamma$ + $\omega\pi\gamma$ + $\rho\eta\gamma$ + $\omega\eta\gamma$ 
vector-meson decay and N$\Delta\gamma(\pi,\rho)$ exchange currents. 
We obtain a good description of the available data.
The N$\Delta\gamma(\pi)$ current is shown to dominate the large two-body 
contributions and closed-form expressions for various non-relativistic 
approximations are analyzed. An experimental sensitivity to the admixture
of pseudo-scalar and pseudo-vector admixture of the NN$\pi$ interaction 
is demonstrated.  We examine the Lorentz invariance of the
NN$\rightleftharpoons$NN $t$-matrices and show a dominantly pseudo-vector
NN$\pi$ coupling renders impulse approximation calculations without boost 
operators to be essentially exact.
Conversely, a similar analysis of the $\Delta{\rm N}\rightleftharpoons$NN 
transitions shows that boost operators and the two-body $N\Delta\gamma$ 
wave function re-orthonormalization meson-recoil currents are required in
NN, $\Delta$N and $\Delta\Delta$ coupled channel $t$-matrix applications.
The need for additional data is stressed.
\end{abstract}
\begin{multicols}{2}\narrowtext
\section{Introduction}\label{sec1}
The realization that meson-exchange currents play a vital role in 
the description of the low-energy pp-bremsstrahlung observables has 
consequences which are only now coming to be understood. 
For example, the traditional objective of pp-bremsstrahlung 
investigations, as indicated in Fig.~1, centers on the capacity of 
experiment to differentiate the accuracy of the off-shell $t$-matrices that 
are predicted by a range of model-dependent NN-interactions. 
This is now recognized \cite{JE95,VH95} as an exceedingly difficult task and,
at best, is contingent on a completely reliable and consistent description 
of the associated meson-exchange currents. As such, a meaningful calculation
of the pp-bremsstrahlung observables requires knowledge of the meson-baryon
form factors, meson-exchange currents and the NN-interaction within a 
fully consistent and microscopic effective theory.

Recognizing the importance of exchange currents in pp-bremsstrahlung  
implies a complete departure form the conventional approach to the problem
and considerably changes the nature of such investigation.
For many years pp-bremsstrahlung was regarded as something of a special 
case in nuclear physics because both meson-exchange currents and
relativistic effects were expected to be small. The principle reason for
this expectation stems from the fact that gauge invariance demands the 
real photon couples only to conserved currents, so that
the n-body parts of the complete hadronic current $J_{[n]}$ 
\par\noindent
\begin{figure}
\vbox to 2.75 true in {\hbox{\hskip 0 mm \epsfxsize=3.4 true in \epsfbox{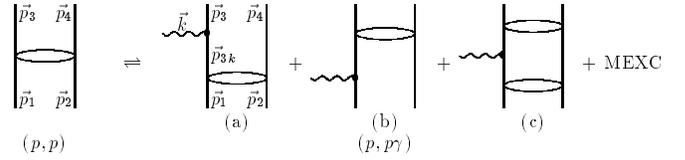}}
\vglue 3 true mm
\caption{ 
Relationship between the on-shell $t$-matrix in pp-scattering, $(p,p)$ and the
off-shell $t$-matrix in pp-bremsstrahlung $(p,p\gamma)$. 
The parameters in the NN-interaction $V$ defining $t$=$V$+$VGt$ (shown as a bubble) 
are fitted to the (N,N) phase shifts. Bremsstrahlung is usually
considered in nucleon-pole dominance, where (a) initial-, (b) final- and (c) 
rescattering-interactions are retained, but all meson-exchange currents are neglected.
Within this assumption, bremsstrahlung has been regarded as the best means of 
testing the off-shell $t$-matrix. However, a sensitivity to off-shell effects 
requires a large photon energy, so that $G$=$(E-H_0)^{-1}$ diminishes the dominant 
nucleon pole contributions and the exchange currents become important.
}}
\end{figure}
\par\noindent
for any given NN-interaction $V_{\rm NN}$ must satisfy,
\begin{eqnarray}\label{eq1.1}
 0 &=& \nabla.\vec{J} + i\left[ {\cal H},J^0 \right]
\nonumber\\
\Rightarrow\quad  0 &=& \cases{
 \nabla.\vec{J}_{[1]} + i\left[ H_0,J_{[1]}^0 \right] & (one body) \cr
 \nabla.\vec{J}_{[2]} + i\left[ V_{\rm NN},  J_{[1]}^0 \right]  & (two body)\cr
    + i\left[ H_0+V_{\rm NN},J_{[2]}^0\right]&}
\end{eqnarray}
To isolate the dominant contributions to the observables, it is useful to
consider only the static limit, where the two-body charge density 
$J_{[2]}^0$ can be ignored. The isospin structure of the exchange currents 
$\vec{J}_{[2]}$ for isovector mesons ($\pi$,$\rho$,..) then reduces to 
$(\vec{\tau}_1\times\vec{\tau}_2)^z$, which vanishes in isospin conserving
processes like pp-bremsstrahlung. Relativistic processes can also expected 
to be small since the dominant $\pi$-exchange contributions to the 
N$\bar{\rm N}$-pair creation and annihilation amplitudes share this isospin 
structure.  Finally, all NN$\gamma$ couplings with meson-recoil terms can be 
neglected since they are exactly canceled by the corresponding wave function 
re-orthonormalization contributions \cite{GA76}. 
All of this information suggests that a static limit description of 
pp-bremsstrahlung involves only the photon coupling to one of the 
protons either before and/or after (but not during) strong interaction. 
The leading-order exchange currents, according to this analysis, begin
with the $\eta$ (549~MeV), $\omega$ (782~MeV) and $\epsilon$ (975~MeV) 
iso-scalar mesons, and can therefore be reasonably neglected.

However, the above analysis is flawed for several reasons. Even within the 
static limit there are purely transverse currents $\vec{J}_{\rm t}$ which
automatically satisfy $\nabla.\vec{J}_{\rm t}$ =0, so that current conservation 
places no constraints on the manifestly gauge invariant $\rho\pi\gamma$, 
$\omega\pi\gamma$ and N$\Delta\gamma$ exchange currents shown in Fig.~2. None 
of these two-body currents can be included simply by introducing the commutators
shown in eq.~(\ref{eq1.1}), yet they all possess non-vanishing isoscalar 
contributions which can be important in pp-bremsstrahlung. In addition, 
as new experiments have succeeded in selecting kinematics that escape the 
consequences of the low-energy theorems \cite{LO58,SA66} and on-shell 
expansions \cite{NY68,FE67}, they necessarily emphasize dynamics where 
the (usually dominant) nucleon-pole contributions of Fig.~1(a-c) are heavily 
suppressed by the Greens' functions accompanying the highly off-shell 
$t$-matrix.  As such, otherwise less important contributions gain considerable 
significance in the observables.  This shows that the pp-bremsstrahlung 
dynamics involves much more than the off-shell $t$-matrix and the impulse 
current, and appears to share the complexity of other observables like 
np-bremsstrahlung and $n+p\rightleftharpoons d+\gamma$.

A very long list of pp-bremsstrahlung calculations have been reported over the 
last 45 years. We will make no attempt to review them all since more recent 
works 
\cite{HF86,VB91,VH91,VH92,HF93a,HF93b,JE93,JE94,KA93,MJ93a,MJ93,MJ94,JO94}
already contain appropriate citation and serve to remove a number of 
questionable approximations. A notable exception to this trend is found in the 
very detailed $r$-space calculations reported some 25 years ago by Brown 
\cite{VB69,VB71}, where the rescattering amplitudes of Fig.~1(c) were retained and 
eq.~(\ref{eq1.1}) was used to constrain the longitudinal 
meson-exchange currents. Noteworthy calculations since that time have generally
been less complete, but find their merit in the application of more reliable
NN-interactions and the exploration of coulomb corrections \cite{MJ93a} 
and relativistic corrections to the impulse current \cite{HF86,VH92,JE93,KA93}. 
\par\noindent
\begin{figure}
\vbox to 5.00 true in {\hbox{\hskip 0 mm \epsfxsize=3.4 true in \epsfbox{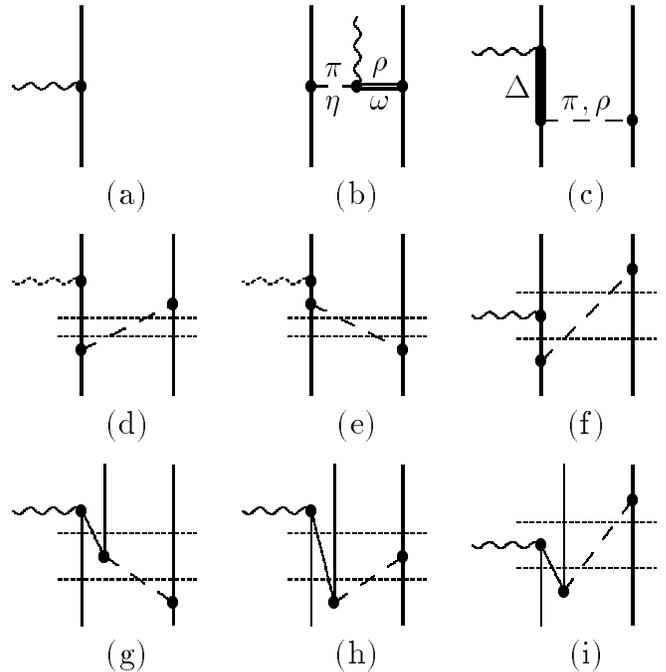}}
\vglue 3 true mm
\caption{ 
Currents included in the present calculations: (a) impulse current 
(b) radiative vector-meson decay currents 
VP$\gamma$ = $\rho\pi\gamma$ + $\omega\pi\gamma$ + $\rho\eta\gamma$ + $\omega\eta\gamma$,
(c) N$\Delta\gamma$ $\pi-$ and $\rho-$ exchange currents,
(d-e) wave function re-orthonormalization and (f) NN$\gamma$ meson-recoil currents,
and (g-i) ${\rm N}\bar{\rm N}$-pair creation and annihilation currents.
In each of (d-f) and (g-i) we show 3 of the 12 time-ordered diagrams
with the energy cuts represented by dotted lines.
None of the exchange currents (b-i) can be obtained from eq~(1.1).
}}
\end{figure}
\par\noindent
Only a few of the most recent pp-bremsstrahlung calculations 
\cite{JE95,JE94,MJ94,JO94} attempt a significant improvement on the standard 
set by Brown, although there are important technical differences in these works 
that we will need to consider.

In the present work we will subject the RuhrPot description of meson-baryon 
interactions to the test of reproducing the pp-bremsstrahlung data below the 
$\pi$-production threshold. The reasons for selecting this effective theory 
are
\begin{itemize}
\item A microscopic definition of the strong form factors is available
      from non-perturbative and self-consistent calculation \cite{JF95}.
      The results are compatible with skyrme\cite{TC86}, non-topological 
      soliton \cite{PA88} and bag-model \cite{BB82} calculations, 
      and moreover, with an analysis of experiment \cite{GK92}.
\item An NN-interaction model using calculated (not fitted) form factors
      has been constructed and gives an excellent description
      of the world data for the NN-scattering phase shifts \cite{RuhrPot}.
\item The extension to include meson-exchange currents in the calculation
      of observables introduces no free parameters whatsoever \cite{JE95,JE94}.
\item A parameter free-extension of the model to define the 3-body interaction
      has been shown to provide an accurate description of the triton
      bonding energy \cite{DH95}.
\end{itemize}
We indicate the relationships between the form factors, the NN interaction and
the exchange currents in Fig.~3.  Although such consistent calculations can 
(in principle) be performed for any effective meson field theory, such work 
has so far only been completed for the RuhrPot description and it appears that 
there are severe difficulties in obtaining similar consistency in other models. 
For example, the RuhrPot form factor calculations have been modified to adopt
the coupling constants of a conventional boson-exchange NN interaction and
yield results \cite{JF95} which can be accurately parameterized as monopoles 
with typical regularization scales of $\Lambda\sim0.8$~GeV.  While consistency 
demands the use of such form factors, conventional boson-exchange models
require\cite{AT89} artificial scales (`cut-offs') of $\Lambda_\pi\ge 1.3$~GeV 
and $\Lambda_\rho\sim 1.8$~GeV. Such an artificial description of the
meson-nucleon vertices necessarily frustrates any attempt to obtain a
realistic description of the meson-exchange currents and the 3N interaction.
\begin{figure}
\vbox to 3.25 true in {\hbox{\hskip 0 mm \epsfxsize=3.4 true in \epsfbox{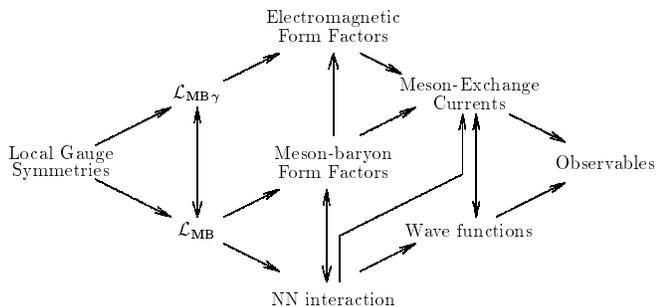}}
\vglue 3 true mm
\caption{ 
In the RuhrPot effective meson field theory, meson-baryon form factors 
are calculated non-perturbatively and the results are used without
adjustment as input for subsequent calculations of electromagnetic 
form factors, the NN interaction and the meson-exchange currents.
Such consistency is necessary, for example, to satisfy gauge invariance
(which relates the meson-exchange currents to the NN interaction) 
and ensure orthonormality of the wave functions (through inclusion 
of wave function re-orthonormalization exchange currents).
}}
\end{figure}

Coupled channel $t$-matrices providing a non-perturbative description
of all possible transitions between NN, $\Delta$N and $\Delta\Delta$ states
have been available for more than 25 years \cite{GA76b,GA79,GA81}
and have already been used to calculate the $\Delta$-isobar contributions 
to pp-bremsstrahlung observables \cite{MJ94,JO94}.
Under these circumstances it may appear curious that we choose to develop
a perturbative description of the N$\Delta\gamma$ $\pi-$ and $\rho-$ 
exchange currents.
However, the coupled channel $t$-matrices used in recent pp-bremsstrahlung 
calculations are obtained by the inconsistent combination of the 
Paris \cite{Paris} NN$\rightleftharpoons$NN and a static limit version of the 
Ried-parameterized Bochum \cite{GA79} NN$\rightleftharpoons\Delta$N interaction.
It is therefore impossible to accurately remove the double counted two-pion 
exchange amplitudes with intermediate N$\Delta$ states, so a free parameter 
is introduced to permit an approximate subtraction procedure. As such,  
this approach discards from the outset any hope of a obtaining a microscopic 
description and the quality of the results must be interpreted in terms of a 
meaningless parameter. We will later show that, even if these inconsistencies 
were to be resolved by fully consistent calculation, such an approach is  
contingent upon a reliable description of boost operators, as well as
(a subset of) the relativistic meson-exchange currents that will be 
developed in this work.

In the present work we develop our earlier 
description \cite{JE95,JE93,JE94} of the RuhrPot meson-baryon interactions 
in the pp-bremsstrahlung data below the $\pi$-production threshold. 
In refs~\cite{JE93} we included the relativistic single and rescattering 
impulse-current amplitudes, and in refs~\cite{JE95,JE94} we introduced
the fully relativistic description of the radiative vector-meson decay 
currents and the non-relativistic description of the N$\Delta\gamma$ $\pi-$ 
and $\rho-$meson exchange currents without recourse to the soft-photon 
approximation. In the present work we investigate a number of important 
extensions. In particular, after describing our model-independent
formalism in sec~\ref{sec2}, we provide in sec~\ref{sec3} the first 
bremsstrahlung calculations including a fully relativistic description of the 
wave function re-orthonormalization and meson-recoil currents that are 
required to ensure the orthonormality of the wave functions is preserved.
We also investigate the Lorentz structure of the NN$\pi$ vertex by providing
the first calculations for the purely relativistic N$\bar{\rm N}$ pair 
creation and annihilation currents. We further present relativistic expressions
for the N$\Delta\gamma$ $\pi-$ and $\rho-$exchange currents and identify the
source of error in various approximations.
Supporting calculation details are supplied in the three appendices.
In sec~\ref{sec4}, after establishing the sensitivity of the selected 
pp-bremsstrahlung  observables to each of these currents and concluding
that a relativistic calculation of the isobar amplitudes is necessary, 
we compare our relativistic results with the complete data set available 
from the 1990 TRIUMF pp-bremsstrahlung experiment \cite{TRIUMF}. 
We obtain a good description of the experimental data and conclude that 
a large pseudoscalar admixture in the NN$\pi$ Lagrangian is ruled out.
Further conclusions and future objectives are given in sec~\ref{sec5}.
\section{Formalism}\label{sec2}
\subsection{Observables}\label{sec2A}
We begin by presenting the model independent expressions we require for the 
calculation observables for the reaction $N +N \rightarrow N + N + \gamma$.
The $S$--matrix from covariant perturbation theory 
\begin{equation}\label{eq2.1}
S_{fi}
=\delta_{fi}-i \int d^4x\langle f\vert J^\mu(x){\cal A}_\mu(x)\vert i\rangle  +\cdots 
\end{equation}
gives the probability amplitude for a transition $\vert i\rangle  \rightarrow \vert f\rangle $ 
as a series involving 0,1,$\ldots$ interactions where the photon field
${\cal A}_\mu(x)$ couples to the hadronic current density 
$J^\mu(x)$.
Since only terms with an odd number of electromagnetic interactions can 
contribute to the production of a single real photon, and each of these 
diminishes by $\alpha\sim 1/137$, we retain only the lowest-order contribution
and define the transition amplitude as,
\begin{equation}\label{eq2.2}
(2\pi)^4{\cal T}_{fi}\delta^{(4)}(P_f-P_i)
= -i \int d^4x\langle f\vert J^\mu(x){\cal A}_\mu(x)\vert i\rangle 
\end{equation}
Following a well trodden path, we integrate the transition amplitude over the 
phase space available to the final state, and divide by the incident flux, so 
that with plane wave normalized to a $\delta$-function, we obtain the Lab-frame
differential cross section as,
\begin{equation}\label{eq2.3}
{d^3\sigma \over d\Omega_3 d\Omega_4 d\theta_\gamma}
   = {(2\pi)^{-5} {1 \over 2} m^3 \over  \vert\vec{p}_1\vert  } 
     {\vert \buildrel {~\sim} \over {\cal M}_{fi} \vert {}^2} J_{\rm ps}
\end{equation}
where, for the pp- and nn-bremsstrahlung reactions we have 
\end{multicols}\widetext
\LeftLineBreak
\begin{equation}\label{eq2.4}
{\vert \buildrel {~\sim} \over {\cal M}_{fi} \vert {}^2}
= 
\cases{ 
   \vert {\cal M}_{fi}^2 \vert &
     if $M_{S_i}$, $M_{S_f}$, $\lambda$ measured \cr
   \displaystyle{ 
     {1 \over 4} \sum_{S_i M_{S_i}} \sum_{S_f M_{S_f}} 
     \sum_{\lambda=1}^{2} \vert {\cal M}_{fi} \vert^2
   } & if $M_{S_i}$, $M_{S_f}$, $\lambda$ not measured \cr
}
\end{equation}
and for np-bremsstrahlung we require,
\begin{equation}\label{eq2.5}
{\vert \buildrel {~\sim} \over {\cal M}_{fi} \vert {}^2} =
\cases{ 
   \displaystyle{ 
     {1 \over 2} \sum_{T_i, T_f}
     \vert {\cal M}_{fi} \vert^2 
   } & if $M_{S_i}$, $M_{S_f}$, $\lambda$ measured  \cr 
   \displaystyle{ 
     {1 \over 8}\sum_{T_i,T_f}\sum_{S_i M_{S_i}}\sum_{S_f M_{S_f}}
                \sum_{\lambda=1}^{2} 
     \vert {\cal M}_{fi} \vert^2 
   } & if $M_{S_i}$, $M_{S_f}$, $\lambda$ not measured \cr
}
\end{equation}
\RightLineBreak
\begin{multicols}{2}
with the invariant amplitude given by,
\begin{equation}\label{eq2.6}
{\cal M}_{fi} 
= i (2\pi)^{15/2} m^{-2}
  \left[ 2\omega E_1 E_2 E_3E_4 \right]^{1/2}
  {\cal T}_{fi}
\end{equation}
We retain only the transverse polarization vectors for the real photon since,
within the Gupta-Bleuler quantization formalism, the longitudinal and scalar 
components can be made to cancel with a gauge transformation, and therefore 
cannot effect the observables. 
The `phase-space factor' $J_{\rm ps}$ appearing in eq~(\ref{eq2.3}) is defined
for arbitrary non-coplanarity $\Phi=(\pi+\phi_3-\phi_4)/2$ as,
\begin{equation}\label{eq2.7}
J_{\rm ps}= { p_3^2 p_4^2 \over E_3 E_4 \vert \cos\theta_\gamma \vert }
   \Bigl| { N  \over k\sin\theta_\gamma\cos\theta_\gamma } \Bigr|^{-1}
\end{equation}
with 
\begin{eqnarray}\label{eq2.8}
N &=& \bigl( p_4\sin\theta_4 -  p_3\sin\theta_3 \bigr)
   \bigl[ 
   \sin(\theta_3 + \theta_4) 
\nonumber \\ &&-
   \bigl( \beta_3\sin\theta_4 + \beta_4\sin\theta_3 \bigr) \cos\theta_\gamma
   \bigr] 
\nonumber \\ &-& 
    k\sin^2\theta_\gamma
   \bigl( \beta_3\cos\theta_4 - \beta_4\cos\theta_3 \bigr) 
\nonumber \\ &+& 
    2 \sin\theta_3 \sin\theta_4 \sin^2\Phi
   \bigl[ 
   \bigl( p_3\cos\theta_3 - p_4\cos\theta_4 \bigr) 
\nonumber \\ &&-
   \bigl( p_3\beta_3 - p_4\beta_4 \bigr) \cos\theta_\gamma  
   \bigr] 
\end{eqnarray}
where we use $\beta_i$=$p_i/E_i$ for laboratory-frame reaction kinematics
$p_1+p_2\rightarrow p_3+p_4+k$.  We realize that in the limit 
$\theta_\gamma\rightarrow 0$, where $\Phi\rightarrow 0$ is guaranteed, $N^{-1}$
possesses singularities but $k\sin\theta_\gamma/N$ remains well defined.  Our 
description of kinematics and phase space is the same as that reported in the 
detailed discussion of ref~\cite{DM68}. As such, it is sufficient to note that 
in ref~\cite{DM68} it was shown that $J_{\rm ps}$ possess a square root 
singularity at the kinematic limit of non-coplanarity, although in the present 
work we avoid the non-relativistic simplifications discussed therein. 

For the calculation of polarized observables it is convenient to denote 
$d\sigma(\pm\hat{i})$ as the cross section of eq~(\ref{eq2.3}) measured with 
the beam polarized in the $\pm\hat{i}$ direction. We choose the  quantization
axis as the beam direction in the lab frame, and define the vector analyzing 
powers as,
\begin{equation}\label{eq2.9}
A_{i} = { d\sigma(+\hat{i}) - d\sigma(-\hat{i}) \over 
	   d\sigma(+\hat{i}) + d\sigma(-\hat{i})}
= { \sum_{T_i T_f\lambda}\hbox{Tr}
\Bigl\{{\cal M}_{fi}(\vec{\sigma}.\hat{i})_{[1]}{\cal M}_{fi}^\dagger\Bigr\}
\over
\sum_{T_i T_f\lambda}\hbox{Tr}
\Bigl\{ {\cal M}_{fi} {\cal M}_{fi}^\dagger \Bigr\} }
\end{equation}
where $i=\hat{x}, \hat{y}$ or $\hat{z}$ in the lab frame. Similarly, the tensor
analyzing powers (sometimes called `spin-correlation coefficients') are given 
by,
\begin{eqnarray}\label{eq2.10}
A_{ij} &=&{d\sigma(\!+\!\hat{i},\!+\!\hat{j})\!+\!d\sigma(\!-\!\hat{i},\!-\!\hat{j}) 
      \!-\!d\sigma(\!+\!\hat{i},\!-\!\hat{j})\!-\!d\sigma(\!-\!\hat{i},\!+\!\hat{j}) 
\over      d\sigma(\!+\!\hat{i},\!+\!\hat{j})\!+\!d\sigma(\!-\!\hat{i},\!-\!\hat{j}) 
      \!+\!d\sigma(\!+\!\hat{i},\!-\!\hat{j})\!+\!d\sigma(\!-\!\hat{i},\!+\!\hat{j}) 
}
\nonumber \\ &=& 
{\sum_{T_i T_f\lambda} 
\hbox{Tr}\Bigl\{ {\cal M}_{fi} 
(\vec{\sigma}.\hat{i})_{[1]} (\vec{\sigma}.\hat{j})_{[2]} 
{\cal M}_{fi}^\dagger \Bigr\}
\over
\sum_{T_i T_f\lambda} 
\hbox{Tr}\Bigl\{ {\cal M}_{fi} {\cal M}_{fi}^\dagger \Bigr\} }
\end{eqnarray}
where, for example, $d\sigma(+\hat{i},-\hat{j})$ is the cross section measured 
with the beam polarized in the $+\hat{i}$ direction and the target polarized in
the $-\hat{j}$ direction.

\subsection{The Hadronic Current}\label{sec2B}
To obtain a microscopic definition of the invariant amplitude we compute the 
Fock-space matrix elements of the photon field and hadronic current densities 
appearing in eq~(\ref{eq2.2}), so that after making use of translational 
invariance and selecting the Lorentz-Heaviside system with natural units, 
eq~(\ref{eq2.6}) can be recast as,
\begin{equation}\label{eq2.11}
{\cal M}_{fi} 
=  (2\pi)^6 m^{-2} \sqrt{E_1 E_2 E_3 E_4}
   \epsilon_\mu (\vec{k}, \lambda) 
   \langle \Psi_f^{(-)}\vert  J^\mu(0) \vert \Psi_i^{(+)}\rangle  
\end{equation}
where the field-theoretic hadronic current is given by,
\begin{equation}\label{eq2.12}
J^\mu(x)=\partial_\nu {\partial {\cal L} \over \partial(\partial_\nu {\cal A}_\mu)}
- {\partial {\cal L} \over \partial {\cal A}_\mu}
\end{equation}
for the Lagrangian ${\cal L}$ describing electromagnetic interactions with the 
interacting meson-baryon system. Direct calculation of eq.~(\ref{eq2.11}) is 
impossible since $\vert \Psi_i^{(+)}\rangle $ and $\vert \Psi_f^{(-)}\rangle $ represent 
complete meson-baryon states and therefore involve nucleon, resonance and meson
degrees of freedom to infinite order. The problem can, however, be approached 
with a Hamilton formalism \cite{GA76} where the total Hilbert space is 
partioned into meson+resonance+anti-nucleon vacuum- and existing subspaces,
\begin{equation}\label{eq2.13}
{\cal H}_{\eta} = \Bigl\{ \vert NN\rangle  \Bigr\}         , \quad
{\cal H}_{\lambda} = \Bigl\{ \vert \hbox{the rest}\rangle  \Bigr\} 
\end{equation}
We will refer to these subspaces as the $\eta$-space and $\lambda$-space
respectively. Defining $\eta$ and $\lambda$ as operators satisfying the 
conventional algebra $\eta^2=\eta$, $\lambda^2=\lambda$ and 
$\eta\lambda=\lambda\eta=0$ and which project out the components of 
${\cal H}_{\eta}$ and ${\cal H}_{\lambda}$ respectively, we apply a unitary 
transformation to decouple the meson-resonance vacuum and existing components 
of wave functions. Although this formalism makes provision for applications
involving (for example) explicit meson and/or $\Delta$ (see App.~A) degrees of 
freedom in the initial and final states, we confine our present application to 
energies below the $\pi$-production threshold, so that the complete interacting
meson-baryon wave function can be written as,
\begin{equation}\label{eq2.14}
\vert \Psi\rangle  = (1+A) {1 \over \sqrt{1 + A^\dagger A}} \vert \cal X\rangle  
\end{equation}
where $\vert {\cal X}\rangle $ is the two-nucleon state vector and 
$A=\lambda A \eta$ is required to satisfy $\lambda( H + [H,A] - AHA )\eta=0$.
In particular, we can expand both $A$ and the Hamiltonian $H$ in powers $n$ of 
the coupling constant, so that with the free-particle energy denoted as $H_0$,
we have,
\begin{equation}\label{eq2.15}
H = H_0 + \sum_{n=1}^{\infty}  H_n,
\qquad A =\sum_{n=1}^{\infty}  A_n
\end{equation}
and note that $\eta H_I\eta$=$\eta H_0\lambda$=$\lambda H_0\eta$=0, to obtain,
\begin{eqnarray}\label{eq2.16}
0 &=& \sum_{n=1}^{\infty} \lambda \Biggl[
      H_n +  [H_0\, , A_n]
      + \sum_{i=1}^{n-1} H_i A_{n-i}
\nonumber \\ &&
      -\sum_{i=1}^{n-2} \sum_{j=1}^{n-i-1} A_i H_j A_{n-i-j} \Biggr]\eta
\end{eqnarray}
We are free to further constrain $A$ by demanding eq.~(\ref{eq2.16}) is 
satisfied at each order of $n$, as would be required for any perturbative 
application.  Since $H_0\,\eta\vert \Psi\rangle $=${\cal E}_i\,\eta\vert \Psi\rangle $,
where ${\cal E}_i$ is the asymptotic energy of the free two-nucleon state, 
we obtain,
\begin{eqnarray}\label{eq2.17}
({\cal E}_i-H_0) A_n &=& \lambda \Biggl[
   H_n 
+ \sum_{i=1}^{n-1} H_i A_{n-i}
\nonumber \\ &&
- \sum_{i=1}^{n-2}\,\, \sum_{j=1}^{n-i-1} A_i H_j A_{n-i-j}
\Biggr]\eta
\end{eqnarray}
Since $A_0$=0, we have
\begin{mathletters}\label{eq2.18}
\begin{eqnarray}
A_1
&=& {\lambda\over {\cal E}_i-H_0} H_1 \eta  
\\
A_2
&=& {\lambda\over {\cal E}_i-H_0} H_2 \eta 
 +  {\lambda\over {\cal E}_i-H_0} H_1 {\lambda\over {\cal E}_i-H_0} H_1 \eta 
\\
\vdots && \nonumber
\end{eqnarray}
\end{mathletters}
Finally, with 
\begin{equation}\label{eq2.19}
H_I= -\int {\cal L} d^3x
\end{equation}
we observe that $A$ is completely determined by the strong interaction 
Lagrangian density defining any model of interest. Combining eqs.~(\ref{eq2.11}) 
and (\ref{eq2.14}), we then obtain,
\begin{equation}\label{eq2.20}
{\cal M}_{fi} = 
   (2\pi)^{6} m^{-2} \sqrt{E_1 E_2 E_3 E_4 }
   \langle {\cal X}_f\vert   \,
   \epsilon_\mu (\vec{k}, \lambda) 
   J_{\hbox{eff}}^\mu(0) 
   \, \vert {\cal X}_i\rangle 
\end{equation}
where
\begin{eqnarray}\label{eq2.21}
&&J_{\hbox{eff}}^\mu(0) =
\eta {1 \over \sqrt{1 + A^\dagger A}} (1+A^\dagger)
J^\mu(0)
(1+A) {1 \over \sqrt{1 + A^\dagger A}} \eta
\nonumber \\ &&=
\eta \bigl[ J^\mu(0) + J^\mu(0) A + A^\dagger J^\mu(0) 
            + A^\dagger J^\mu(0) A
\nonumber \\ &&
            - {1\over 2} J^\mu(0) A^\dagger A 
            - {1\over 2} A^\dagger A J^\mu(0) + \cdots\bigl] \eta
\end{eqnarray}
is the effective meson-baryon current density. Eq.~(\ref{eq2.21}) provides 
a time-ordered relativistic description of the impulse- and 
meson-exchange currents implied by the strong interaction Lagrangian density 
defining any model of interest. This provides, without approximation, 
a non-covariant three-vector representation where all particles are confined 
to their mass shells and energy need not be conserved at individual vertices.
An intuitive understanding of the processes 
embedded in the effective current density can be obtained by noting that the 
operator $A$ is always associated with transitions from the $\eta-$space into 
the $\lambda-$space, so that with our present definitions, $A$ serves to 
create meson+resonance+anti-nucleon existing states and $A^\dagger$ serves to 
restore purely two-nucleon states. In the second-order expansion of this 
current we observe direct terms $J$, initial- and final-state interaction terms
$JA$ + $A^\dagger J$, meson-recoil terms $A^\dagger J A$, and wave function 
re-orthonormalization terms $J A^\dagger A$ and $A^\dagger A J$. The wave 
function re-orthonormalization terms result directly from the requirement 
that the transformation used to obtain eq.~(\ref{eq2.14}) is unitary - or 
equivalently, from the fact that we insist upon working with orthonormal 
wave functions. This point has been discussed in considerable detail elsewhere 
\cite{GA76}.

We select a momentum-space representation and perform a $t$-matrix expansion 
of the two-nucleon wave functions according to the standard procedure.  With 
the photon field quantized in the Gupta-Bleuler formalism, we require only the 
transverse polarization vectors. Since these have a vanishing time component, 
we require only the spacial parts of the effective current density, so that,
\begin{mathletters}\label{eq2.22}
\begin{eqnarray}
{\cal M}_{fi} 
&=& N \vec{\epsilon} (\vec{k}, \lambda) 
\langle \widetilde{\vec{p}_3 \vec{p}_4;\alpha_f}\vert 
\vec{J}_{\hbox{eff}}(0) 
\vert \widetilde{\vec{p}_1 \vec{p}_2;\alpha_i}\rangle 
\nonumber \\ &&
\\ &+& N \vec{\epsilon} (\vec{k}, \lambda) 
\langle \widetilde{\vec{p}_3 \vec{p}_4;\alpha_f}\vert 
\vec{J}_{\hbox{eff}}(0) 
G_i t^{(+)} 
\vert \widetilde{\vec{p}_1 \vec{p}_2;\alpha_i}\rangle 
\nonumber \\ &&
\\ &+& N \vec{\epsilon} (\vec{k}, \lambda) 
\langle \widetilde{\vec{p}_3 \vec{p}_4;\alpha_f}\vert 
 t^{(-)\dag} G_f
\vec{J}_{\hbox{eff}}(0) 
\vert \widetilde{\vec{p}_1 \vec{p}_2;\alpha_i}\rangle 
\nonumber \\ &&
\\ &+& N \vec{\epsilon} (\vec{k}, \lambda) 
\langle \widetilde{\vec{p}_3 \vec{p}_4;\alpha_f}\vert 
 t^{(-)\dag} G_f
\vec{J}_{\hbox{eff}}(0) 
 G_i t^{(+)} 
\vert \widetilde{\vec{p}_1 \vec{p}_2;\alpha_i}\rangle 
\Bigr\}
\nonumber \\ &&
\end{eqnarray}
\end{mathletters}
where $G_i$ and $G_f$ are $\eta$-space Green's functions describing the 
propagation of two-nucleon states and
\begin{equation}\label{eq2.23}
N = - (2\pi)^{6} m^{-2} \sqrt{E_1 E_2 E_3 E_4}
\end{equation}
The four terms shown in eq.~(\ref{eq2.22}) will be referred to as
`direct', `initial-state', `final-state' and `rescattering'
amplitudes respectively.
\section{Model Definition and Calculation Details}\label{sec3}
\subsection{The RuhrPot Lagrangian}\label{sec3A}
We adopt the strong-interaction Lagrangian densities,
\begin{eqnarray}\label{eq3.1A}
{\cal L}_{{\rm NN}\pi}&=&
- i g_{{\rm NN}\pi} 
\bar{\psi}[
  \lambda\gamma^5
 - (1-\lambda)\displaystyle{1\over 2m}\gamma^5\gamma^{\mu}(i\partial_\mu)]
\psi \vec{\pi}.\vec{\tau}
\nonumber \\
{\cal L}_{{\rm NN}\eta}&=&
 - {g_{{\rm NN}\eta} \over 2m} \bar{\psi} \gamma^5\gamma^{\mu}\partial_\mu \psi \eta
\nonumber \\
{\cal L}_{{\rm NN}\rho}&=&- g_{{\rm NN}\rho} \bar{\psi} \gamma^\mu \psi \vec{\rho}_\mu.\vec{\tau}
\nonumber \\
{\cal L}_{{\rm NN}\omega}&=&- g_{{\rm NN}\omega} \bar{\psi}\gamma^\mu \psi \omega_\mu 
\nonumber \\
{\cal L}_{{\rm NN}\delta}&=&
- g_{{\rm NN}\delta} 
\bar{\psi} \psi \vec{\delta}.\vec{\tau}
\nonumber \\
{\cal L}_{{\rm NN}\epsilon}&=&
- g_{{\rm NN}\epsilon} 
\bar{\psi} \psi \epsilon
\nonumber \\
{\cal L}_{{\rm N}\Delta\pi}&=&- {g_{{\rm N}\Delta\pi}\over 2m} 
\bar{\psi}^\mu \vec{\tau}_{{\rm N}\Delta} \psi \partial_\mu \vec{\pi} 
+ \hbox{h.c.}
\nonumber \\
{\cal L}_{{\rm N}\Delta\rho}&=&- i {g_{{\rm N}\Delta\rho}\over 2m} 
\bar{\psi}^\mu\gamma^5\gamma^\nu\vec{\tau}_{{\rm N}\Delta}\psi
\vec{\rho}_{\mu\nu} + \hbox{h.c.}
\end{eqnarray}
where $\rho_{\mu\nu}=\partial_\mu \rho_\nu -\partial_\nu \rho_\mu$, 
and we have denoted
$g_{{\rm N}\Delta\rho}$=$\mu_{{\rm N}\Delta} g_{\rho}$,
$g_{{\rm NN}\rho} $=$ {1\over 2} g_{\rho}$ and 
$g_{{\rm NN}\omega} $=$ {1\over 2} g_{\omega}$, where $g_{\rho}$ and 
$g_{\omega}$ 
the strong charges for the $\rho$ and $\omega$ gauge fields.
For the electromagnetic-interaction we use the Lorentz-Heaviside system with
natural units, where the charge of the proton is defined as 
$e_p = +\sqrt{4\pi\alpha}$ with $\alpha\sim$1/137.04, and we adopt the 
Lagrangians,
\begin{eqnarray}\label{eq3.2}
{\cal L}_{{\rm NN}\gamma} &=&
- e_{\rm N} \bar{\psi} \gamma^\mu \psi {\cal A}_\mu 
+ {e_p\kappa_{\rm N} \over 2m}\bar{\psi}\sigma^{\mu\nu}\partial_\nu
{\cal A}_\mu \psi 
\nonumber \\
{\cal L}_{{\rm PV}\gamma} &=& - {e_p g_{{\rm PV}\gamma}\over 2 m_{\rm V} }
\epsilon^{\mu\nu\sigma\tau} F_{\mu\nu} 
\vec{\phi}^{\rm V}_\sigma . \partial_\tau \vec{\phi}^{\rm P}
\nonumber \\
{\cal L}_{{\rm N}\Delta\gamma} &=& - i { e_p \over 2m} \mu_{{\rm N}\Delta}
\bar{\psi}^\mu\gamma^5\gamma^\nu\tau_{{\rm N}\Delta}^3\psi F_{\mu\nu} 
+ \hbox{h.c.}
\end{eqnarray}
where $F_{\mu\nu}=\partial_\mu A_\nu -\partial_\nu A_\mu$, 
$\vec{\phi}^{\rm V}$=$\vec{\rho}$ or $\omega$, 
$\vec{\phi}^{\rm P}$=$\vec{\pi}$ or $\eta$,
$e_{\rm N}$=$e_p{(1+\tau^3)\over 2}$ and 
$\mu_{\rm N}$=${(1+\kappa^{\rm is})\over 2}
+{(1+\kappa^{\rm iv})\over 2}\tau^3$ = 1+$\kappa_{\rm N}$
with $\kappa^{\rm is}$=$-0.12$ and $\kappa^{\rm iv}$=$3.706$.

These Lagrangians describe the bare coupling of mass renormalized fields.
The form factors describing the coupling constant renormalization have been
calculated \cite{JF95} as a coupled set of integral equations yielding
results which can be accurately parameterized as monopoles with the regularization
scales shown in Table~1. 
These renormalized couplings are implemented by replacing the
Lagrangians with the vertex functions describing their dressed counterparts,
\begin{eqnarray}\label{eq3.1}
\Gamma_{{\rm NN}\pi}&=& 
- ig_{{\rm NN}\pi}F_{{\rm NN}\pi}   
\bar{\psi}[
  \lambda\gamma^5
 - (1-\lambda)\displaystyle{1\over 2m}\gamma^5\gamma^{\mu}(i\partial_\mu)]
\psi \vec{\pi}.\vec{\tau}
\nonumber \\
\Gamma_{{\rm NN}\eta}&=&
-{g_{{\rm NN}\eta}F_{{\rm NN}\eta}  \over 2m} \bar{\psi} \gamma^5\gamma^{\mu}\partial_\mu \psi \eta
\nonumber \\
\Gamma_{{\rm NN}\rho}&=&- g_{{\rm NN}\rho} 
\bar{\psi} [F^{(1)}_{{\rm NN}\rho} \gamma^\mu - {\kappa_\rho F^{(2)}_{{\rm NN}\rho} \over 2m}\sigma^{\mu\nu} \partial_\nu]
\psi \vec{\rho}_\mu.\vec{\tau}
\nonumber \\
\Gamma_{{\rm NN}\omega}&=& - g_{{\rm NN}\omega} 
\bar{\psi}[F^{(1)}_{{\rm NN}\omega}\gamma^\mu - {\kappa_\omega F^{(2)}_{{\rm NN}\omega}\over 2m}\sigma^{\mu\nu} \partial_\nu]
\psi \omega_\mu 
\nonumber \\
\Gamma_{{\rm NN}\delta}&=&-
g_{{\rm NN}\delta} F_{{\rm NN}\delta} 
\bar{\psi} \psi \vec{\delta}.\vec{\tau}
\nonumber \\
\Gamma_{{\rm NN}\epsilon}&=&-
g_{{\rm NN}\epsilon} F_{{\rm NN}\epsilon} 
\bar{\psi} \psi \epsilon
\nonumber \\
\Gamma_{{\rm N}\Delta\pi}&=&- {g_{{\rm N}\Delta\pi}F_{{\rm N}\Delta\pi}\over 2m}  
\bar{\psi}^\mu \vec{\tau}_{{\rm N}\Delta} \psi \partial_\mu \vec{\pi} 
+ \hbox{h.c.}
\nonumber \\
\Gamma_{{\rm N}\Delta\rho}&=&- i {g_{{\rm NN}\rho}G_{{\rm N}\Delta\rho}\over 2m}  
{g_{{\rm N}\Delta\pi} \over g_{{\rm NN}\pi}}
\bar{\psi}^\mu\gamma^5\gamma^\nu\vec{\tau}_{{\rm N}\Delta}\psi
\vec{\rho}_{\mu\nu} + \hbox{h.c.}
\end{eqnarray}
where 
$G_{{\rm N}\Delta\rho}$=$F^{(1)}_{{\rm N}\Delta\rho}$+$\kappa_\rho F^{(2)}_{{\rm N}\Delta\rho}$ 
and we normalize all form factors as $F(0)=1$.
As in other exchange current applications \cite{DR89} the 
experimentally unknown values of $g_{{\rm N}\Delta\rho}=20.73$ and 
$\mu_{{\rm N}\Delta}$=3.993 are fixed according to SU(6) \cite{MA84,CD81} 
and vector-meson dominance as,
\begin{equation}\label{eq3.3}
\mu_{{\rm N}\Delta}=\mu_{\rm N}^{{\rm iv}}{g_{{\rm N}\Delta\pi} \over g_{{\rm NN}\pi}} 
\qquad
g_{{\rm N}\Delta\rho}= g_{{\rm NN}\rho}(1+\kappa_\rho) 
{g_{{\rm N}\Delta\pi} \over g_{{\rm NN}\pi}}
\end{equation}
where $\mu_{\rm N}^{{\rm iv}} = {1\over 2}G_{{\rm M}}^{{\rm V}}(0)=2.353$.
Note that the tensor couplings $\kappa_\rho$ and $\kappa_\omega$ 
are absent from the Lagrangians of eq.~(\ref{eq3.1A}) but appear in the dressed
vertex functions of eq.~(\ref{eq3.1}) since they are directly
computed from the loop-integrals appearing in the form factor calculation \cite{JF95}.
A similar consideration for the electromagnetic form factors is obviously not required 
for the real photon.

The NN$\alpha$ properties are taken from the fit of the RuhrPot two-nucleon 
interaction \cite{RuhrPot} to the NN-scattering data. The form factor scales 
adopted in ref~\cite{RuhrPot} were actually calculated within a non-relativistic 
framework \cite{SD91,KG83}, but the recent relativistic calculation \cite{JF95} 
has confirmed the parameterizations.  We acknowledge some 
ambiguity in the signs of the PV$\gamma$ coupling constants \cite{MC74,JT89} 
but adopt $g_{\rho\pi\gamma}$=0.53, $g_{\omega\pi\gamma}$=2.58,
$g_{\rho\eta\gamma}$=1.39 and $g_{\omega\eta\gamma}$=0.15,
as reported in ref~\cite{PDG}. 
We use the experimental result $g_{{\rm N}\Delta\pi}$ = 28.85, which is consistent 
with the Chew-Low \cite{GB75} and strong-coupling \cite{AG76} models and, moreover,
with the form factor calculations \cite{JF95}.
We will not fiddle with these values 
in order to optimize selected experimental results since this would spoil 
the consistency between the calculation of the meson-baryon 
form-factors, the NN-interaction and the exchange currents.  
\end{multicols}\widetext
\vskip -0.5 true cm
\LeftLineBreak
\begin{table}
{\offinterlineskip \tabskip=0pt \halign{ \strut
\vrule \quad#&\quad 
\vrule \quad#&\quad 
\vrule \quad#&\quad 
\vrule \quad#&\quad 
\vrule \quad#&\quad 
\vrule \quad#&\quad 
\vrule \quad#&\quad 
\vrule #\cr
\noalign{\hrule}
 $\beta$ & 
 $m_{\beta}$~(MeV) & 
 $g_{{\rm NN}\beta}$ & 
 $\kappa_{\beta}$~(GeV)$^{-2}$ & 
 $\Sigma_{\beta}$ &  
 $\kappa_{\Sigma_{\beta}}$ &  
 $g_{\Delta{\rm N}\beta}$ &\cr
\noalign{\hrule}
 $\pi$       & 136.5 & 12.922 &  -    & 49.516 & -      &  28.85 &\cr
 $\eta$      & 548.8 &  6.015 &  -    & -      & -      &  -     &\cr
 $\rho$      & 776.9 &  1.651 & 6.400 & 0.0124 & 28.105 &  20.73 &\cr
 $\omega$    & 782.4 &  4.945 & 1.088 & 12.379 & 0.4334 &  -     &\cr
 $\delta$    & 983.0 &  6.043 & -     & -      & -      &  -     &\cr
 $\epsilon$  & 975.0 & 10.567 & -     & 5.6911 & -      &  -     &\cr
\noalign{\hrule}
}} 
\vskip 0.5 true cm
\caption{RuhrPot parameters adopted in the present calculation. All NN-meson
form factors are taken from direct calculation. For the $\epsilon$ meson
this requires a meson scale of $\Lambda_1$=0.6~GeV, whereas all other mesons
require $\Lambda_1$=0.8~GeV. (Further details can be found in [23,28]).
We adopt the experimental results
$g_{\rho\pi\gamma}$=0.53, $g_{\omega\pi\gamma}$=2.58,
$g_{\rho\eta\gamma}$=1.39, $g_{\omega\eta\gamma}$=0.15 and
$\kappa^{is}$=-0.12, $\kappa^{iv}$=3.706. SU(6) and vector
dominance indicate $\mu_{{\rm N}\Delta}$=3.993.
}
\end{table}
\vskip -0.5 true cm
\RightLineBreak
\begin{multicols}{2}

\subsection{Impulse and Exchange Currents}\label{sec3B}
We describe here the impulse and meson-exchange currents 
$J_{\rm eff}$, as required in eq.~(\ref{eq2.22}). We adopt a
partition of Hilbert spaces into meson+resonance+anti-nucleon -vacuum and 
-existing parts, as described in section~\ref{sec2}. In the present work we 
confine our attention to leading-order exchange currents involving the 
electromagnetic coupling to the NN, $\bar{\rm N}$N, 
PV=$\rho\pi$, $\omega\pi$, $\rho\eta$ and $\omega\eta$ 
and N$\Delta$ currents, so that,
\begin{equation}\label{eq3.4}
J_{\rm eff}  = J^{{\rm N}{\rm N}}_{\rm eff} 
                   + J^{\bar{\rm N}{\rm N}}_{\rm eff}
                   + J^{\rm PV}_{\rm eff}
                   + J^{{\rm N}\Delta}_{\rm eff}
\end{equation}
The effective current can therefore be derived unambiguously from 
eqs.~(\ref{eq3.2}), (\ref{eq3.1}), (\ref{eq2.12}), (\ref{eq2.18}-\ref{eq2.19}) 
and (\ref{eq2.21}).  Throughout we describe the momenta of a meson with mass 
$m_\beta$ with,
\begin{eqnarray}\label{eq3.5}
&&\vec{q}_{1}\!=\!\vec{p}_{3}-\vec{p}_{1},              \hskip 2 mm
  \omega_1   \!=\!\sqrt{\vec{q}_{1}^{\,2} + m_\beta^2}, \hskip 2 mm 
  q_{1}^0    \!=\! E_3 - E_1,                           \hskip 2 mm 
  Q_{1}^2    \!=\! -q_{1}^2 
\nonumber \\
&&\vec{q}_{2}\!=\!\vec{p}_{4}-\vec{p}_{2},              \hskip 2 mm 
 \omega_2    \!=\!\sqrt{\vec{q}_{2}^{\,2} + m_\beta^2}, \hskip 2 mm 
 q_{2}^0     \!=\! E_4 - E_2,                           \hskip 2 mm 
 Q_{2}^2     \!=\! -q_{2}^2
\nonumber \\ &&
\end{eqnarray}
and, denoting the nucleon and $\Delta$-isobar masses as $m$ and $m_{\Delta}$,
we condense our notation with,
\begin{eqnarray}\label{eq3.6}
\vec{p}_{ik} &=&\cases{\vec{p}_{i}-\vec{k} & for $i$=1, 2 \cr
                       \vec{p}_{i}+\vec{k} & for $i$=3, 4 \cr}
\nonumber \\ 
E_{       ik}=\sqrt{\vec{p}_{ik}+m^2         }, 
&&     \quad {\cal E}_{ik}= E_{ik} + m,
\nonumber \\ 
E_{\Delta ik}=\sqrt{\vec{p}_{ik}+m_{\Delta}^2}, 
&&     \quad {\cal E}_{\Delta ik}= E_{\Delta ik} + m_{\Delta}
\end{eqnarray}
\subsubsection{Impulse and Exchange currents with the Relativistic 
               NN$\gamma$ vertex}\label{sec3B1}
For the partition  of Hilbert spaces defined in section~\ref{sec2B}, 
all contributions involving a vertex where the photon couples to the 
nucleon current must satisfy 
$\lambda J_{\rm NN} \eta = \eta J_{\rm NN} \lambda = 0$, 
so that eq.~(\ref{eq3.4}) requires,
\begin{figure}
\vbox to 5.0 true in {\hbox{\hskip 0 mm \epsfxsize=3.4 true in \epsfbox{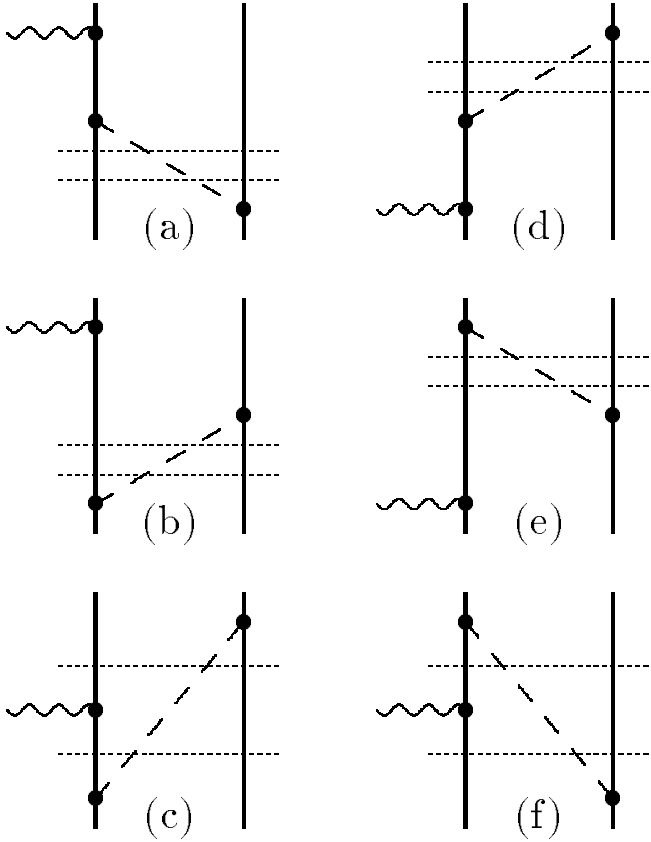}}
\vglue 3 true mm
\narrowtext
\caption{ 
NN$\gamma$ wave function re-orthonormalization 
and meson-recoil exchange currents. These currents are necessary to 
preserve the orthonormality of the initial- and final-state wave functions.
}}
\end{figure}
\par\bigskip
\begin{eqnarray}\label{eq3.7}
\langle \vec{p}_3 \vec{p}_4\vert  
\left[J^{\rm NN}_{\rm eff}\right]^\mu  
\vert \vec{p}_1 \vec{p}_2\rangle  
&=& 
  J^\mu_{{\rm NN}\gamma}[1](\vec{p}_{1},\vec{p}_{3}) \delta(\vec{p}_4-\vec{p}_2)
\nonumber \\
&+& J^\mu_{{\rm NN}\gamma}[2](\vec{p}_{2},\vec{p}_{4}) \delta(\vec{p}_3-\vec{p}_1)
\nonumber \\
&+& \langle \vec{p}_3 \vec{p}_4\vert  J^\mu_{\rm wfrr} \vert \vec{p}_1 \vec{p}_2\rangle  
\end{eqnarray}
where the first two terms describe the impulse currents for nucleons 1 and 2, 
and the last term denotes the wave function re-orthonormalization and meson 
recoil currents,
\begin{eqnarray}\label{eq3.8}
&&\langle \vec{p}_3 \vec{p}_4\vert  J^\mu_{\rm wfrr} 
   \vert \vec{p}_1 \vec{p}_2\rangle  
= -g_{\sigma\tau}  \sum_{\beta} \Bigl[
\nonumber \\ &&
D^{\rm NN\beta}_{\rm abc} 
J^\mu_{{\rm NN}\gamma}[1](\vec{p}_{3k},\vec{p}_{3}) 
H^{\sigma}_{{\rm NN}\beta}[1](\vec{p}_{1},\vec{p}_{3k}) 
H^{\tau}_{{\rm NN}\beta}[2](\vec{p}_{2},\vec{p}_{4})  
\nonumber \\ +&&
D^{\rm NN\beta}_{\rm def}
H^{\sigma}_{{\rm NN}\beta}[1](\vec{p}_{1k},\vec{p}_{3}) 
H^{\tau}_{{\rm NN}\beta}[2](\vec{p}_{2},\vec{p}_{4})  
J^\mu_{{\rm NN}\gamma}[1](\vec{p}_{1k},\vec{p}_{1}) 
\Bigr] 
\nonumber \\ + && (1,3 \rightleftharpoons 2,4)
\end{eqnarray}
where $\beta$ = $\pi$, $\eta$, $\rho$, $\omega$, $\delta$ and $\epsilon$, and 
the factor $ -g_{\sigma\tau}$ resulting from the contraction of the vector 
meson polarizations and all references to the Lorentz  indices $\sigma$ and 
$\tau$ are to be ignored for the scalar mesons. Explicit expressions for the 
vertex functions $H_{{\rm NN}\beta}$ and current $J_{{\rm NN}\gamma}$ are 
supplied in app.~B. The propagator functions are labeled in correspondence
to Fig.~4 and are 
defined as,
\begin{eqnarray}\label{eq3.9}
     D^\beta_{\rm abc}  &=&
   {-{1\over 2} \over [E_3+E_4-E_{3k}-E_2-\omega_2] [E_1-E_{3k}-\omega_2] }
\nonumber \\ &&
 + {1\over [E_4-E_2-\omega_2][E_1-E_{3k}-\omega_2]}
\nonumber \\ &&
 +  {-{1\over 2} \over [E_3-E_1-\omega_2] [E_2-E_4-\omega_2] }
\nonumber \\ D^\beta_{\rm def}  &=&
   {-{1\over 2} \over [E_3-E_{1k}-\omega_2][E_1+E_2-E_{1k}-E_4-\omega_2] }
\nonumber \\ &&
 + {1\over [E_3-E_{1k}-\omega_2][E_2-E_4-\omega_2]}
\nonumber \\ &&
 +  {-{1\over 2} \over [E_4-E_2-\omega_2] [E_1-E_3-\omega_2] }
\end{eqnarray}
In the static limit we note that 
$D^\beta_{\rm abc}= D^\beta_{\rm def}= 0$ 
so that the wave function renormalization and meson recoil exchange currents 
$J^\mu_{\rm wfrr}$ simply vanish. The same conclusion can be reached within
the soft-photon approximation in the barycentric frame.

Since we will avoid these approximations, we are forced to accept
that a relativistic description of the photon coupling to
the positive-frequency components of the impulse current 
$J^\mu_{{\rm NN}\gamma}$ necessarily leads to an effective 
current density $J_{\rm eff}$ comprising both one- and two-body operators.
These two-body contributions have, to date, never
been explicitly included in any of the bremsstrahlung calculations 
that seek to include the relativistic components of the
NN$\gamma$ vertex.
\subsubsection{Pair Currents}\label{sec3B2}
The sum of one-body impulse-currents $J^\mu_{{\rm NN}\gamma}$ and
the wave function re-orthonormalization and meson-recoil exchange
currents $J^\mu_{\rm wfrr}$ do not exhaust the requirements needed 
to obtain a relativistic description of the photon coupling to the 
nucleon current density since the off-shell nucleon comprises 
a linear super-position of positive and (so far neglected) negative
frequency components. In the Feynman-St\"uckelberg approach, the 
negative-frequency components of the off-shell nucleon field are 
interpreted as anti-particles, so we are led to introduce the photon 
coupling to the N$\bar{\rm N}$-pair creation and annihilation currents. 

Within our partition  of Hilbert spaces, the photon coupling to the
N$\bar{\rm N}$-pair creation and annihilation vertices must satisfy
$\eta J_{\bar{\rm N}N} \eta = 0$, so that 
eq.~(\ref{eq3.4}) requires,
\begin{eqnarray}\label{eq3.10}
&&\langle \vec{p}_3 \vec{p}_4\vert  
\left[ J^{\bar{\rm N}{\rm N}}_{\rm eff}\right]^\mu
\vert \vec{p}_1 \vec{p}_2\rangle  
= -g_{\sigma\tau}  \sum_{\beta} \Bigl[
\nonumber \\ &&
D^{{\rm N}\bar{\rm N}\beta}_{\rm abc} 
H^{\sigma}_{{\rm N}\bar{\rm N}\beta}[1](\vec{p}_{1},\vec{p}_{3k}) 
H^{\tau}_{{\rm NN}\beta}[2](\vec{p}_{2},\vec{p}_{4})  
J^\mu_{\bar{\rm N}{\rm N}\gamma}[1](\vec{p}_{3k},\vec{p}_{3}) 
\nonumber \\ +&&
D^{{\rm N}\bar{\rm N}\beta}_{\rm def}
J^\mu_{{\rm N}\bar{\rm N}\gamma}[1](\vec{p}_{1k},\vec{p}_{1}) 
H^{\sigma}_{\bar{\rm N}{\rm N}\beta}[1](\vec{p}_{1k},\vec{p}_{3}) 
H^{\tau}_{{\rm NN}\beta}[2](\vec{p}_{2},\vec{p}_{4})  
\Bigr]
\nonumber \\ + && (1,3 \rightleftharpoons 2,4)
\end{eqnarray}
\par\noindent
\begin{figure}
\vbox to 5.5 true in {\hbox{\hskip 0 mm \epsfxsize=3.4 true in \epsfbox{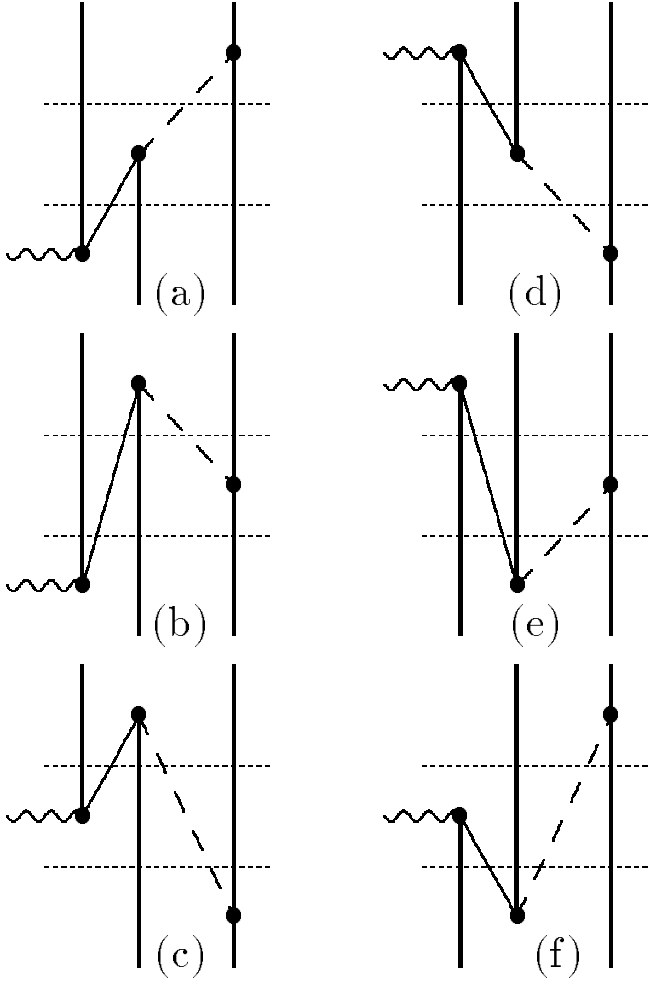}}
\vglue 3 true mm
\caption{ 
${\rm N}\bar{\rm N}$ pair creation and annihilation
meson-exchange currents. These currents are necessary for a relativistic
description of the NN$\gamma$ vertex.
}}
\end{figure}
\par\bigskip\noindent
where $\beta$ = $\pi$, $\eta$, $\rho$, $\omega$, $\delta$ and $\epsilon$,
and the factor $ -g_{\sigma\tau}$ and all references to the Lorentz 
indices $\sigma$ and $\tau$ are to be ignored for the scalar mesons. 
Explicit expressions for the pair-creation and -annihilation currents 
$J_{{\rm N}\bar{\rm N}\gamma}$ and $J_{\bar{\rm N}{\rm N}\gamma}$
and all required vertex functions are described in app.~B. 
\par
The propagator functions are labeled in correspondence to Fig.~5 and are defined as,
\begin{eqnarray}\label{eq3.11}
D^\beta_{\rm abc} 
&=&{1\over [-E_{3k}-E_1-\omega_2][ E_4 - E_{3k} - E_1 - E_2]}
\nonumber \\ &&
 + {1\over [E_4 - E_2 - \omega_2][E_4 - E_{3k} - E_1 - E_2]}
\nonumber \\ && 
+ {1\over [-E_{3k} - E_1 - \omega_2] [E_2 - \omega_2 - E_4]}
\nonumber \\ 
D^\beta_{\rm def} 
&=&{1\over [E_2 -E_{1k} - E_3 - E_4] [-E_{1k} - E_3 - \omega_2]}
\nonumber \\ &&
+{1\over [E_2 -E_{1k} - E_3 - E_4] [E_2 - \omega_2 - E_4]}
\nonumber \\ && 
+ {1\over [E_4 - \omega_2 - E_2] [- E_{1k} -E_3 - \omega_2]}
\end{eqnarray}
We will adopt these expressions for our numerical work. Nonetheless, it is 
interesting to consider the corresponding result under various approximations.
If we demand energy conservation across the current matrix elements, then
eq.~(\ref{eq3.11}) reduces to,
\begin{eqnarray}\label{eq3.12}
D^\beta_{\rm abc} &=& {\!-\!2\omega_2\over (E_3 \!+\! k \!+\! E_{3k})(q_2^2 \!-\! m_\beta^2)} 
\,\,{\buildrel \rm nr \over \sim}\,\, {2 \over \omega_2 (2m \!+\! k)} 
\,\,{\buildrel \rm spa\over \sim}\,\, {1 \over \omega_2 m} \nonumber \\
D^\beta_{\rm def} &=& {\!-\!2\omega_2\over (E_1 \!-\! k \!+\! E_{1k})(q_2^2 \!-\! m_\beta^2)}
\,\,{\buildrel \rm nr \over \sim}\,\, {2 \over \omega_2 (2m \!-\! k)} 
\,\,{\buildrel \rm spa\over \sim}\,\, {1 \over \omega_2 m} 
\nonumber \\ &&
\end{eqnarray}
where we also provide the static-limit and soft-photon reductions.  
The corresponding $\pi$-exchange contribution under such approximations
are,
\end{multicols}\widetext
\LeftLineBreak
\begin{eqnarray}\label{eq3.13}
\langle \vec{p}_3 \vec{p}_4\vert  
\left[ J^{\bar{\rm N}{\rm N}}_{\rm eff}\right]^\mu
\vert \vec{p}_1 \vec{p}_2\rangle  
& \,\,{\buildrel \rm nr  \over \sim}\,\, &
{\lambda e_p g_{{\rm NN}\pi}^2 F_{{\rm NN}\pi}^2(\vec{q}_2) 
    \over  (2\pi)^6 4m(\vec{q}_2^2+m_\pi^2)}
\vec{\sigma}_1
(\vec{\sigma}_2.\vec{q}_2)
\Biggl\{ 
  { (1 + \tau_1^z)\vec{\tau}_1.\vec{\tau}_2 \over (2m - k)}
- { \vec{\tau}_1.\vec{\tau}_2 (1 + \tau_1^z) \over (2m + k)}   
\Biggr\} 
+ (1\rightleftharpoons 2)
\nonumber \\
\langle \vec{p}_3 \vec{p}_4\vert  
\left[ J^{\bar{\rm N}{\rm N}}_{\rm eff}\right]^\mu
\vert \vec{p}_1 \vec{p}_2\rangle  
& \,\,{\buildrel \rm spa \over \sim}\,\, &
{-\lambda e_p g^2_{{\rm NN}\pi} F_{{\rm NN}\pi}^2(\vec{q}_2)
\over (2\pi)^6 4m^2 (\vec{q}_2^2+m_\pi^2)} \vec{\sigma}_1 (\vec{\sigma}_2.\vec{q}_2) 
(i\vec{\tau}_1\times\vec{\tau}_2)^3
+(1\rightleftharpoons 2)
\end{eqnarray}
\RightLineBreak
\begin{multicols}{2}
Both of these results 
scale linearly with the parameter $\lambda$ controlling the admixture of
ps- and pv-couplings in the NN$\pi$ Lagrangian, but the well-known isovector 
structure of the non-relativistic pair currents holds only for soft photons.
Since the data with which we will compare our results was planned to maximize
the photon energy, we anticipate a non-negligible contribution fom the
isoscalar components of eq~(\ref{eq3.10}). This offers the possibility of
studying $\lambda$ without the complications of describing the many 
leading-order exchange currents that contribute to np-bremsstrahlung.

\subsubsection{PV$\gamma$ Currents}\label{sec3B3}
The $\omega$ meson (782.4~MeV) decays as $\omega\rightarrow\pi^0\gamma$ 
with an 8.7\% branching ratio and indicates the coupling constant 
of $g_{\omega\pi\gamma}$=2.58. As such, the $\omega\pi\gamma$ exchange
currents can be expected to make a non-trivial contribution to both
\begin{figure}
\vbox to 5.0 true in {\hbox{\hskip 0 mm \epsfxsize=3.4 true in \epsfbox{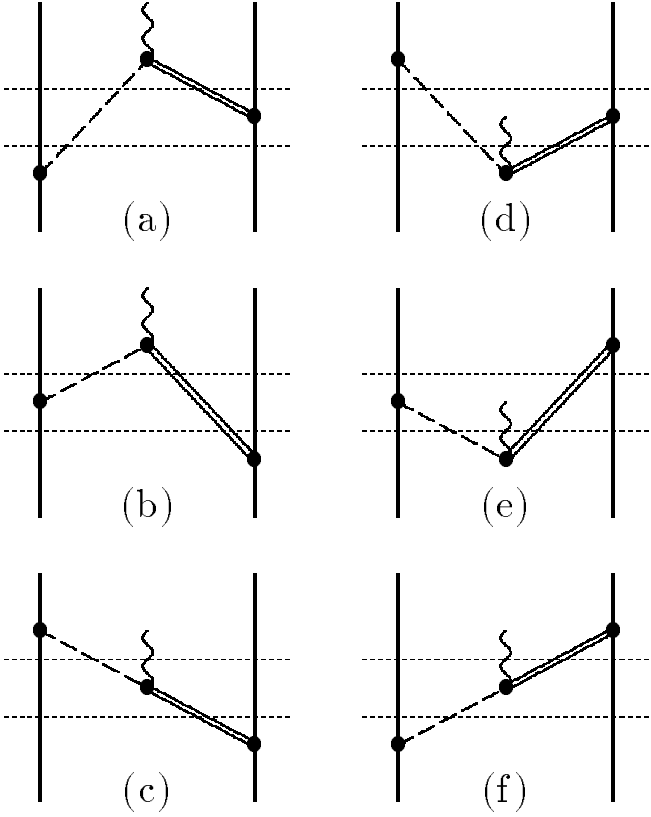}}
\vglue 3 true mm
\narrowtext
\caption{ 
VP$\gamma$ = $\rho\pi\gamma$ + $\omega\pi\gamma$ + $\rho\eta\gamma$ 
+ $\omega\eta\gamma$ exchange currents.  When energy is conserved 
across these current matrix elements, the time-ordered graphs exactly 
sum to form the corresponding Feynman  diagrams.
}}
\end{figure}
\par\bigskip\noindent
the pp- and np-bremsstrahlung observables.
Analogous arguments indicate that the $\rho\pi\gamma$ contributions
will be large in np-bremsstrahlung, and perhaps also of some lesser
importance in pp-bremsstrahlung.

Our desire to preserve complete consistency between the form factors,
NN-interaction and the exchange currents leads us to introduce 
all leading-order exchange currents describing the photon coupling to the 
decay of all vector-meson mesons present in the form factor and
NN-interaction calculations. We therefore include the ${\rm PV}\gamma$ = 
$\rho\pi\gamma$ + $\omega\pi\gamma$ + $\rho\eta\gamma$ + $\omega\eta\gamma$
exchange currents as shown in Fig.~6.  Each of these vertices
satisfies $\eta J_{\bar{\rm PV}} \eta = 0$, so that after making use of
eqs.~(\ref{eq3.2}), (\ref{eq2.12}) and (\ref{eq2.21}), eq.~(\ref{eq3.4}) 
requires,
\begin{eqnarray}\label{eq3.14}
&&\langle \vec{p}_3 \vec{p}_4\vert  
\left[ J^{\rm VP}_{\rm eff}\right]^\mu 
\vert \vec{p}_1 \vec{p}_2\rangle  
=
{\sqrt{4\omega_{\rm P}(\vec{q}_1)\omega_{\rm V}(\vec{q}_2)}  
\over (2\pi)^3m_{\rm V} [q_1^2-m_{\rm P}^2] [q_2^2-m_{\rm V}^2] }
\nonumber \\ &&\times
  H_{\rm NNP}[1](\vec{p}_{1},\vec{p}_{3}) 
  \epsilon_\nu(\hat{q}_2,\lambda_{\rm V})
  H^\nu_{\rm NNV}[2](\vec{p}_{2},\vec{p}_{4})
  J^\mu_{{\rm VP}\gamma}(q_1,q_2)
\nonumber \\ &&
+ (1,3 \rightleftharpoons 2,4)
\end{eqnarray}
where  $J_{{\rm VP}\gamma}$ is given in App.~B. We require only the 3-vector 
current, for which the relativistic form is,
\begin{eqnarray}\label{eq3.15}
&& \langle \vec{p}_3 \vec{p}_4\vert  
\vec{J}^{\rm VP}_{\rm eff}
\vert \vec{p}_1 \vec{p}_2\rangle  
= 
- { e_p g_{{\rm VP}\gamma} F_{{\rm VP}\gamma}
  \sqrt{4\omega_{\rm P}(\vec{q}_1)\omega_{\rm V}(\vec{q}_2)}  
  \over (2\pi)^3m_{\rm V} [q_1^2-m_{\rm P}^2] [q_2^2-m_{\rm V}^2] } 
\nonumber \\ \times  &&
H_{\rm NNP}[1](\vec{p}_{1},\vec{p}_{3}) 
\Bigl\{
H^0_{\rm NNV}[2](\vec{p}_{2},\vec{p}_{4}) [\vec{q}_1\times\vec{q}_2]
\nonumber \\  &&
\qquad +
\vec{H}_{\rm NNV}[2](\vec{p}_{2},\vec{p}_{4})\times [(q_2)_0 \vec{q}_1 - (q_1)_0 \vec{q}_2]
\Bigr\}
\nonumber \\  &&
+ (1,3\rightleftharpoons 2,4)
\end{eqnarray}
We will not resort to the non-relativistic limit, but we realize it implies 
$q_1^0$=$q_2^0$=0, so that we recover the well-known result for the
emission of a real photon of momentum  $\vec{k}$,
\begin{eqnarray}\label{eq3.16}
&&\langle \vec{p}_3 \vec{p}_4\vert  
\vec{J}^{\rm VP}_{\rm eff}
\vert \vec{p}_1 \vec{p}_2\rangle  
\,\,{\buildrel \rm nr \over \sim}\,\, 
{ i e_p g_{{\rm VP}\gamma} g_{\rm NNV} g_{\rm NNP} 
\over 
(2\pi)^6 2m_{\rm V}m [\vec{q}_1^2+m_{\rm P}^2] [\vec{q}_2^2+m_{\rm V}^2] 
}
\nonumber \\ &&
\times
(\vec{\sigma}_1.\vec{q}_1)
 (\vec{q}_1\times\vec{q}_2)
(\vec{\tau}_1)_{\rm P}(\vec{\tau}_2)_{\rm V}
+ (1,3\rightleftharpoons 2,4)
\end{eqnarray}
where
$(\vec{\tau}_1)_{\rm P}(\vec{\tau}_2)_{\rm V}$ = $\vec{\tau}_1.\vec{\tau}_2$, 
$\vec{\tau}_1^0$, $\vec{\tau}_2^0$ and $\openone$ for the 
$\rho\pi\gamma$, $\omega\pi\gamma$, $\rho\eta\gamma$  and $\omega\eta\gamma$ 
currents respectively.
\subsubsection{N$\Delta\gamma$ Exchange Currents}\label{sec3B4}
All contributions involving a vertex where the photon couples to the 
N$\Delta$ current must satisfy 
$\eta J_{\Delta{\rm N}} \eta = \eta J_{{\rm N}\Delta} \eta = 0$, 
so that eq.~(\ref{eq3.4}) requires,
\begin{eqnarray}\label{eq3.17}
&&\langle \vec{p}_3 \vec{p}_4\vert  
\vec{J}^{{\rm N}\Delta}_{\rm eff} 
\vert \vec{p}_1 \vec{p}_2\rangle   
= -g_{\sigma\tau} \sum_{\beta}\Bigl[
\nonumber \\ && 
D^\beta_{\rm abc} 
    \vec{J}_{{\rm N}\Delta\gamma}[1](\vec{p}_{3k},\vec{p}_{3}) 
    H^\sigma_{\Delta{\rm N}\beta}[1](\vec{p}_{1},\vec{p}_{3k}) 
    H^\tau_{{\rm NN}\beta}[2](\vec{p}_{2},\vec{p}_{4}) 
\nonumber \\ + &&
D^\beta_{\rm def} 
    H^\sigma_{{\rm N}\Delta\beta}[1](\vec{p}_{1k},\vec{p}_{3}) 
    \vec{J}_{\Delta{\rm N}\gamma}[1](\vec{p}_{1},\vec{p}_{1k}) 
    H^\tau_{{\rm NN}\beta}[2](\vec{p}_{2},\vec{p}_{4}) 
\Bigr]
\nonumber \\ + &&  (1,3 \rightleftharpoons 2,4)
\end{eqnarray}  
where $\beta$=$\vec{\pi}$ or $\vec{\rho}$ and
the factor $ -g_{\sigma\tau}$ resulting from the contraction of the 
$\rho$-meson polarizations and all references to the Lorentz indices $\sigma$ 
and $\tau$ are to be ignored for $\beta$=$\pi$.  Explicit expressions 
for the vertex functions and currents shown in eq.~(\ref{eq3.17}) are supplied 
in app.~B. We ignore all negative frequency resonance contributions. 
The propagator functions are labeled in correspondence to 
Fig.~7 and are defined in analogy to the previous sections.
Using energy conservation for the current matrix elements and
introducing the $\Delta$ decay width $\Gamma_\Delta$ via \cite{EW88,HA75} 
$E_{\Delta k}\rightarrow E_{\Delta k}-i\Gamma_\Delta/2$,
the propagators reduce to,
\begin{eqnarray}\label{eq3.18}
D^\beta_{\rm abc}  &=&
{ 2\omega_2 \over (q_2^2-m_\beta^2)(E_3+k-E_{\Delta3k}+i\Gamma_\Delta/2)}
\nonumber \\
D^\beta_{\rm def} &=&
{ 2\omega_2 \over (q_2^2-m_\beta^2)(E_1-k-E_{\Delta1k}+i\Gamma_\Delta/2)}
\end{eqnarray}
An exact calculation of eq.~(\ref{eq3.17}) can be achieved with the use
of the vertex functions and currents given in App.~B. 
However, at present there exists some uncertainty in the coupling 
constants $g_{{\rm N}\Delta\rho}$ and $\mu_{{\rm N}\Delta}$, so that such a 
rigorous procedure is of limited interest.
We simplify matters by  dropping terms of order $p^2/(E+m)^2$ 
\par\noindent
\begin{figure}
\vbox to 5.6 true in {\hbox{\hskip 0 mm \epsfxsize=3.4 true in \epsfbox{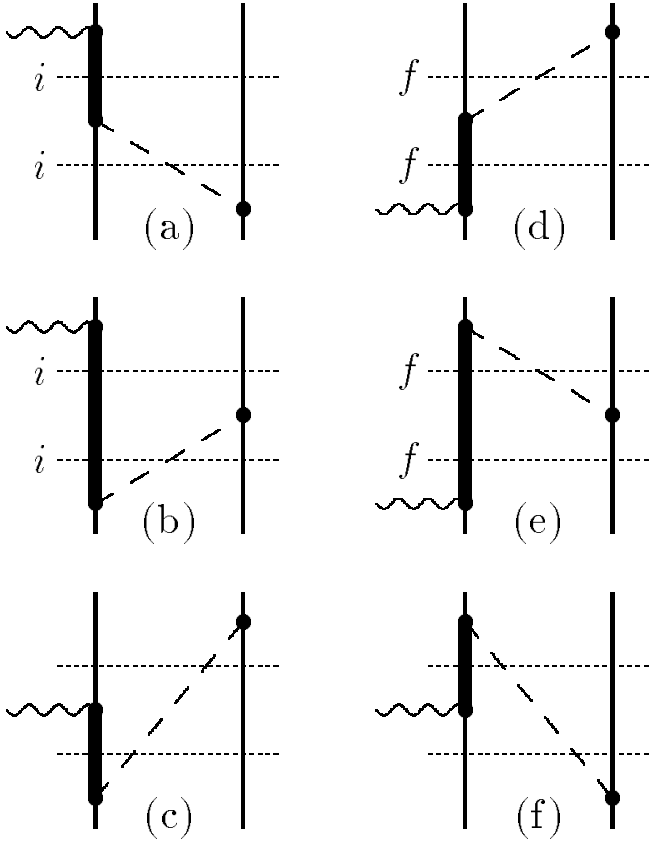}}
\vglue 3 true mm
\narrowtext
\caption{ 
N$\Delta\gamma$ $\pi-$ and $\rho-$meson exchange currents.
In the RuhrPot model the $\pi-$exchange contributions represent the largest of 
the two-body currents in pp-bremsstrahlung but the $\rho-$exchange contributions 
are very small.  When energy is conserved across these current matrix elements,
the time-ordered graphs exactly sum to form the corresponding Feynman  diagrams.
}}
\end{figure}
\par\bigskip\noindent
beyond leading 
order - an approximation which does not involve any $p/m-$expansion and should 
be accurate to within a few per-cent at energies below the $\pi$-production 
threshold. This is surely adequate for the first-order $S_{fi}$-matrix 
described in sec~\ref{sec2A}.
\par\noindent
We introduce the condensed notation,
\begin{eqnarray}\label{eq3.19}
\vec{\cal P}_{i} &=& {\vec{p}_{i}\over {\cal E}_i}, \quad
\vec{\cal P}_{\Delta k i}= {\vec{p}_{ik}\over {\cal E}_{\Delta ik}}, \quad
\vec{\cal Q}_{ij} = \vec{p}_{ik}\left({E_j\over m_{\Delta}}\right)-\vec{p}_{j},\quad
\nonumber \\
\vec{\cal K}_{i}&=&
\vec{p}_{ik}\Bigl({2E_{\Delta ik}-E_i\over m_{\Delta}}\Bigr) -\vec{p}_{i},\quad
\nonumber \\ 
\vec{\cal L}_{i}&=&
\left({\vec{p}_{ik}\over{\cal E}_{\Delta ik}}
     -{\vec{p}_{i} \over{\cal E}_{i}}\right)(m+m_{\Delta})
\qquad i=1,2,3,4.
\end{eqnarray}
and proceed to calculate separately the contributions from $\pi$- and 
$\rho$-exchange,
\begin{equation}\label{eq3.20}
\vec{J}^{{\rm N}\Delta}_{\rm eff} =
\vec{J}^{{\rm N}\Delta}_{\rm eff}(\pi) +
\vec{J}^{{\rm N}\Delta}_{\rm eff}(\rho) 
\end{equation}
\subsubsection{N$\Delta\gamma$ $\pi-$Exchange Currents}\label{sec3B5}
The effective current describing the excitation of intermediate 
$\Delta$-isobars through $\pi$-exchange can now be obtained from 
eqs.~(\ref{eq3.17}) and App.~B. For the real photon, we require 
only the spacial current, which takes the form,
\begin{eqnarray}\label{eq3.21}
&&\langle \vec{p}_3 \vec{p}_4\vert  
\vec{J}^{{\rm N}\Delta}_{\rm eff}(\pi) 
\vert \vec{p}_1 \vec{p}_2\rangle   
\nonumber \\
&=&
N_1\vec{\sigma}_2.(\vec{\cal P}_{4}\!-\! \vec{\cal P}_{2}) 
\Bigl[ 2 \tau_2^3 \!-\! (i\tau_1\!\times\!\tau_2)^3\Bigr] 
\nonumber \\ && \times 
\Bigl[ 
  (\vec{\cal K}_1\!+\!\vec{\cal L}_1)\!\times\!\vec{\cal Q}_{13} 
\!+\! (i\vec{\sigma}_1\!\times\!\vec{\cal Q}_{13})\!\times\!\vec{\cal L}_1 
\nonumber \\ && 
\!+\! i\vec{\sigma}_1\!\times\![(\vec{\cal K}_1\!-\!\vec{\cal L}_1)\!\times\!\vec{\cal Q}_{13}] 
\!-\! 2i\vec{\sigma}_1 [(\vec{\cal K}_1\!-\!\vec{\cal L}_1).\vec{\cal Q}_{13}]
\Bigr]
\nonumber \\  &\!+\!&
N_3 \vec{\sigma}_2.(\vec{\cal P}_{4}\!-\! \vec{\cal P}_{2}) 
\Bigl[ 2\tau_2^3 \!+\! (i\tau_1\!\times\!\tau_2)^3\Bigr] 
\nonumber \\ && \times 
\Bigl[
 (\vec{\cal K}_3\!+\!\vec{\cal L}_3)\!\times\!\vec{\cal Q}_{31} 
\!-\!(i\vec{\sigma}_1\!\times\!\vec{\cal Q}_{31})\!\times\!\vec{\cal L}_3 
\nonumber \\ && 
\!-\!i\vec{\sigma}_1\!\times\![(\vec{\cal K}_3\!-\!\vec{\cal L}_3)\!\times\!\vec{\cal Q}_{31}] 
\!+\! 2i\vec{\sigma}_1 [(\vec{\cal K}_3\!-\!\vec{\cal L}_3).\vec{\cal Q}_{31}]
\Bigr] 
\nonumber \\ &\!+\!& (1,3 \rightleftharpoons 2,4)
\end{eqnarray}
where,
\begin{eqnarray}\label{eq3.22}
N_{0}&& = {-ie_p \mu_{{\rm N}\Delta} g_{{\rm N}\Delta\pi}g_{{\rm NN}\pi}  
          \over (2\pi)^6 72 m^2}
\Bigl[
 {{\cal E}_{1}{\cal E}_{2}{\cal E}_{3}{\cal E}_{4}\over 16 E_1 E_2 E_3 E_4}
\Bigr]^{1\over 2}
{ F_{{\rm NN}\pi}F_{{\rm N}\Delta\pi} \over \omega_{\pi}(\vec{q}_{2}^2)},
\nonumber \\
N_{1} &&= {N_0 {\cal E}_{\Delta 1k}D^\pi_{\rm def} \over 2E_{\Delta 1k}},
\quad\qquad
N_{3} = {N_0 {\cal E}_{\Delta 3k}D^\pi_{\rm abc} \over 2E_{\Delta 3k}}
\end{eqnarray}
We will adopt eq.~(\ref{eq3.21}) for our numerical applications
and use it to consider the merit of various approximations that are
conventionally adopted to recover a simplified operator structure.

The first level of approximation involves taking the static limit and
ignoring the N-$\Delta$ mass difference in eq.~(\ref{eq3.19}). We call this
the {\it vertex static limit} approximation and note that it is
equivalent to casting eq.~(\ref{eq3.21}) as,
\begin{eqnarray}\label{eq3.23}
&&\langle \vec{p}_3 \vec{p}_4\vert  
\vec{J}^{{\rm N}\Delta}_{\rm eff}(\pi) 
\vert \vec{p}_1 \vec{p}_2\rangle   
\nonumber \\
&=&
(N_1+N_3){(\vec{\sigma}_2.\vec{q}_2)\over 2m} \bigl[
   4 \tau_2^3 \vec{q}_2 
 - (i\vec{\tau}_1\times\vec{\tau}_2)^3(i\vec{\sigma}_1\times\vec{q}_2)
\bigr]\times\vec{k}
\nonumber \\ &-&
2(N_1-N_3)  {(\vec{\sigma}_2.\vec{q}_2)\over 2m} \bigl[
   \tau_2^3(i\vec{\sigma}_1\times\vec{q}_2)
 - (i\vec{\tau}_1\times\vec{\tau}_2)^3 \vec{q}_2
\bigr]\times\vec{k}
\nonumber \\ &&
+  (1,3\rightleftharpoons 2,4)
\end{eqnarray}
The substantial simplification results primarily because the vertex static 
limit approximation indicates
${\cal K}_1$=${\cal L}_1$=$-\vec{k}$, and ${\cal K}_3$=${\cal L}_3$=$\vec{k}$,
so that all operator structures involving ${\cal K}_i - {\cal L}_i$ immediately 
vanish.  However, if we consider the static limit with the more reasonable 
approximation $m_\Delta\sim {4\over 3}m$, then we find
\begin{eqnarray}\label{eq3.24}
{\cal K}_1+{\cal L}_1&\sim&         - {17\over 8}\vec{k}-{1 \over 24}\vec{p}_1,
\qquad
{\cal K}_1-{\cal L}_1 \sim          - {3 \over 8}\vec{k}+{13\over 24}\vec{p}_1
\nonumber \\
{\cal K}_3+{\cal L}_3&\sim&\phantom{-}{17\over 8}\vec{k}-{1 \over 24}\vec{p}_3,
\qquad
{\cal K}_3-{\cal L}_3 \sim \phantom{-}{3 \over 8}\vec{k}+{13\over 24}\vec{p}_3
\nonumber \\ &&
\end{eqnarray}
so that the ${\cal K}_i + {\cal L}_i$ terms surviving in the vertex static 
limit are rather well approximated, but the neglected 
${\cal K}_i - {\cal L}_i$ terms are poorly represented.

The second level of approximation involves 
casting the complete expression in the static limit. This is the 
approximation we adopted in earlier work \cite{JE95,JE94}.  At the small 
momentum transfers relevant to the present numerical application, 
this {\it complete static limit} approximation will differ little from the 
vertex static limit and, if we further set $\Gamma_\Delta=0$
and drop the form factor dependence,
it is equivalent to casting eq.~(\ref{eq3.21}) in the simpler form,
\begin{eqnarray}\label{eq3.25}
&&\langle \vec{p}_3 \vec{p}_4\vert  
\vec{J}^{{\rm N}\Delta}_{\rm eff}(\pi) 
\vert \vec{p}_1 \vec{p}_2\rangle   
=
{-ie_p\mu_{{\rm N}\Delta}g_{{\rm N}\Delta\pi}g_{{\rm NN}\pi}\over(2\pi)^6 36m^3}
\nonumber \\ &&\times 
(\vec{\sigma}_2.\vec{q}_2)\Biggl\{ 
{ 4(m_{\Delta}-m)\vec{q}_2 + 2|\vec{k}| (i\vec{\sigma}_1\times\vec{q}_2) 
\over[(m_{\Delta}-m)^2-|\vec{k}|^2](\vec{q}_2^2+m_\pi^2)}\tau_2^3
\qquad\qquad \nonumber \\ && 
 \qquad\quad +
{ (m_{\Delta}-m) (i\vec{\sigma}_1\times\vec{q}_2) + 2|\vec{k}| \vec{q}_2
\over[(m_{\Delta}-m)^2-|\vec{k}|^2](\vec{q}_2^2+m_\pi^2)}(i\vec{\tau}_1\times\vec{\tau}_2)^3
\Biggr\}\times\vec{k}
\nonumber \\ &&
+ (1,3\rightleftharpoons 2,4)
\end{eqnarray}
The third and final approximation we consider is obtained by neglecting 
the $|\vec{k}|$-dependence in the complete static limit description of 
the baryon propagators. This {\it soft-photon approximation} gives
the conventional description \cite{DR89} of the N$\Delta\gamma(\pi)$ exchange 
current for the radiation of a photon of momentum $\vec{k}$,
\begin{eqnarray}\label{eq3.26}
&&\langle \vec{p}_3 \vec{p}_4\vert  
\vec{J}^{{\rm N}\Delta}_{\rm eff}(\pi) 
\vert \vec{p}_1 \vec{p}_2\rangle   
=
{-ie_p \mu_{{\rm N}\Delta} g_{{\rm N}\Delta\pi}  g_{{\rm NN}\pi} \over 
(2\pi)^6 36 m^3 (m_{\Delta}-m)}
\nonumber \\ &&
(\vec{\sigma}_2.\vec{q}_2)\Biggl\{ 
    {  4\tau_2^3\vec{q}_2 \over (\vec{q}_2^2+m_\pi^2)}
  + { (i\vec{\tau}_1\times\vec{\tau}_2)^3(i\vec{\sigma}_1\times\vec{q}_2) 
      \over (\vec{q}_2^2+m_\pi^2)}
\Biggr\}\times\vec{k}
\nonumber \\ &&
+ (1,3\rightleftharpoons 2,4)
\end{eqnarray}
We anticipate that this result will differ from the complete static limit 
whenever the photon energy is comparable to the N-$\Delta$ mass difference. 
\subsubsection{N$\Delta\gamma$ $\rho-$Exchange Currents}\label{sec3B6}
Within the soft-photon approximation, it is well known \cite{DR89}
that the N$\Delta\gamma$ $\rho$-exchange currents are small compared 
to the corresponding $\pi$-exchange currents, although the destructive 
interference between the two makes it necessary to include both.
In the present work, we anticipate the $\rho$-exchange contributions
to be less important than in other works because the RuhrPot model suggests
a very weak NN$\rho$ coupling constant.
(In particular, RuhrPot \cite{RuhrPot} uses $g^2_{{\rm NN}\rho}/4\pi$=0.2169
and $\kappa_\rho$=6.4, whereas (for example) Bonn~B \cite{Bonn} adopts 
$g^2_{{\rm NN}\rho}/4\pi$=0.92  and $\kappa_\rho$=6.1).

Since the N$\Delta\gamma$ $\rho$-exchange currents are expected to be small, we
proceed by taking the {\it vertex static limit} approximation from the outset.
In complete analogy to the derivation of eq.~(\ref{eq3.21}), we obtain,
\begin{eqnarray}\label{eq3.27}
\vec{J}^{{\rm N}\Delta}_{\rm eff}(\rho) =
\vec{J}^{{\rm N}\Delta}_{\rm eff}(\rho;{\rm E}) +
\vec{J}^{{\rm N}\Delta}_{\rm eff}(\rho;{\rm M}) 
\end{eqnarray}
where,
\end{multicols}\widetext
\LeftLineBreak
\begin{eqnarray}
\label{eq3.28}
\langle \vec{p}_3 \vec{p}_4\vert  
\vec{J}^{{\rm N}\Delta}_{\rm eff}(\rho;{\rm E}) 
&& \vert \vec{p}_1 \vec{p}_2\rangle   
= \bigl\{ \bigl[ 
 4N_{\rm E}^{(+)}(\vec{p}_{2}+\vec{p}_{4})\times\vec{q}_2
+2N_{\rm E}^{(-)} i\vec{\sigma}_1\times
               [(\vec{p}_{2}+\vec{p}_{4})\times\vec{q}_2]
\bigr] \tau_2^3 
\nonumber \\ && \bigl[ 
+2N_{\rm E}^{(-)}(\vec{p}_{2}+\vec{p}_{4})\times\vec{q}_2
+N_{\rm E}^{(+)} i\vec{\sigma}_1\times
               [(\vec{p}_{2}+\vec{p}_{4})\times\vec{q}_2]
\bigr] (i\vec{\tau}_1\times\vec{\tau}_2)^3 
\bigr\}\times \vec{k}
+ (1,3\rightleftharpoons 2,4)
\end{eqnarray}
and
\begin{eqnarray}
\label{eq3.29}
\langle \vec{p}_3 \vec{p}_4\vert  
\vec{J}^{{\rm N}\Delta}_{\rm eff}(\rho;{\rm M}) 
&&\vert \vec{p}_1 \vec{p}_2\rangle    
= \bigr\{ \bigl[ 
 4N_{\rm M}^{(+)}\bigl[(i\vec{\sigma}_2\times\vec{q}_2)\times \vec{q}_2\bigr]
-2N_{\rm M}^{(-)}\bigl[\vec{\sigma}_1\times 
                      [(\vec{\sigma}_2\times\vec{q}_2)\times \vec{q}_2]\bigr]
\bigr]\tau_2^3 
\nonumber \\ 
+&&\bigl[ 
2N_{\rm M}^{(-)}\bigl[(i\vec{\sigma}_2\times\vec{q}_2)\times \vec{q}_2\bigr]
-N_{\rm M}^{(+)}\bigl[\vec{\sigma}_1\times
                     [(\vec{\sigma}_2\times\vec{q}_2)\times \vec{q}_2]\bigr]
\bigr] (i\vec{\tau}_1\times\vec{\tau}_2)^3 
\bigr\}\times \vec{k}
+ (1,3\rightleftharpoons 2,4) 
\end{eqnarray}
with,
\begin{eqnarray}\label{eq3.30}
N_{\rm M}^{(+)} &=&
N_0 [F_{{\rm NN}\rho}^{(1)}(Q_2^2) + \kappa_\rho F_{{\rm NN}\rho}^{(2)}(Q_2^2)]
[ D^\rho_{\rm def} + D^\rho_{\rm abc}], \qquad
N_{\rm E}^{(+)} = 
N_0 F_{{\rm NN}\rho}^{(1)}(Q_2^2) [ D^\rho_{\rm def} + D^\rho_{\rm abc}],
\nonumber \\
N_{\rm M}^{(-)} &=&
N_0 [F_{{\rm NN}\rho}^{(1)}(Q_2^2) + \kappa_\rho F_{{\rm NN}\rho}^{(2)}(Q_2^2)]
[ D^\rho_{\rm def} - D^\rho_{\rm abc}], \qquad
N_{\rm E}^{(-)} = 
N_0 F_{{\rm NN}\rho}^{(1)}(Q_2^2) 
     [ D^\rho_{\rm def} - D^\rho_{\rm abc}],
\nonumber \\
N_0 
\hskip 0.1 true cm 
&{\buildrel \rm SU(6) \over = } &
\hskip 0.1 true cm 
{e_p g_{{\rm NN}\rho}^2G_{\rm M}^{\rm V}(0)G_{\rm M}^{{\rm N}\Delta\rho}(Q_2^2) 
\over 100m^3(2\pi)^6\omega_\rho(\vec{q}_2)}
\end{eqnarray}
where we normalize $G_{\rm M}^{{\rm N}\Delta\rho}(0)= 1 + \kappa_\rho$ and
adopt the propagators $D^\rho$ of eq.~(\ref{eq3.18}).
The {\it complete static limit} and {\it soft-photon} approximations follow in
analogy to the procedures used to develop eqs.~(\ref{eq3.25}) and 
(\ref{eq3.26})), the latter resulting in,
\begin{mathletters}
\begin{eqnarray}\label{eq3.31}
\langle \vec{p}_3 \vec{p}_4\vert  
\vec{J}^{{\rm N}\Delta}_{\rm eff}(\rho;{\rm E}) 
\vert \vec{p}_1 \vec{p}_2\rangle   
&=&
{4N_0 (1 +\kappa_\rho)
\over (m_{\Delta}-m)}
\Biggl\{ 
 { 4\tau_2^3  (\vec{p}_{4}+\vec{p}_{2})\times\vec{q}_2
   \over (\vec{q}_2^2+m_\rho^2)}
-{ (\vec{\tau}_1\times\vec{\tau}_2)^3 \vec{\sigma}_1\times[(\vec{p}_{4}+\vec{p}_{2})\times\vec{q}_2]
   \over (\vec{q}_2^2+m_\rho^2)}
\Biggr\}\times\vec{k}
\nonumber \\ &&
+ (1\rightleftharpoons 2)
\\ 
\langle \vec{p}_3 \vec{p}_4\vert  
\vec{J}^{{\rm N}\Delta}_{\rm eff}(\rho;{\rm M}) 
\vert \vec{p}_1 \vec{p}_2\rangle   
&=&
{4N_0 (1 +\kappa_\rho)^2
\over (m_\Delta-m)}
\Biggl\{ 
 { 4\tau_2^3 \bigl[(i\vec{\sigma}_2\times\vec{q}_2)\times \vec{q}_2\bigr]
  \over (\vec{q}_2^2+m_\rho^2) }
-{ (\vec{\tau}_1\times\vec{\tau}_2)^3\bigl[\vec{\sigma}_1\times[(i\vec{\sigma}_2\times\vec{q}_2)\times \vec{q}_2]\bigr]
 \over (\vec{q}_2^2+m_\rho^2) }
\Biggr\}\times \vec{k}
\nonumber \\ &&
+ (1\rightleftharpoons 2)
\end{eqnarray}
\end{mathletters}
\RightLineBreak
\begin{multicols}{2}\narrowtext
where $\vec{J}^{{\rm N}\Delta}_{\rm eff}(\rho;{\rm M})$ is the conventional 
result, and $\vec{J}^{{\rm N}\Delta}_{\rm eff}(\rho;{\rm E})$ is an additional 
piece (resulting from the convection current part of the NN$\rho$ vertex) 
which is usually ignored on the basis that it is smaller than 
$\vec{J}^{{\rm N}\Delta}_{\rm eff}(\rho;{\rm M})$ by a factor of about
1+$\kappa_\rho \sim$ 7.
\subsection{Direct, Single- and Rescattering Amplitudes}\label{sec3C}
A model-independent expression for the complete invariant amplitude was given 
in eq.~(\ref{eq2.22}), where we developed a decomposition into the four terms 
describing direct, single ({\it i.e.} initial- and final-state) and rescattering
amplitudes shown in Fig.~1. The left-hand side of eq~(\ref{eq2.22}) is defined 
by the model-independent expressions of eqs.~(\ref{eq2.2}) and (\ref{eq2.6}), 
and is Lorentz invariant by construction. However, since the right-hand side 
of eq.~(\ref{eq2.22}) is determined by a model-dependent calculation, there can
be no {\it a priori} guarantee that each of the direct, single and 
rescattering amplitudes are individually Lorentz invariant unless the wave 
functions are calculated in a manifestly covariant formalism \cite{JT75}.

Since the wave functions are usually constructed from an NN-interaction 
$t$-matrix that is defined only in the barycentric frame, and two distinct 
barycentric frames appear in eq.~(\ref{eq2.22}), we cast the entire expression
into the (maximally symmetric) average barycentric frame, so that the momenta 
satisfy,
\begin{equation}\label{eq3.32}
{\rm A-frame:}\hskip 1 true cm
  \vec{p}_{1}+\vec{p}_{2}-{\scriptstyle{1\over 2}}\vec{k} 
= 0
= \vec{p}_{3}+\vec{p}_{4}+{\scriptstyle{1\over 2}}\vec{k} 
\end{equation}
and we acknowledge that a formal solution of the initial-, final- and 
rescattering amplitudes requires the application of boost operators
\cite{LF61,RK74,JF75,WG80}.

In the following we will make use of the fact that the effective
current operator of eq.~(\ref{eq3.4}) is a totally symmetric under
interchange of particles 1 and 2. As such, we obtain in an
arbitrary frame,
\begin{eqnarray}\label{eq3.33}
&&
\langle  \vec{p}_3 \vec{p}_4; (s_1 s_2) S_f; (t_1 t_2)T_f\vert 
J_{\rm eff}
\vert  \vec{p}_1 \vec{p}_2; (s_1 s_2) S_i; (t_1 t_2)T_i\rangle 
\nonumber \\
&=&
\langle  \vec{p}_4 \vec{p}_3; (s_2 s_1) S_f; (t_2 t_1)T_f\vert 
J_{\rm eff}
\vert  \vec{p}_2 \vec{p}_1; (s_2 s_1) S_i; (t_2 t_1)T_i\rangle 
\nonumber \\
&=&
\langle  \vec{p}_4 \vec{p}_3; (s_1 s_2) S_f; (t_1 t_2)T_f\vert 
J_{\rm eff}
\vert  \vec{p}_2 \vec{p}_1; (s_1 s_2) S_i; (t_1 t_2)T_i\rangle 
\nonumber \\ && \quad \times (-1)^{(S_i+S_f+T_i+T_f)}
\end{eqnarray}
Denoting $\vert \alpha\rangle  = \vert (s_1 s_2) S; (t_1 t_2)T\rangle $,
we define the anti-symmeterized states as,
\begin{equation}\label{eq3.34}
\vert \widetilde{\vec{p}_1 \vec{p}_2;\alpha}\rangle 
= {1\over \sqrt{2}}\left\{
   \vert \vec{p}_1 \vec{p}_2;\alpha\rangle 
- (-1)^{(S+T)} \vert \vec{p}_2 \vec{p}_1;\alpha\rangle  \right\}
\end{equation}
It is easy to see that parity conservation is consistent with 
the Fermi statistics requirement $L+S+T$=odd in the barycentric frame.
\subsubsection{Direct Amplitudes}\label{sec3C1}
The direct amplitudes appearing eq.~(\ref{eq2.22}a) can be simplified 
with use of eqs.~(\ref{eq3.33}) and (\ref{eq3.34})  to give,
\begin{eqnarray}\label{eq3.35}
{\cal M}_{fi}^{\rm D} &= &
N \vec{\epsilon}(\vec{k},\lambda).
\bigl[\langle \vec{p}_{3}\vec{p}_{4};\alpha_f\vert \vec{J}_{\rm eff}\vert \vec{p}_{1}\vec{p}_{2};\alpha_i\rangle  
\nonumber \\ &&
-(-1)^{(S_i+T_i)}
      \langle \vec{p}_{3}\vec{p}_{4};\alpha_f\vert \vec{J}_{\rm eff}\vert \vec{p}_{2}\vec{p}_{1};\alpha_i\rangle 
\bigr]
\end{eqnarray}
where $N$ is defined in eq.~(\ref{eq2.23}) and $\vec{J}_{\rm eff}$ is given by 
eq.~(\ref{eq3.4}).  However, momentum conservation demands that the direct 
terms cannot involve the 1-body part of the effective current density, so in 
the present numerical results we include the photon coupling to 
\begin{itemize}
\item NN~currents with recoil and wave function renormalization currents.
\item N$\bar{\rm N}$~creation and annihilation currents 
\item $\rho\pi\gamma$, $\omega\pi\gamma$, 
      $\rho\eta\gamma$ and $\omega\eta\gamma$ exchange currents.
\item $\Delta$N currents with $\pi$ and $\rho$ exchange.
\end{itemize}
as described by eqs.~(\ref{eq3.8}), (\ref{eq3.10}), (\ref{eq3.15}),
(\ref{eq3.21}), (\ref{eq3.28}) and (\ref{eq3.29})
\subsubsection{Single-Scattering Amplitudes}\label{sec3C2}
The initial-state interaction amplitudes appearing eq.~(\ref{eq2.22}b) 
are given by,
\begin{eqnarray}\label{eq3.36}
{\cal M}_{fi}^{\rm I}
=&&  N \vec{\epsilon}(\vec{k},\lambda).
\langle \widetilde{\vec{p}_{3}\vec{p}_{4}};\alpha_f\vert 
 \vec{J}_{\rm eff}(0)] G_i t^{(+)} 
\vert \widetilde{\vec{p}_{1}\vec{p}_{2}};\alpha_i\rangle 
\nonumber \\ =&& N \vec{\epsilon}(\vec{k},\lambda). 
\sum_{\alpha} \!\!\int\!\!\!\!\int\!\! d\vec{p}_{1}' d\vec{p}_{2}'
\langle \vec{p}_{3}\vec{p}_{4};\alpha_f\vert   \vec{J}_{\rm eff}(0) 
\vert \vec{p}_{1}'\vec{p}_{2}';\alpha\rangle 
\nonumber \\ && \qquad\qquad
\times \langle \widetilde{\vec{p}_{1}'\vec{p}_{2}'};\alpha\vert  G_i t^{(+)}
\vert \widetilde{\vec{p}_{1}\vec{p}_{2}};\alpha_i\rangle 
\end{eqnarray}
A formal specification of this amplitude follows by inserting the
full effective current density of eq.~(\ref{eq3.4}) and defining
boost the operators needed to cast the $t$-matrix in the initial-state
barycentric frame.  None of the bremsstrahlung calculations known to us
has attempted either of these tasks. Instead, the current has 
been truncated to include only one-body contributions and boost operators
are ignored under the assumption that the part of the invariant 
amplitude resulting from initial-state correlations alone is itself 
individually Lorentz invariant.

The first approximation could be removed with a straightforward application of 
the expressions provided in earlier sections, but since we anticipate the 
impulse current to be significantly larger than the summed exchange currents,
our present numerical applications share the conventional approximation of 
retaining only the impulse current in the single-scattering amplitudes. 
The second approximation will be justified in section~\ref{sec4A}, 
where we provide a perturbative analysis that indicates the NN$\gamma$ 
impulse-current contributions to the single-scattering terms are close to 
invariant under Lorentz transformation into the barycentric frames. 

We therefore cast the initial-state correlation amplitudes into the 
initial-state barycentric frame,  where
\begin{equation}\label{eq3.37}
{\rm I-frame:}\hskip 1 true cm
  \vec{p}_{1}+\vec{p}_{2} = 0 = \vec{p}_{3}+\vec{p}_{4}+\vec{k} 
\end{equation}
so that eq.~(\ref{eq2.22}b) becomes,
with the kinematical notation of eq.~(\ref{eq3.6}),
\begin{eqnarray}\label{eq3.38}
{\cal M}_{fi}^{\rm I} =&&
N \vec{\epsilon}(\vec{k},\lambda). \sum_{\alpha} 
\langle \alpha_f\vert   
  \vec{J}_{{\rm NN}\gamma}[1](\vec{p}_{3k},\vec{p}_{3}) 
\vert \alpha\rangle 
\nonumber \\ &&  \qquad\times
{\langle \widetilde{\vec{p}_{3k};\alpha}\vert 
  t^{(+)}
 \vert \widetilde{\vec{p}_{1};\alpha_i}\rangle 
\over 2E_1-2E_{3k} + i\eta}
\nonumber \\ +&& 
N \vec{\epsilon}(\vec{k},\lambda). \sum_{\alpha} 
\langle \alpha_f\vert   
  \vec{J}_{{\rm NN}\gamma}[2](\vec{p}_{4k},\vec{p}_{4}) 
\vert \alpha\rangle 
\nonumber \\ &&  \qquad\times
{\langle \widetilde{\vec{p}_{3};\alpha}\vert 
  t^{(+)}
 \vert \widetilde{\vec{p}_{1};\alpha_i}\rangle 
\over 2E_1-2E_{3} + i\eta}
\end{eqnarray}
where we have denoted
$\langle \vec{p}', -\vec{p}' ;\alpha_f\vert  t^{(+)}
 \vert \vec{p} , -\vec{p}  ;\alpha_i\rangle $ = 
$\langle \vec{p}' ; \alpha_f\vert  t^{(+)} \vert \vec{p};\alpha_i\rangle $.
In complete analogy, we cast the final-state correlation amplitudes
into the final-state barycentric frame,  where
\begin{equation}\label{eq3.39}
{\rm F-frame:}\hskip 1 true cm
  \vec{p}_{1}+\vec{p}_{2} -\vec{k}  = 0 = \vec{p}_{3}+\vec{p}_{4} 
\end{equation}
so that eq.~(\ref{eq2.22}c) becomes,
\begin{eqnarray}\label{eq3.40}
{\cal M}_{fi}^{\rm F} =&& 
N \vec{\epsilon}(\vec{k},\lambda). \sum_{\alpha} 
{\langle \widetilde{\vec{p}_{3};\alpha_f}\vert 
  t^{(-)\dagger}
 \vert \widetilde{\vec{p}_{1k};\alpha}\rangle 
\over 2E_3-2E_{1k} + i\eta}
\nonumber \\ &&  \qquad\times
\langle \alpha\vert   
  \vec{J}_{{\rm NN}\gamma}[1](\vec{p}_{1},\vec{p}_{1k}) 
\vert \alpha_i\rangle 
\nonumber \\ +&& 
N \vec{\epsilon}(\vec{k},\lambda). \sum_{\alpha} 
{\langle \widetilde{\vec{p}_{3};\alpha_f}\vert 
  t^{(-)\dagger}
 \vert \widetilde{\vec{p}_{1};\alpha}\rangle 
\over 2E_3-2E_{1} + i\eta}
\nonumber \\ &&  \qquad\times
\langle \alpha\vert   
  \vec{J}_{{\rm NN}\gamma}[2](\vec{p}_{2},\vec{p}_{2k}) 
\vert \alpha_i\rangle 
\end{eqnarray}
Simple kinematics establishes that the radiation of a real photon 
implies an off-shell $t$-matrix, so that we are free to immediately 
take the limits $\eta\rightarrow$ 0 in eqs.~(\ref{eq3.38}) and 
(\ref{eq3.40}).

\subsubsection{Rescattering Amplitudes}\label{sec3C3}
From the results of impulse approximation calculations 
\cite{VB91,VH91,JE93} we already know that the impulse contributions to
the rescattering amplitudes constitute a correction of $\le$ 15\% to the
corresponding single-scattering amplitudes, so for simplicity we neglect 
from the outset all 2-body currents in the rescattering amplitudes 
of eq.~(\ref{eq2.22}d). Hence, in the A-frame of eq.~(\ref{eq3.32}) 
we obtain,
\begin{eqnarray}\label{eq3.41}
&&{\cal M}_{fi}^{\rm R}
=  N
\sum_{\alpha\alpha'}\!\!\int\!\!\!\!\int\!\! d\vec{p}_{1}' d\vec{p}_{2}'
\delta^{(3)}(\vec{p}_1'+\vec{p}_2'-\vec{p}_{1}-\vec{p}_{2})
\Bigg\{ \nonumber\\
\phantom{+} &&
   \langle \widetilde{ \vec{p}_{3},\vec{p}_{4};\alpha_f }\vert  
    t^{(-)\dag} G_f 
   \vert \vec{p}_{1k}', \vec{p}_2'; \alpha_b \rangle 
   \langle \alpha'\vert  J_{{\rm NN}\gamma}[1](\vec{p}_1',\vec{p}_{1k}') \vert \alpha\rangle 
\nonumber \\ && \qquad\qquad\times
   \langle \vec{p}_1', \vec{p}_2'; \alpha_a \vert  
    G_i t^{(+)} 
   \vert \widetilde{\vec{p}_{1}, \vec{p}_{2};\alpha_i }\rangle 
\nonumber \\ + &&
   \langle \widetilde{ \vec{p}_{3},\vec{p}_{4};\alpha_f }\vert  
    t^{(-)\dag} G_f \vert \vec{p}_1', \vec{p}_{2k}'; \alpha_b \rangle 
   \langle \alpha'\vert  J_{{\rm NN}\gamma}[2](\vec{p}_2',\vec{p}_{2k}') \vert \alpha\rangle 
\nonumber \\ && \qquad\qquad\times
   \langle \vec{p}_1', \vec{p}_2'; \alpha_a \vert  
    G_i t^{(+)} 
   \vert \widetilde{\vec{p}_{1}, \vec{p}_{2};\alpha_i }\rangle 
\Bigg\} 
\end{eqnarray}
The initial- and final-state barycentric frames differ by the photon momentum, 
so that no frame can be found where both $t$-matrices are expressed in their 
barycentric frame. We therefore introduce a boost operator $\chi$ satisfying 
\cite{LF61,RK74,JF75,WG80},
\begin{eqnarray}\label{eq3.42}
\vert \vec{p}_{a},\vec{p}_{b}\rangle  &&=
\bigl\{ 1 - i\chi(\vec{\cal P}) \bigr\}\;
\vert +\vec{p}_{}\rangle   \vert \vec{\cal P}\rangle  + {\cal O}(1/m^4), 
\nonumber \\ &&
\vec{p}_{} = {1\over 2}(\vec{p}_{a} - \vec{p}_{b}),
\qquad \vec{\cal P} =      (\vec{p}_{a} + \vec{p}_{b})
\end{eqnarray}
Eq.~(\ref{eq3.41}) is manifestly symmetric under interchange of particles 1 
and 2.  However, for computation purposes, it proves convenient to make use 
of eqs.~(\ref{eq3.33}), (\ref{eq3.34}) and (\ref{eq3.42}) to formally 
re-express eq.~(\ref{eq3.41}) in terms of the photon coupling only to 
nucleon 1,
\end{multicols}\widetext
\LeftLineBreak
\begin{eqnarray}\label{eq3.43}
&&{\cal M}_{fi}^{\rm R}
= 2N(-1)^{(S_f+S_i+T_f+T_i)} \sum_{\alpha,\alpha'} 
\vec{\epsilon}. \int d\vec{p}_{}
\langle \widetilde{\vec{p}_{3}+{1\over 4}\vec{k}};\alpha_f\vert  [1+i\chi(-{\vec{k}\over 2})] 
   t^{(-)\dag} G_f 
[1-i\chi(-{\vec{k}\over 2})]\vert \vec{p} + {1\over 4}\vec{k};\alpha' \rangle  
\nonumber \\ &&\times 
   \langle \alpha'\vert  \vec{J}_{{\rm NN}\gamma}[1] (-\vec{p}-{\vec{k}\over 2}, -\vec{p}+{\vec{k}\over 2}) \vert \alpha\rangle 
\langle +\vec{p}_{} - {1\over 4}\vec{k};\alpha\vert  [1+i\chi(+{\vec{k}\over 2})]
   G_i t^{(+)}
[1-i\chi(+{\vec{k}\over 2})] \vert \widetilde{\vec{p}_{1} - {1\over 4}\vec{k}}; \alpha_i\rangle 
\end{eqnarray}
\RightLineBreak
\begin{multicols}{2}\narrowtext
where, since the $t$-matrix conserves spin- and isospin, the sums over 
the intermediate-state quantum numbers ($\alpha$=$S,M_S,T,M_T$) are 
restricted such that $S$=$S_i$, $S'$=$S_f$, $T$=$T_i$, $T'$=$T_f$ with
conserved isospin projection $M_{T}$. For reasons already indicated
in the discussion of the single-scattering amplitudes, 
the boost operators can be neglected in the present work.
A recipe for performing the numerical integration over the pole-structures
of eq.~(\ref{eq3.43}) is given in App.~C.
\section{Results and Discussion}\label{sec4}
\subsection{Impulse Contributions}\label{sec4A}
In the present work we are primarily interested to investigate a consistent
calculation of the dominant isoscalar meson-exchange currents in 
pp-bremsstrahlung. An important precursor to this lies in establishing that the
well-known discrepancy between impulse approximation calculations and 
experimental data cannot be resolved by selecting a different (phase-equivalent)
NN-interaction. Although a qualitative similarity exists between the results of
recent impulse approximation calculations for pp-bremsstrahlung observables
\cite{HF86,VB91,VH91,VH92,HF93a,HF93b,JE93,KA93,MJ93}
the calculation differences are generally 
not confined to the differing NN-potentials. Some of these calculations 
describe the photon coupling to the one-body current via a Foldy-Wouthuysen 
transformations \cite{HF86,HF93a,HF93b,KA93,MJ93} or direct Pauli reduction 
\cite{VH91,VH92,JE93}, whereas others adopt a static limit description 
\cite{VB91}. Some include the rescattering amplitudes of Fig.~1c 
\cite{VB91,VH91,VH92,JE93,MJ93} whereas others don't 
\cite{HF86,HF93a,HF93b,KA93}. Further differences are found in the use of 
relativistic or non-relativistic two-nucleon propagators and/or the application
of (guessed) off-shell minimal relativity factors \cite{VH91,VH92}.

We avoid all of these uncertainties by  presenting the results
of calculations that are identical apart from the 
definition of the potential used to generate the $t$-matrix elements.
We also extend the list of commonly compared potentials to include
the Bonn \cite{Bonn}, Paris \cite{Paris}, Nijmegen \cite{Nijmegen} and
RuhrPot \cite{RuhrPot}.
For the present comparisons we adopt the impulse approximation, so that 
the effective current density developed in section~\ref{sec3B} 
reduces to the sum of (one-body) impulse currents,
$\vec{J}_{\rm eff}\sim\vec{J}_{{\rm NN}\gamma}[1]+\vec{J}_{{\rm NN}\gamma}[2]$,
as defined in App.~B, and is therefore common to all potentials. 
As such, we retain initial-state, final-state and 
rescattering amplitudes with the two-nucleon Green's functions described by the
relativistic Lippmann-Schwinger propagators. Partial-waves are
summed to $J_{\rm max}=8$ and no form of soft-photon approximation
is adopted at any stage. 

In Fig.~8 we compare such calculations with the TRIUMF coplanar 
pp-bremsstrahlung data at $E_{\rm lab}$=280~MeV \cite{TRIUMF}. The cross section
geometries are 
\par\noindent
\begin{figure}
\vbox to 7.5 true in {\hbox{\hskip -15 mm \epsfxsize=4.0 true in \epsfbox{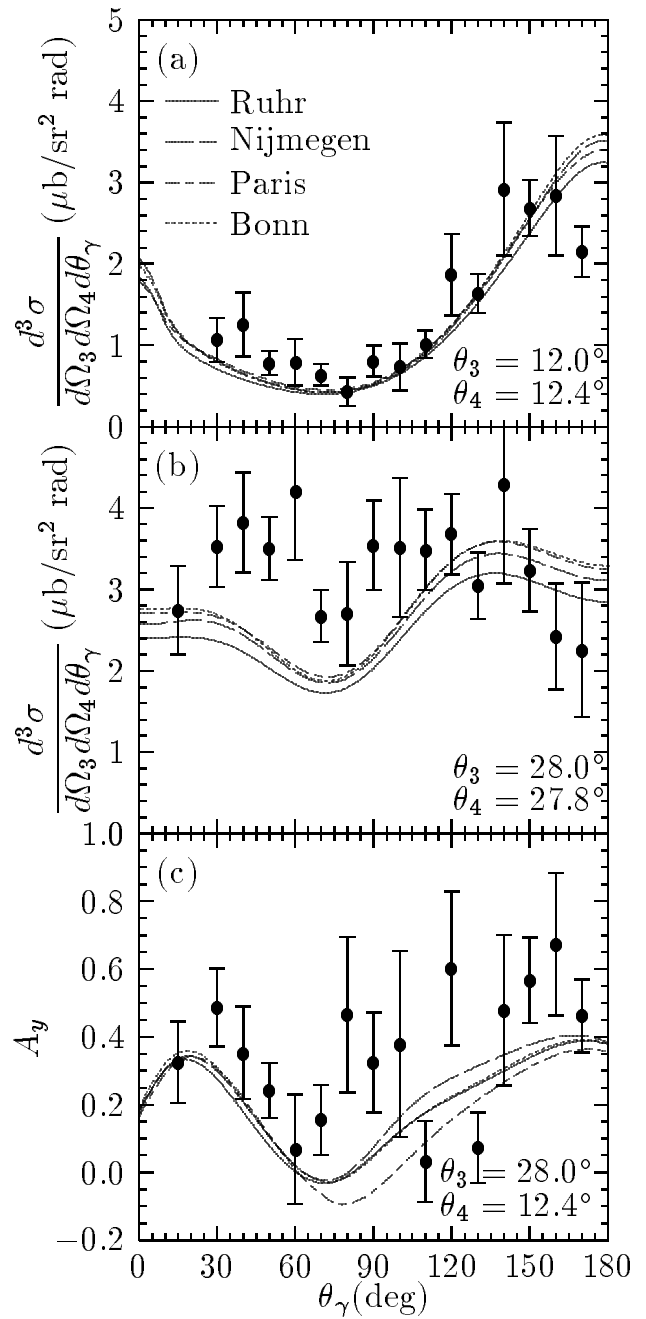}}
\vglue 3 true mm
\caption{ 
Comparison of impulse approximation calculations using Ruhr, 
Nijmegen, Paris and Bonn potentials and coplanar pp-bremsstrahlung 
data at $E_{\rm lab}$=280~MeV. The differences between the model results are
smaller than their collective discrepancy with experiment. 
}}
\end{figure}
\par\noindent
selected to sample both small and large proton emission angles. 
The analyzing power geometry is selected on the basis that it is the result 
most different from zero, and therefore presumably the most reliably measured.
Some differences exist between our analyzing power results and those reported 
elsewhere \cite{VH92} - primarily due to differences in the rescattering calculation, 
as discussed in App.~C. The essential conclusion here is that impulse 
approximation calculations using Bonn~B, Nijmegen, Paris and RuhrPot 
wave functions are almost indistinguishable, but exhibit a collective 
discrepancy with experiment. Given that the purpose of this experiment was to 
distinguish the predictions of such potentials, the differences between theory 
and experiment are large.

The final task remaining here is to establish that the impulse current 
contributions to the single-scattering amplitudes given in eqs~(\ref{eq3.38}) 
and (\ref{eq3.40}) can be accurately described without boost operators. 
This approximation is common to all momentum-space bremsstrahlung calculations 
known to us, yet it appears to have never been verified. Some authors 
\cite{VH91,VH92,JO94} 
have sought a solution to the problem by arguing that the non-relativistic 
$t$-matrix can be made Lorentz invariant simply by attaching 
`minimal relativity' factors \cite{KE74}, so that the Lippmann-Schwinger (LS) 
equation can be cast in a form that is apparently identical to the 
Blankenbecler-Sugar (Bbs) equation.  However, although both of these integral 
equations describe the NN interaction in ladder approximation, they are not 
formally identical because the LS kernel is constructed in a time-ordered 
(non-covariant) relativistic framework, whereas the BbS-kernel represents one 
of an infinite number of arbitrary three-dimensional reductions of the 
covariant Bethe-Salpeter (BS) equation.
As such, the Lippmann-Schwinger $t$-matrix with minimal relativity
factors should not be confused with a covariant definition of the NN-interaction
\cite{JT75}.  The most serious flaw in the use of minimal relativity factors
is, however, that the off-shell factors are completely unknown and must be 
simply guessed \cite{KE74}. We regard this procedure as unreasonably arbitrary,
particularly since the guessed minimal relativity factors contradict the well 
known form of the two-nucleon boost operators \cite{LF61,RK74,JF75,WG80}.

Since the $t$-matrix is defined in the conventional way as $t$=$V$+$VGt$ 
but is available only in the barycentric frame, we consider the arbitrary-frame 
perturbative reduction of these amplitudes by constructing a toy-model 
one-boson exchange NN-interaction as,
\begin{eqnarray}\label{eq4.1}
&&\langle \vec{p}_1^{\;'},\vec{p}_2^{\;'}\vert  V \vert \vec{p}_1,\vec{p}_2\rangle 
\nonumber \\
 =&& {1\over 2} \sum_{\beta} 
\eta H_{{\rm NN}\beta}[1](\vec{p}_1,\vec{p}_1^{\;'}) \lambda
\Bigl[ {1\over {\cal E}_i - H_0} 
+ {1\over {\cal E}_f - H_0} \Bigr]
\nonumber \\ &&
\times \lambda H_{{\rm NN}\beta}[2](\vec{p}_2,\vec{p}_2^{\;'}) \eta
+ (1\rightleftharpoons 2)
\end{eqnarray}
where $\beta$=$\pi$,$\eta$,$\rho$ and $\omega$, and $H_{{\rm NN}\beta}[i]$ is
the interaction energy for a meson coupling to nucleon $i$, as defined in 
App.~B. In Fig.~9 we present the results including only the initial- and 
final-state interaction amplitudes when the RuhrPot $t$-matrix is replaced 
with our toy NN-interaction, the complete expression being cast into the 
A-, I- and F-frames of eq.~(\ref{eq3.32}), (\ref{eq3.37}), (\ref{eq3.39}).
We find that the neglect of boost operators represents an error of about 
3\% or less. We conclude that eq.~(\ref{eq3.38}) and (\ref{eq3.40})
are essentially exact descriptions of the impulse contributions to the
single-scattering amplitudes. As such, the application of minimal relativity 
factors \cite{VH91,VH92,JO94} may 
\par\noindent
\begin{figure}
\vbox to 5.5 true in {\hbox{\hskip -15 mm \epsfxsize=4.0 true in \epsfbox{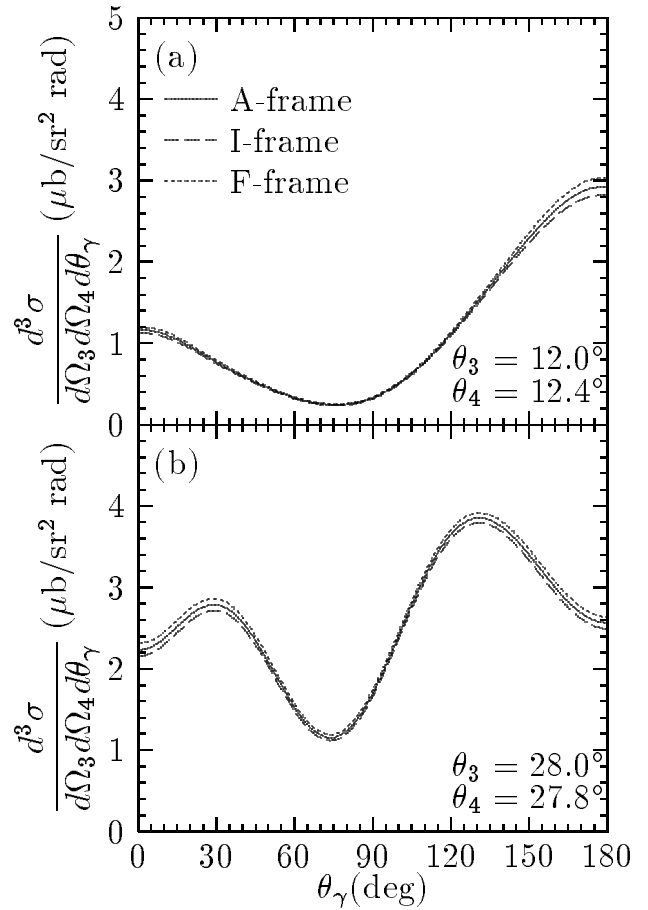}}
\vglue 3 true mm
\caption{ 
Initial- and final-state interaction amplitudes in impulse approximation 
when the $t$-matrix is replaced with a toy-model boson exchange potential
and evaluated in the A-, I-, and F-frames.
This justifies the neglect of boost operators in the present work.
}}
\end{figure}
\par\noindent
need to be reconsidered.
%
%
\subsection{Exchange Currents and the Relativistic NN$\gamma$ Vertex}
\label{sec4B}
In section~\ref{sec3B} we showed that a relativistic description of the part of
the effective current operator involving the photon coupling to the nucleon 
current comprises not only the impulse current, but also two-body contributions 
from the meson-recoil, wave function re-orthonormalization and pair currents. 
Although relativistic corrections to 1-body NN$\gamma$ currents have received 
considerable attention in recent pp-bremsstrahlung works 
\cite{HF86,VH92,HF93a,HF93b}, the 2-body contributions remain to be 
investigated.

The wave function re-orthonormalization and meson-recoil contributions are not 
expected to be individually small, but in section~\ref{sec3B} we recalled 
\cite{GA76} that their non-relativistic limits cancel exactly, leaving only 
purely relativistic effects in the pp-bremsstrahlung observables. In the 
present numerical applications we retain these contributions in the spirit of 
exploring the relativistic aspects of the NN$\gamma$ currents. 

By far the most interesting aspect of the relativistic NN$\gamma$ currents
is found in the different manifestations of the N$\bar{\rm N}$ creation 
and annihilation pair currents given by various Dyson-equivalent chiral 
Lagrangians. In particular, the simplest meson-theoretic Lagrangian 
satisfying PCAC and the chiral commutation relations is the renormalizable 
$\sigma$-model, for which the nucleon-meson interactions are of the form,
\begin{equation}\label{eq4.2}
{\cal L} = - g\bar{\psi}[\sigma + i\gamma^5\vec{\tau}.\vec{\pi}']\psi
\end{equation}
With a chiral transformation
$\psi\rightarrow(1+\vec{\xi}^2)^{-{1\over 2}}
 \exp[i\gamma^5\vec{\tau}.\vec{\pi}')]\psi$,
Weinberg \cite{SW67} showed that, for a suitably constrained $\vec{\xi}$ and 
redefined pion fields $\vec{\pi}=(2m/g)\vec{\xi}$, the Lagrangian tranforms to
give NN$\pi$ and NN$\pi\pi$ interactions with,
\begin{equation}\label{eq4.3}
{\cal L}_{\rm pv} = - {g\over 2m} \bar{\psi}
\left[1\!+\!{g^2\vec{\pi}^2\over4m^2}\right]\!\!\!
\left[\gamma^5\gamma^\mu\vec{\tau}.\partial_\mu\vec{\pi}  \!+\!
      {g\over 2m} \vec{\tau}.(\vec{\pi}\!\times\!\partial_\mu\vec{\pi}) \right]\!\!\psi
\end{equation}
Similarly, the chiral transformation
$\psi\rightarrow(1+\vec{\xi}^2)^{-{1\over 2}}
 \exp[i(g/2m)\vec{\tau}.(\vec{\pi}'\times\vec{\xi})]\psi$
has been shown by Wess and Zumino \cite{JW67} to yield NN$\pi$ and NN$\pi\pi$ 
interactions with,
\begin{equation}\label{eq4.4}
{\cal L}_{\rm ps} = - {g\over 2m}\bar{\psi}
\left[1+{g^2\vec{\pi}^2\over4m^2}\right]
\left[i\gamma^5\vec{\tau}.\vec{\pi}  -
      {g\over 2m} \vec{\pi}^2 \right]\psi
\end{equation}
In an elegent summary of these chiral Lagrangians, Gross \cite{FG82} noted that
both ${\cal L}_{\rm ps}$ and ${\cal L}_{\rm pv}$ give the correct $\pi-N$ 
scattering lengths and that the NN$\pi\pi$ interactions have $\rho$ and 
$\sigma$ quantum numbers.

Our immediate interest lies in the pseudo-vector (pv) and pseudo-scalar (ps) 
NN$\pi$ couplings, both of which are included via the hybrid form of the 
NN$\pi$ Lagrangian we adopt in eq.~(\ref{eq3.1}) with $0\leq\lambda\leq 1$.
It is a trivial exercise to show that the $\pi$-coupling to the 
positive-frequency components of the nucleon current described in 
eq.~(\ref{eq3.1})  is independent of $\lambda$, so that non-relativistic 
calculations cannot differentiate the ps- and pv-couplings. In relativistic 
applications $\lambda=0$ is commonly preferred, presumably because the presence 
of the derivative in the pv-coupling provides for the gauge coupling of a       
photon to the NN$\pi$ vertex, so that the purely isovector non-relativistic
seagull exchange currents can be included even when the microscopic structure 
of the NN$\pi$ form factor is ignored. However, there are no formal arguments 
to rule out $\lambda\neq 0$ and the subject is still under investigation 
\cite{FG89,FG90}.

In Fig.~10 we examine the relativistic NN$\gamma$ currents in pp-bremsstrahlung
by comparing the results of calculations with RuhrPot wave functions which 
neglect (IA) or retain (IA+MEXC) the wave function re-orthonormalization, 
meson-recoil and N$\bar{\rm N}$ pair-creation 
\par\noindent
\begin{figure}
\vbox to 7.8 true in {\hbox{\hskip -15 mm \epsfxsize=4.0 true in \epsfbox{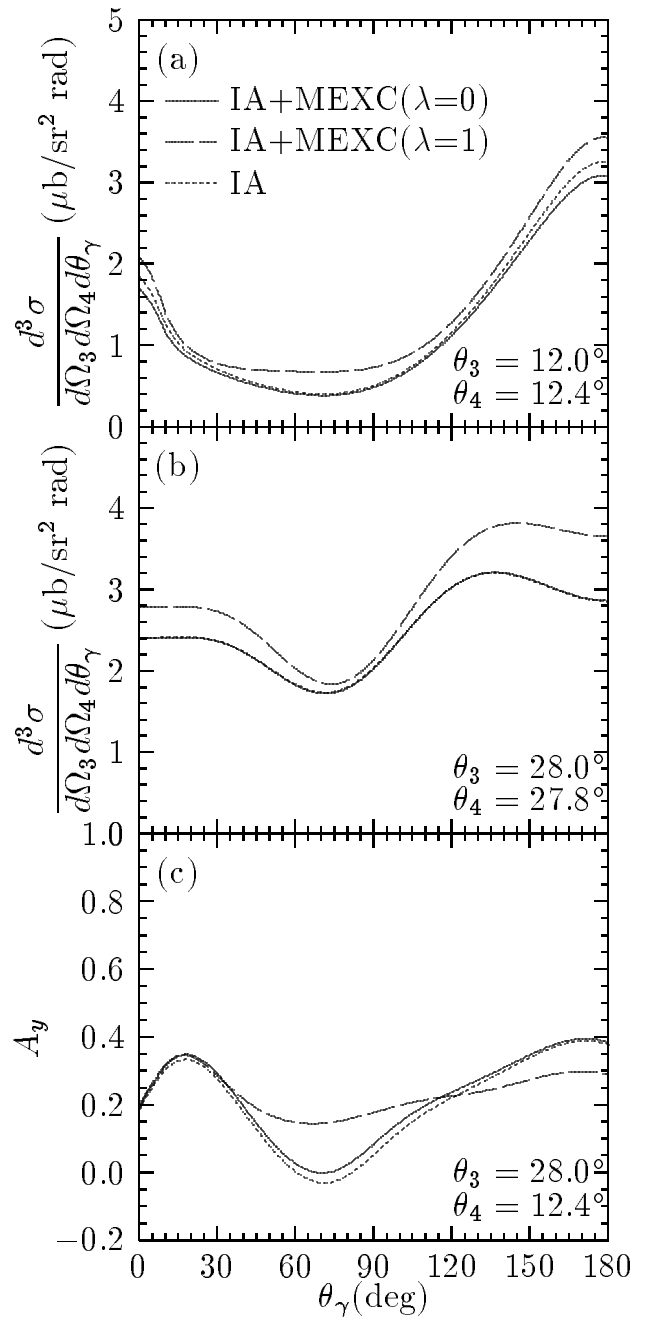}}
\vglue 3 true mm
\caption{ 
RuhrPot results in impulse approximation (IA) as in Fig.~8 compared to
corresponding results when the wave function re-orthonormalization,
meson-recoil and  ${\rm N}\bar{\rm N}$-pair currents (IA+MEXC) are included
with pseudo-vector ($\lambda$=0) or pseudoscalar ($\lambda$=1) NN$\pi$
interactions. $\lambda$ can be determined from pp-bremsstrahlung experiments!
}}
\end{figure}
\par\noindent
and annihilation currents with 
purely pv-type ($\lambda$=0) and ps-type ($\lambda$=1) NN$\pi$ couplings.  

The first observation is the most important one: $\lambda$ can be determined 
by pp-bremsstrahlung  experiments! This is surely the cleanest probe of the 
Lorentz structure of the NN$\pi$ vertex yet considered.  We reserve our 
prediction for $\lambda$ until we have included the other exchange currents 
given in section~\ref{sec3}. 

A secondary observation, which we anticipated in ref~\cite{JE93}, is that
the 2-body contributions are small for $\lambda=0$. In particular, 
the $\lambda$=0 NN$\gamma$ MEXC contributions to the cross section at 
(LEP) $\theta_3$=28.0$^\circ$ and (HEP) $\theta_4$=27.8$^\circ$ 
are almost entirely negligible. However, the effects are visible in the 
analyzing powers, as well as in the cross section at
$\theta_3$=12.0$^\circ$ and  $\theta_4$=12.4$^\circ$, the latter being
reduced at $\theta_\gamma$=0$^\circ$ and 180$^\circ$ by
0.14 and 0.17 $\mu b/sr^2/rad$ (i.e. 5\% and 8\%) respectively.
While these effects are certainly small, they are already comparable to
the model differences shown in Fig.~8.

\subsection{Radiative Vector-Meson Decay Currents}\label{sec4C}
In Fig.~11 we compare the results of calculations using RuhrPot wave functions 
and the relativistic impulse current which either neglect (IA) or retain 
(IA+VP$\gamma$) the relativistic 
$\rho\pi\gamma$, $\omega\pi\gamma$, $\rho\eta\gamma$ and $\omega\eta\gamma$ 
exchange currents \cite{MC74,JT89}. The RuhrPot contributions are uniformly 
small, although we observe a reduction in the cross section at 
(LEP) $\theta_3$=12.0$^\circ$ and (HEP) $\theta_4$=12.4$^\circ$ at 
$\theta_\gamma$=20$^\circ$ by about 0.1 $\mu b/sr^2/rad$ (i.e. 10\%).

We have used the fully relativistic expressions of eq~(\ref{eq3.15}) for the 
numerical applications of Fig.~11, but to identify the dominant behaviour of 
these exchange currents it is sufficient to consider only the static limit, 
neglect the $\omega$- and $\rho$-meson mass difference and neglect the 
$\eta$-meson contributions altogether. In the non-relativistic limit, 
this allows us to write,
\begin{eqnarray}\label{eq4.5}
\langle \vec{p}_3 \vec{p}_4\vert  
\vec{J}^{\rho\pi}_{\rm eff}
\vert \vec{p}_1 \vec{p}_2\rangle  
&& \,\,{\buildrel \rm nr \over \sim}\,\, 
 g_{\rho\pi\gamma} g_{{\rm NN}\rho}
[\hat{J}(\vec{q}_1,\vec{q}_2) + \hat{J}(\vec{q}_2,\vec{q}_1)]
\vec{\tau}_1.\vec{\tau}_2
\nonumber \\
\langle \vec{p}_3 \vec{p}_4\vert  
\vec{J}^{\omega\pi}_{\rm eff}
\vert \vec{p}_1 \vec{p}_2\rangle  
&& \,\,{\buildrel \rm nr \over \sim}\,\, 
 g_{\omega\pi\gamma} g_{{\rm NN}\omega}
[\hat{J}(\vec{q}_1,\vec{q}_2) + \hat{J}(\vec{q}_2,\vec{q}_1)]
\vec{\tau}_1^z
\nonumber \\ &&
\end{eqnarray}
where 
\begin{equation}\label{eq4.6}
\hat{J}(\vec{q}_1,\vec{q}_2)
= { i e_p  g_{{\rm NN}\pi} (\vec{\sigma}_1.\vec{q}_1)(\vec{q}_1\times\vec{q}_2) \over 
(2\pi)^6 2m_{\rm V} m [\vec{q}_1^2+m_\pi^2] [\vec{q}_2^2+m_{\rm V}^2] }
\end{equation}
Since $g_{{\rm NN}\pi}$, $g_{\rho\pi\gamma}$ and $g_{\omega\pi\gamma}$ are 
essentially fixed by experiment, the freedom in the vector-meson decay
exchange currents is limited to the model-dependent values adopted
for the experimentally unknown coupling constants $g_{{\rm NN}\rho}$ and 
$g_{{\rm NN}\omega}$. We have adopted the RuhrPot NN-interaction values of 
$g_{{\rm NN}\rho}$=1.65 and $g_{{\rm NN}\omega}$=4.95, and note that these 
values are consistent with the broken SU(3) requirement 
$g^2_{{\rm NN}\omega}/g^2_{{\rm NN}\rho}$=9. The matrix elements of both isopin
operators in eq.~(\ref{eq4.5}) reduce to unity in pp--bremsstrahlung, so that 
the $\omega\pi\gamma$ contribution completely dominates the vector-meson decay 
current contributions and the corresponding $\rho\pi\gamma$ currents are some 
15 times smaller. 

The magnitude of each PV$\gamma$ exchange current is, of course, dependent on 
the choice of NN-interaction. For example, the Bonn~B model requires 
$g_{{\rm NN}\rho}$=3.36 and 
\par\noindent
\begin{figure}
\vbox to 7.8 true in {\hbox{\hskip -15 mm \epsfxsize=4.0 true in \epsfbox{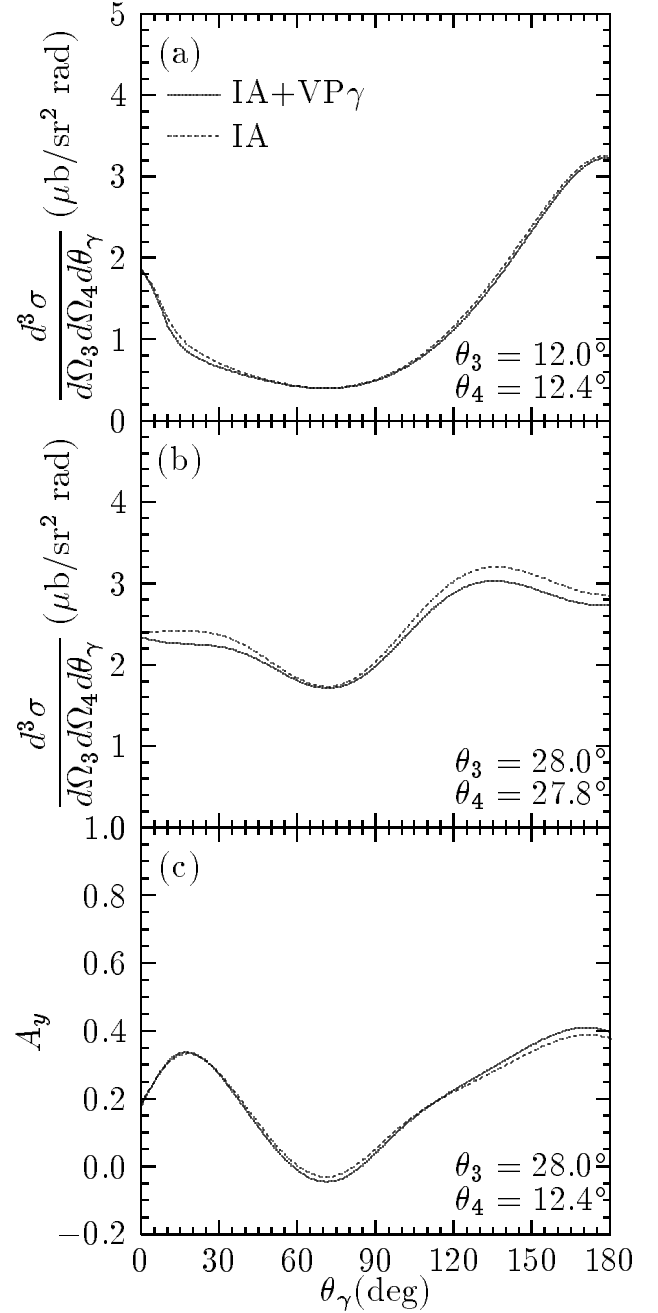}}
\vglue 3 true mm
\caption{ 
RuhrPot results in impulse approximation (IA) as in Fig.~8 compared to
corresponding results when the VP$\gamma$=
$\rho\pi\gamma$ + $\omega\pi\gamma$ + $\rho\eta\gamma$ + $\omega\eta\gamma$
exchange currents (IA+VP$\gamma$) are included. The RuhrPot model
has relatively weak NN$\rho$ and NN$\omega$ couplings, so these exchange 
currents are small, but comparable to the model differences shown in Fig.~8.
}}
\end{figure}
\par\noindent
$g_{{\rm NN}\omega}$=17.5, so that
$g^2_{{\rm NN}\omega}/g^2_{{\rm NN}\rho}$=27, which severely contradicts
the broken SU(3) prediction of 9. Although the  $\rho\pi\gamma$ and 
$\omega\pi\gamma$ exchange currents have never been included in Bonn model 
calculations for bremsstrahlung observables, it is easy to see that these 
currents alone would be respectively more that 4 and 12 times larger than 
the corresponding results of the present calculation.
\par\noindent
\begin{figure}
\vbox to 7.4 true in {\hbox{\hskip -15 mm \epsfxsize=4.0 true in \epsfbox{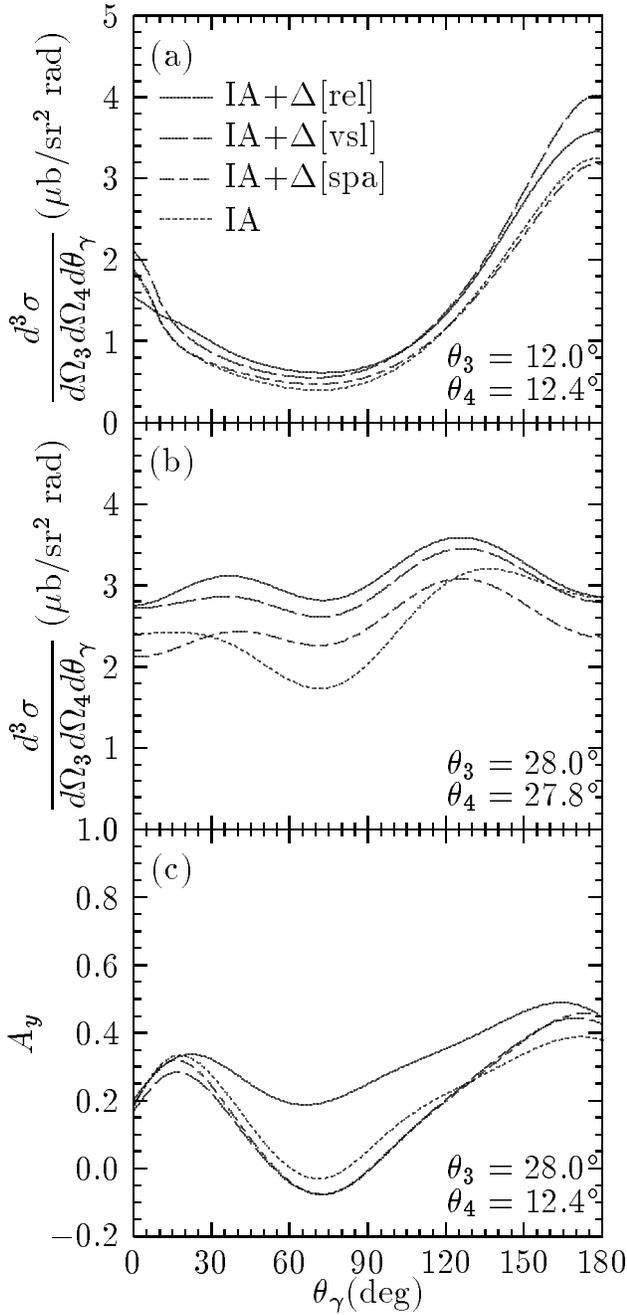}}
\vglue 3 true mm
\caption{ 
RuhrPot results in impulse approximation (IA) as in Fig.~8 compared to
corresponding results with relativistic (rel), vertex-static limit
(vsl) and soft-photon approximation (spa) N$\Delta\gamma$ $\pi-$ 
and $\rho-$exchange currents. A relativistic description of the
N$\Delta\gamma$ exchange currents is necessary.
}}
\end{figure}
\par\noindent
%
\subsection{$\Delta$-isobar Currents}\label{sec4D}
In Fig.~12 we compare the results of calculations using RuhrPot wave functions 
and the relativistic impulse current which either neglect (IA) or retain 
(IA+$\Delta$) the relativistic  N$\Delta\gamma$ $\pi-$ and $\rho-$ exchange 
currents, as prescribed by the relativistic [rel], vertex-static limit [vsl] 
and soft-photon approximation [spa] expressions devel-
\par\noindent
\begin{figure}
\vbox to 7.6 true in {\hbox{\hskip -15 mm \epsfxsize=4.0 true in \epsfbox{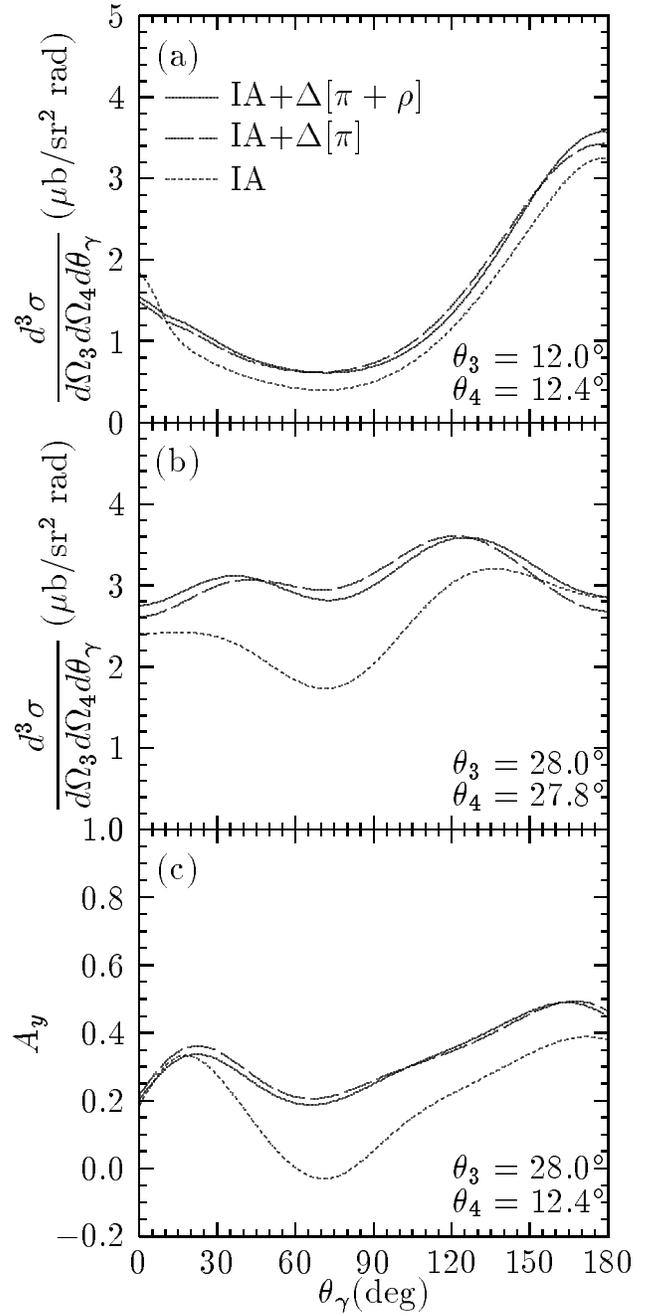}}
\vglue 3 true mm
\caption{ 
RuhrPot results in impulse approximation (IA) as in Fig.~8 compared to
corresponding results with relativistic N$\Delta\gamma$ exchange currents
including $\pi+\rho-$exchange [$\pi+\rho$] and $\pi-$exchange [$\pi$] only.
The RuhrPot model has a relatively weak NN$\rho$ coupling, so that the $\rho-$exchange 
contribution to the N$\Delta\gamma$ currents is small.
}}
\end{figure}
\par\noindent
oped in section~\ref{sec3B}. 
We do not show results for the complete static limit since these results turn 
out to be almost indistinguishable from the vertex static limit. In Fig.~13 we 
decompose the contributions to the relativistic N$\Delta\gamma$ currents into 
$\pi-$ and $\rho-$ exchange contributions. Collectively, these figures show 
that the RuhrPot description of the N$\Delta\gamma$ exchange currents are very 
large and require 
\par\noindent
\begin{figure}
\vbox to 6.0 true in {\hbox{\hskip -15 mm \epsfxsize=4.0 true in \epsfbox{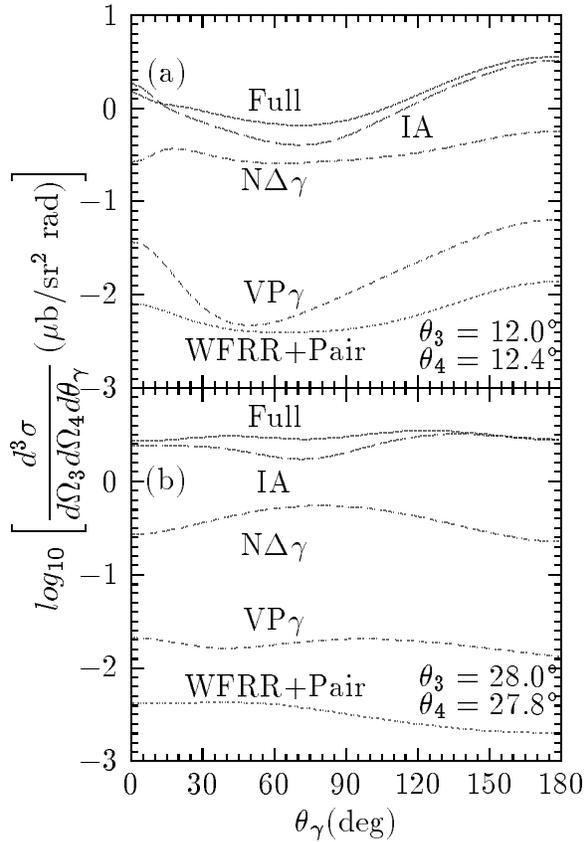}}
\vglue 3 true mm
\caption{ 
Contributions from individual currents to the cross section. These are
(IA) = impulse current including initial-, final- and rescattering correlations,
(N$\Delta\gamma$) with summed $\pi-$ and $\rho-$exchange, 
(VP$\gamma$) = $\rho\pi\gamma$ + $\omega\pi\gamma$ + $\rho\eta\gamma$ + $\omega\eta\gamma$,
(WFRR+Pair) = NN$\gamma$ wave function re-orthonormalization and meson recoil and
$\bar{\rm N}{\rm N}$-pair creation and annihilation currents with $\lambda=0$.
(Full) denotes the sum of all these currents (with interferences).
}}
\end{figure}
\par\noindent
a relativistic description, but that the $\rho-$ exchange 
contribution is comparatively small. 
%
\subsection{Comparison of Exchange Currents}\label{sec4E}
In Fig.~14 we examine the individual contributions of each of the
1- and 2-body currents developed in the previous sections with the 
complete effective current of the present work.
We observe that the N$\Delta\gamma$ exchange currents are substantially
larger than all other contributions  and we recall from Fig.~13 that,
in the RuhrPot description, these are completely dominated by the 
$\pi-$exchange contributions. We observe minor but non-negligible 
contributions from the VP$\gamma$ currents, and we recall from
sec~\ref{sec4C} that the magnitude of these contributions is
essentially determined by the NN$\omega$ coupling constant in the
NN interaction.

\subsection{Comparison With Experiment}\label{sec4F}
In Figs.~15-18 we compare results of calculations using RuhrPot wave functions 
and the relativistic impulse current which either neglect (IA) or retain 
(IA+MEXC) all of the relativistic meson-exchange currents developed in the
present work. These include the wave function re-orthonormalization and
meson-recoil currents, the N$\bar{\rm N}$ creation and annihilation
pair-currents (for both ($\lambda=0$) pv- and ($\lambda=1$) ps- 
NN$\pi$-couplings), the $\rho\pi\gamma$, $\omega\pi\gamma$, 
$\rho\eta\gamma$ and $\omega\eta\gamma$ exchange currents and the 
N$\Delta\gamma$ currents with $\pi-$ and $\rho-$ exchange contributions. 
We observe that the sensitivity to the admixture of ps $(\lambda=1)$ and 
pv $(\lambda=0)$ couplings survives when the currents are consistently combined.

The inclusion of the relativistic exchange currents can be seen to provide a 
reasonable description of the cross section data in Figs.~15-16, but Figs.~17-18 
show that a persistent discrepancy with experiment remains for small $\theta_3$ 
and large $\theta_4$.
The impulse approximation results for the cross section data give 
$\chi^2_{\rm ia}$/datum = 5.8, whereas the complete exchange current calculations 
with ps and pv NN$\pi$ couplings giving $\chi^2_{\rm ps}$/datum = 6.1 and 
$\chi^2_{\rm pv}$/datum = 4.7 respectively. 
Moreover, Fig.~15-16 shows that adopting a ps NN$\pi$ coupling 
produces structure in the cross section that is simply absent in the data.
As such, the present calculations indicate that the data favours $\lambda\sim 0$,
although Figs.~17-18 show that some of the most serious discrepancies with 
experiment cannot be resolved in terms of the NN$\pi$ coupling alone.

The importance of our conclusions on the Lorentz structure of the NN$\pi$ 
vertex are contingent on a reliable data set, so we vigorously stress the need 
for more precise measurements of all observables where the ps- and
pv-couplings give very different results.

\subsection{Some problems with non-perturbative descriptions}\label{sec4G}
Strong interaction transitions between NN, $\Delta$N and $\Delta\Delta$ states 
can be described non-perturbatively with a coupled channel $t$-matrix \cite{GA76b,GA79,GA81}
and indeed have already been used to calculate pp-bremsstrahlung observables 
\cite{MJ94,JO94}. Under these circumstances it may appear curious that we have chosen
to present a perturbative description of such amplitudes as meson-exchange currents.
There are, however, a number of difficulties in applications of these coupled
channel $t$-matrices, the most serious of which appears to arise from the 
inconsistencies that exist between the 
Paris \cite{Paris} NN$\rightleftharpoons$NN  and the Ried parameterized 
version of the static-limit Bochum \cite{GA76b,GA79,GA81} 
$\Delta{\rm N}\rightleftharpoons$NN interaction. 
In particular, the conflicting definitions of the NN$\pi$ and 
N$\Delta\pi$ coupling constants and form factors makes it impossible to 
reliably remove the 
\end{multicols}\widetext
\begin{figure}
\vbox to 8.0 true in {\hbox{\hskip -15 mm \epsfxsize=7.5 true in \epsfbox{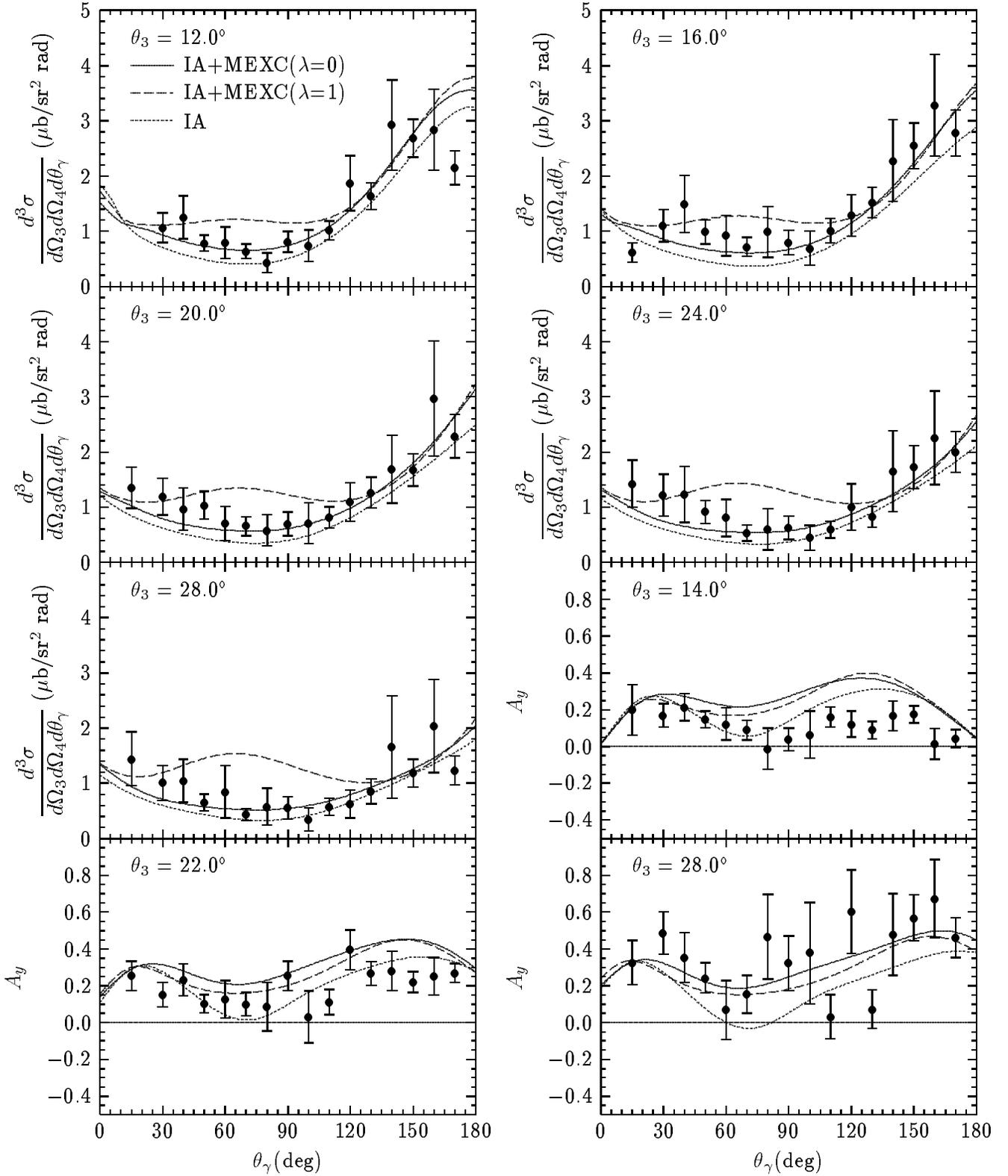}}
\vglue 3 true mm
\caption{ 
Coplanar pp-bremsstrahlung data at $E_{\rm lab}$=280 MeV and $\theta_4=12.4^\circ$ 
compared to RuhrPot calculations including (IA+MEXC) or excluding (IA) relativistic 
meson-exchange currents with pv ($\lambda$=0) or ps ($\lambda$=1) NN$\pi$ interactions.
The exchange currents include wave function re-orthonormalization
and meson-recoil currents, N$\bar{\rm N}$ pair creation and annihilation currents,
$\rho\pi\gamma$ + $\omega\pi\gamma$ + $\rho\eta\gamma$ + $\omega\eta\gamma$ vector-meson decay 
currents and N$\Delta\gamma(\pi,\rho,\Gamma_\Delta$=115~MeV$)$ exchange currents.
}}
\end{figure}
\vfill\eject
\begin{figure}
\vbox to 8.0 true in {\hbox{\hskip -15 mm \epsfxsize=7.5 true in \epsfbox{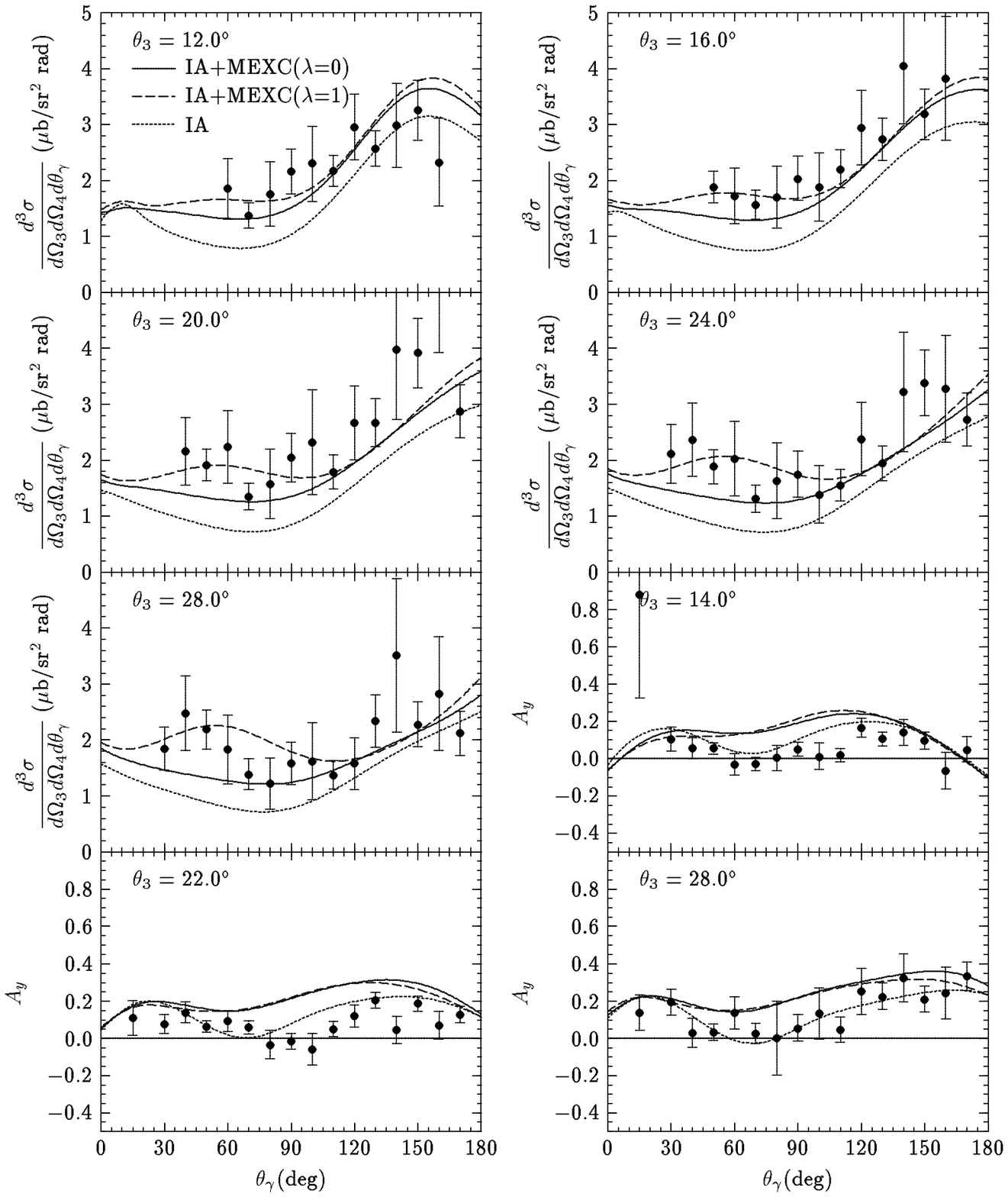}}
\vglue 3 true mm
\caption{ 
Same as Fig.~15 except that $\theta_4=17.3^\circ$.
}}
\end{figure}
\vfill\eject
\begin{figure}
\vbox to 8.0 true in {\hbox{\hskip -15 mm \epsfxsize=7.5 true in \epsfbox{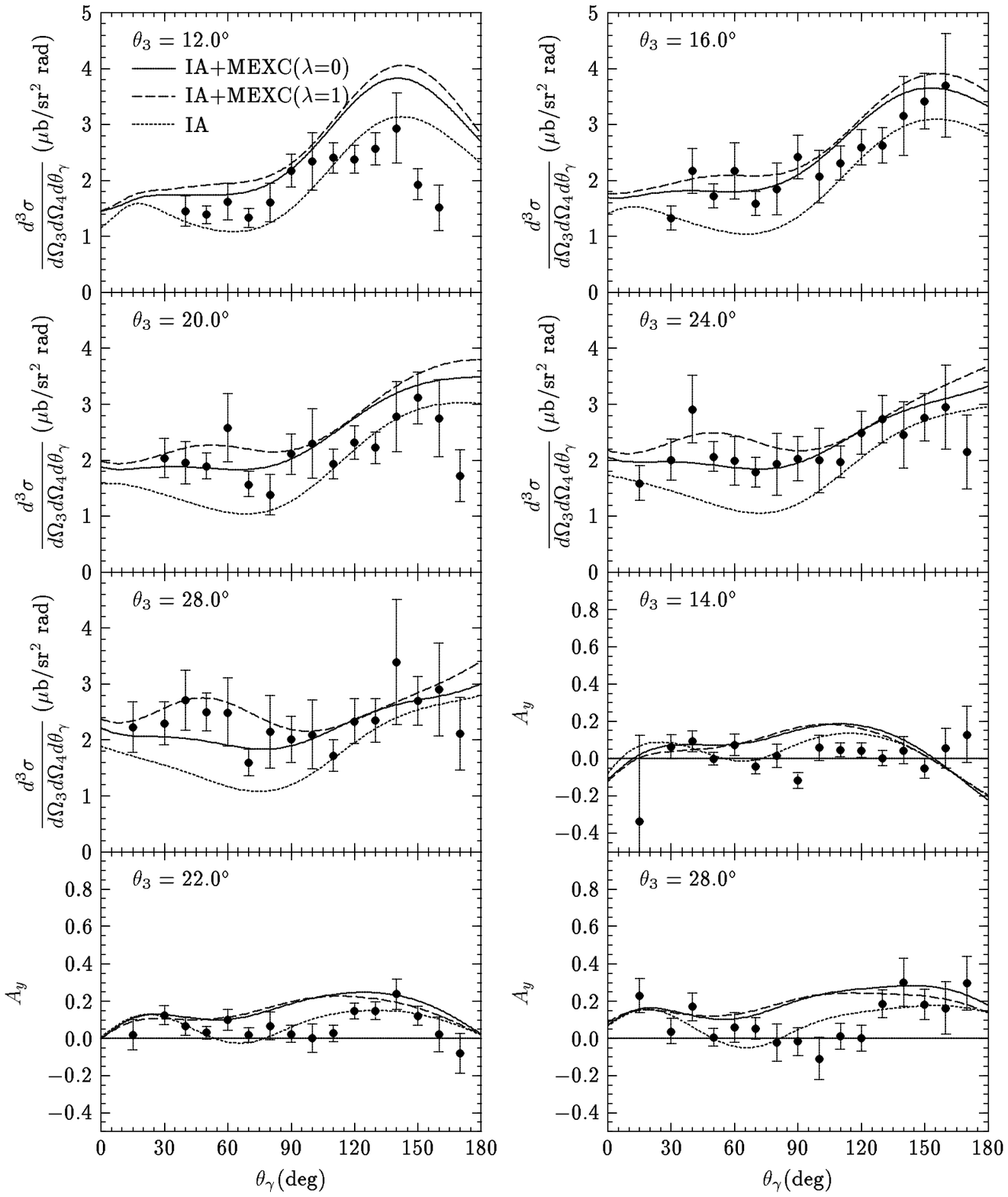}}
\vglue 3 true mm
\caption{ 
Same as Fig.~15 except that $\theta_4=21.2^\circ$.
}}
\end{figure}
\vfill\eject
\begin{figure}
\vbox to 8.0 true in {\hbox{\hskip -15 mm \epsfxsize=7.5 true in \epsfbox{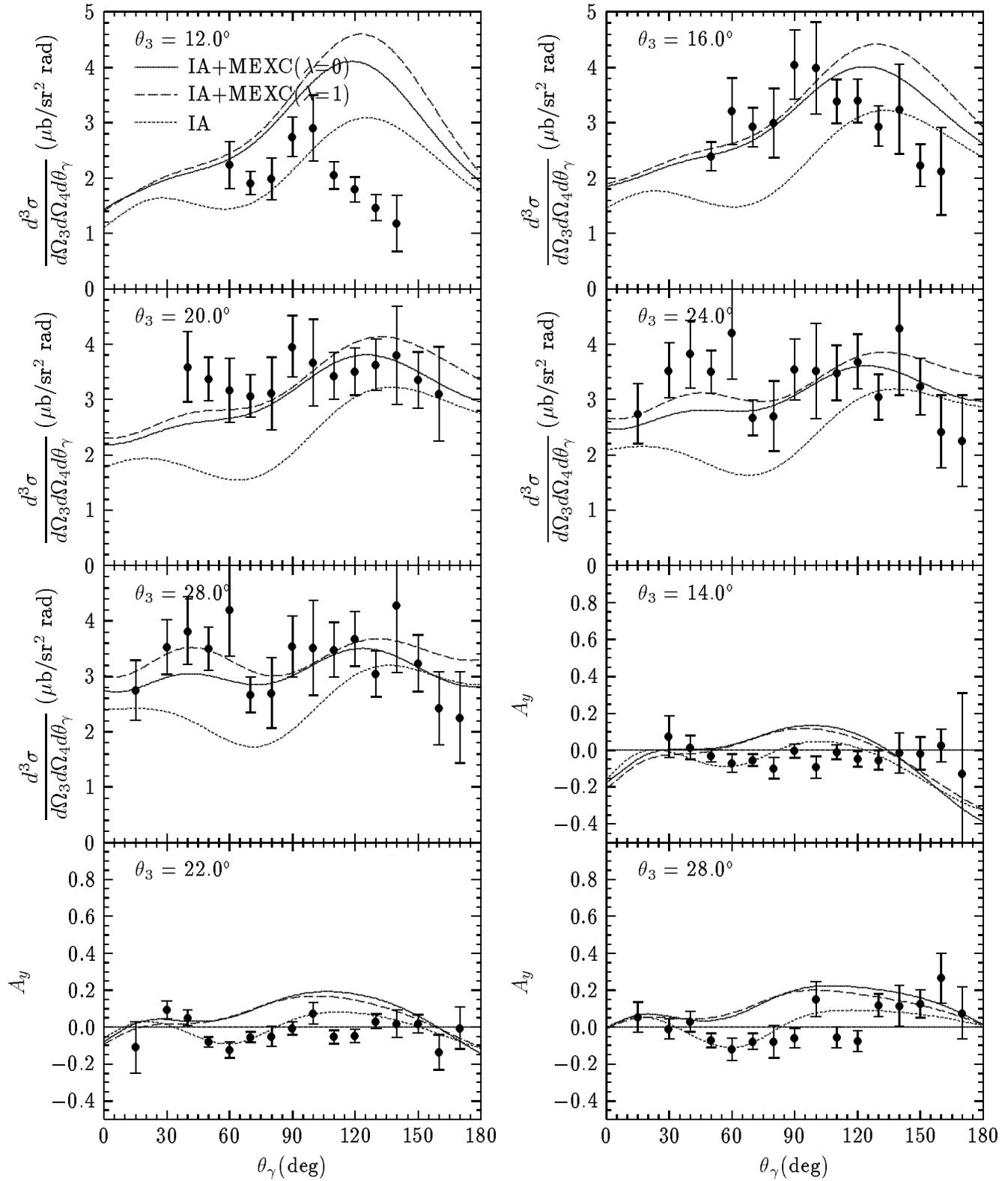}}
\vglue 3 true mm
\caption{ 
Same as Fig.~15 except that $\theta_4=27.8^\circ$.
}}
\end{figure}
\vfill\eject
\begin{multicols}{2}
double-counted NN$\rightleftharpoons$NN amplitudes 
involving intermediate N$\Delta$ states, so that a free parameter is 
introduced to simulate the necessary subtraction.  Any specification of
the two-body currents would suffer a similar ambiguity. Finally, the
assumed Lorentz invariance of the ${\rm NN}\rightleftharpoons\Delta{\rm N}$ 
transition $t$-matrices remains to be investigated.

This leads us to consider a generalization of our formalism to obtain a fully 
consistent and microscopic description of the non-perturbative transitions 
between the NN, $\Delta$N and $\Delta\Delta$ states. More precisely, we will 
identify two minimum requirements of such an approach that appear to have been 
been neglected to date.

Recalling from sec.~\ref{sec2B} the freedom to choose any desired partition of 
the total Hilbert space, we now modify our earlier choice so that the $\Delta$ 
degrees of freedom are included in the $\eta$-space. (Details can be found in 
App~A). Within this approach, the leading order contributions involve not only
the NN$\gamma$ impulse, wave function re-orthonormalization and meson-recoil 
terms, but also N$\Delta\gamma$ initial- and final-state correlations terms 
and N$\Delta\gamma$ wave function re-orthonormalization and meson-recoil terms, 
as shown in Fig.~19. Our earlier specification of the NN$\gamma$ one- and 
two-body currents remains unchanged and will not be further discussed here. 
The additional leading-order contributions involving $\Delta$ isobars are given by,
\begin{eqnarray}\label{eq4.7}
&&\langle \vec{p}_3 \vec{p}_4\vert  
\vec{J}^{{\rm N}\Delta}_{11} 
\vert \vec{p}_1 \vec{p}_2\rangle   
= -g_{\sigma\tau} \sum_{\beta}\Bigl[
\nonumber \\ 
 && D^\beta_1
    H^\sigma_{{\rm N}\Delta\beta}[1](\vec{p}_{1k},\vec{p}_{3}) 
    \vec{J}_{\Delta{\rm N}\gamma}[1](\vec{p}_{1},\vec{p}_{1k}) 
    H^\tau_{{\rm NN}\beta}[2](\vec{p}_{2},\vec{p}_{4}) 
\nonumber \\
+&& D^\beta_3
    \vec{J}_{{\rm N}\Delta\gamma}[1](\vec{p}_{3k},\vec{p}_{3}) 
    H^\sigma_{\Delta{\rm N}\beta}[1](\vec{p}_{1},\vec{p}_{3k}) 
    H^\tau_{{\rm NN}\beta}[2](\vec{p}_{2},\vec{p}_{4}) 
\Bigr]
\nonumber \\ +&& (1,3 \rightleftharpoons 2,4)
\end{eqnarray}  
where $\beta$=$\vec{\pi}$ or $\vec{\rho}$ 
and the factor $ -g_{\sigma\tau}$ and all references to the Lorentz 
indices $\sigma$ and $\tau$ are to be ignored for the scalar mesons. 
The exact form of $D^\beta$ is given in app~A, but here it is sufficient 
to note the non-relativistic limit,
\begin{eqnarray}\label{eq4.8}
&&D^\beta_1 
\,\,{\buildrel \rm nr\over \sim}\,\,
D^\beta_3
\,\,{\buildrel \rm nr\over \sim}\,\,
 {-1 \over 2\omega_\beta(\vec{q}_2) (m_{\Delta}-m +\omega_\beta)}
\nonumber \\ && \qquad\qquad\times
   \Biggl[ 1 + {(m_{\Delta}-m)^2\over
\omega_\beta(\vec{q}_2)[m_{\Delta}-m+\omega_\beta(\vec{q}_2)]} \Biggr]
\end{eqnarray}
The first term in the brackets of eq~(\ref{eq4.8}) represents the 
N$\Delta\gamma$ initial- and final-state strong-interaction amplitudes, and 
the second term gives the N$\Delta\gamma$ wave function re-orthonormalization 
and meson-recoil currents. Unlike their analogous NN$\gamma$ terms, where only 
one-body currents survive in the static limit, the N$\Delta\gamma$ wave 
function re-orthonormalization and meson-recoil currents do {\it not} vanish 
in static limit - even for soft photons (cf. ref~\cite{JO94}).
Indeed noting that $m_{\Delta}-m\sim 2m_\pi$ shows that these contributions
make an important contribution to the intermediate-state $\Delta$N amplitudes 
in the low energy observables. 

\begin{figure}
\vbox to 5.2 true in {\hbox{\hskip 0 mm \epsfxsize=3.4 true in \epsfbox{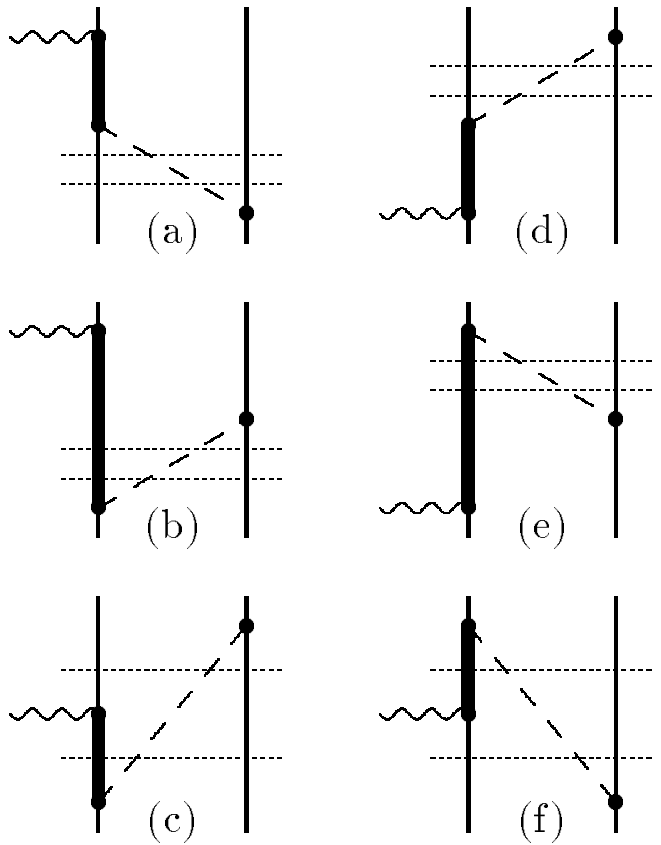}}
\vglue 3 true mm
\narrowtext
\caption{ 
N$\Delta\gamma$ wave function re-orthonormalization 
and meson-recoil exchange currents. These currents are necessary to 
preserve the orthonormality of the initial- and final-state wave functions
described by NN, $\Delta$N coupled channel transition $t$-matrices.
}}
\end{figure}
\par\noindent
Since the coupled channel $t$-matrices have been calculated only in the 
barycentric frame, it remains to either calculate their boost operators, or 
demonstrate that they are effectively Lorentz invariant. In sec~\ref{sec4A} 
(see Fig.~9) we demonstrated such invariance for a toy-model boson-exchange
potential and commented on the reliability of guessing minimal relativity 
factors.  We anticipate that a similar approximate invariance will probably 
hold for the coupled channel NN$\rightleftharpoons$NN 
$t$-matrices involving intermediate $\Delta$N-states. 
However, in Fig.~20 we observe that the corresponding N$\Delta\gamma$ initial- 
and final-state interaction amplitudes calculated in the average barycentric 
frame of eq.~(\ref{eq3.32}) are seriously different from the corresponding 
results evaluated in the I- and F-frames of eqs.~(\ref{eq3.37}) and 
(\ref{eq3.39}).  This shows that the leading-order contributions to the 
coupled channel $\Delta{\rm N}\rightleftharpoons$NN and 
NN$\rightleftharpoons\Delta$N $t$-matrices are poorly approximated under the 
assumption that they are Lorentz invariant. 

We conclude that a non-perturbative description of the $\Delta$ isobar 
amplitudes via coupled channel $t$-matrix calculations involves two 
complications:
\begin{figure}
\vbox to 6.0 true in {\hbox{\hskip -15 mm \epsfxsize=4.0 true in \epsfbox{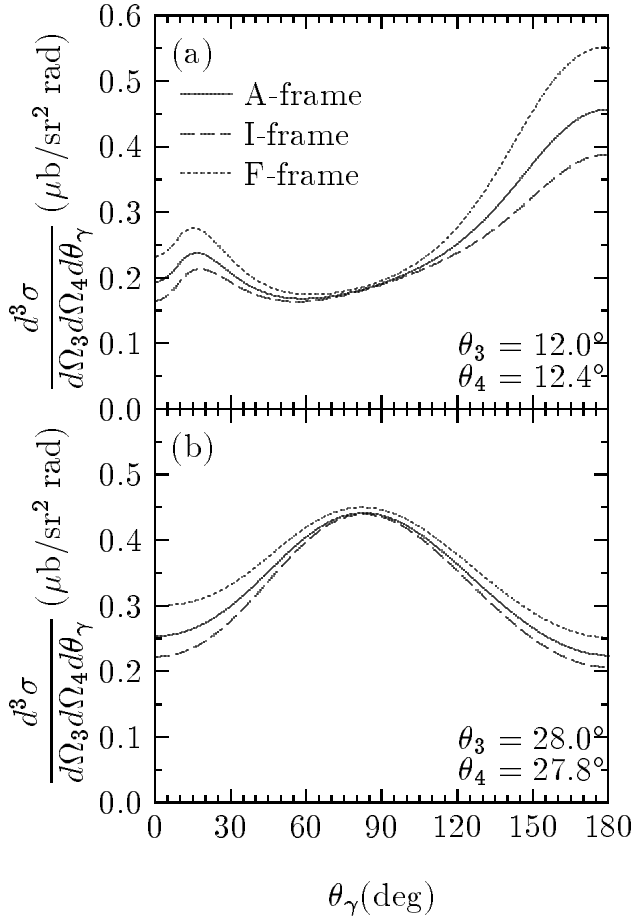}}
\vglue 3 true mm
\caption{ 
Comparison of perturbative N$\Delta$ initial- and final-state correlation
amplitudes calculated in the (average barycentric) A-frame and compared to
corresponding results that are obtained in the initial- and final-state 
barycentric frames. The discrepancies show the need for boost operators
and indicate errors of about 20\% result when the 
NN$\protect\rightleftharpoons\Delta$N $t$-matrix is assumed to be Lorentz invariant.
}}
\end{figure}
\par\noindent
\begin{itemize}
\item
There are non-vanishing contributions from the N$\Delta\gamma$ wave function 
re-orthonormalization and meson-recoil terms. The neglect of these two-body 
currents contributes serious errors at low energies.
\item
The initial- and final-state N$\Delta\gamma$ interactions are poorly 
approximated in the absence of boost operators. The neglect of boost operators 
contributes serious errors at high energies.
\end{itemize}
As such, a meaningful specification of non-perturbative $\Delta$ contributions
requires the calculation of two-body currents and boost operators. 
This has not been recognized in the past. 
Although the first requirement can easily be satisfied by retaining a subset of 
the exchange currents we have presented in this work, there are outstanding problems
that need to be solved if boost operators are to be defined beyond ${\cal O}(1/m^4)$.  
Such developments stand as a challenge for future theoretical work but can only be 
approached within a model providing a consistent and microscopic description of all 
meson-baryon dynamics 
%

\subsection{Relativistic Effects}\label{sec4H}
In earlier work \cite{JE95} we presented selected pp-bremsstrahlung 
observables calculated using RuhrPot wave functions and the relativistic 
impulse current. The re-scattering contributions of Fig.~1c were retained,
as were the relativistic $\rho\pi\gamma$, $\omega\pi\gamma$, $\rho\eta\gamma$ 
and $\omega\eta\gamma$ exchange currents. We also included the N$\Delta\gamma$ 
currents with $\pi-$ and $\rho-$ exchange in the complete static limit,
as defined in sec~\ref{sec3B5}. No form of soft-photon approximation was 
adopted at any stage.

In the present work we have extend the exchange currents to include the wave
function re-orthonormalization and meson-recoil currents of eq.~(\ref{eq3.8})
and the N$\bar{\rm N}$-pair creation and annihilation currents (for both ($\lambda=0$)
pv- and ($\lambda=1$) ps- NN$\pi$-couplings) of eq.~(\ref{eq3.10})
that are required for a truly relativistic description of the NN$\gamma$ vertex.
We have also replaced the complete static limit description of the dominant
N$\Delta\gamma(\pi)$ exchange current with the relativistic up-grade of
eq.~(\ref{eq3.21}).

In Fig.~21 we consider the magnitude of these purely relativistic effects by
comparing with our earlier descriptions of the meson-exchange currents.
We consider here only the pv NN$\pi$ coupling ($\lambda=0$). From Figs~10,12 and 13 we 
already know that the largest difference between these exchange current results stems 
from the relativistic corrections to the N$\Delta\gamma(\pi)$ contribution. 
In addition, from Fig.~12 and sections~\ref{sec3B5} and \ref{sec4D} we note that 
this difference results from the neglect of the N$-\Delta$ mass difference in the
complete static limit exchange current operators used in ref.~\cite{JE95}.
Although we find no need to change the qualitative conclusions reported in 
ref~\cite{JE95}, a comparison of pp-bremsstrahlung calculations with experimental
data near the $\pi$-production threshold clearly requires a relativistic description 
of the isoscalar meson-exchange currents.

\section{Conclusions}\label{sec5}
We have presented a parameter-free and relativistic extension of the
RuhrPot meson-baryon model to define the dominant isoscalar meson-exchange
currents. These include the first relativistic calculations for the wave 
function re-orthonormalization, meson-recoil and N$\bar{\rm N}$-pair creation 
and annihilation currents in pp-bremsstrahlung. We also included the fully 
relativistic $\rho\pi\gamma$, $\omega\pi\gamma$, $\rho\eta\gamma$ and 
$\omega\eta\gamma$ currents and a relativistic upgrade to our earlier 
N$\Delta\gamma(\pi,\rho)$ exchange currents \cite{JE95,JE94}.

The results of these calculations show that the meson-exchange contributions
to the pp-bremsstrahlung  observables below the $\pi-$production threshold are
large. As 
%
\end{multicols}\widetext
\begin{figure}
\vbox to 7.2 true in {\hbox{\hskip -15 mm \epsfxsize=7.5 true in \epsfbox{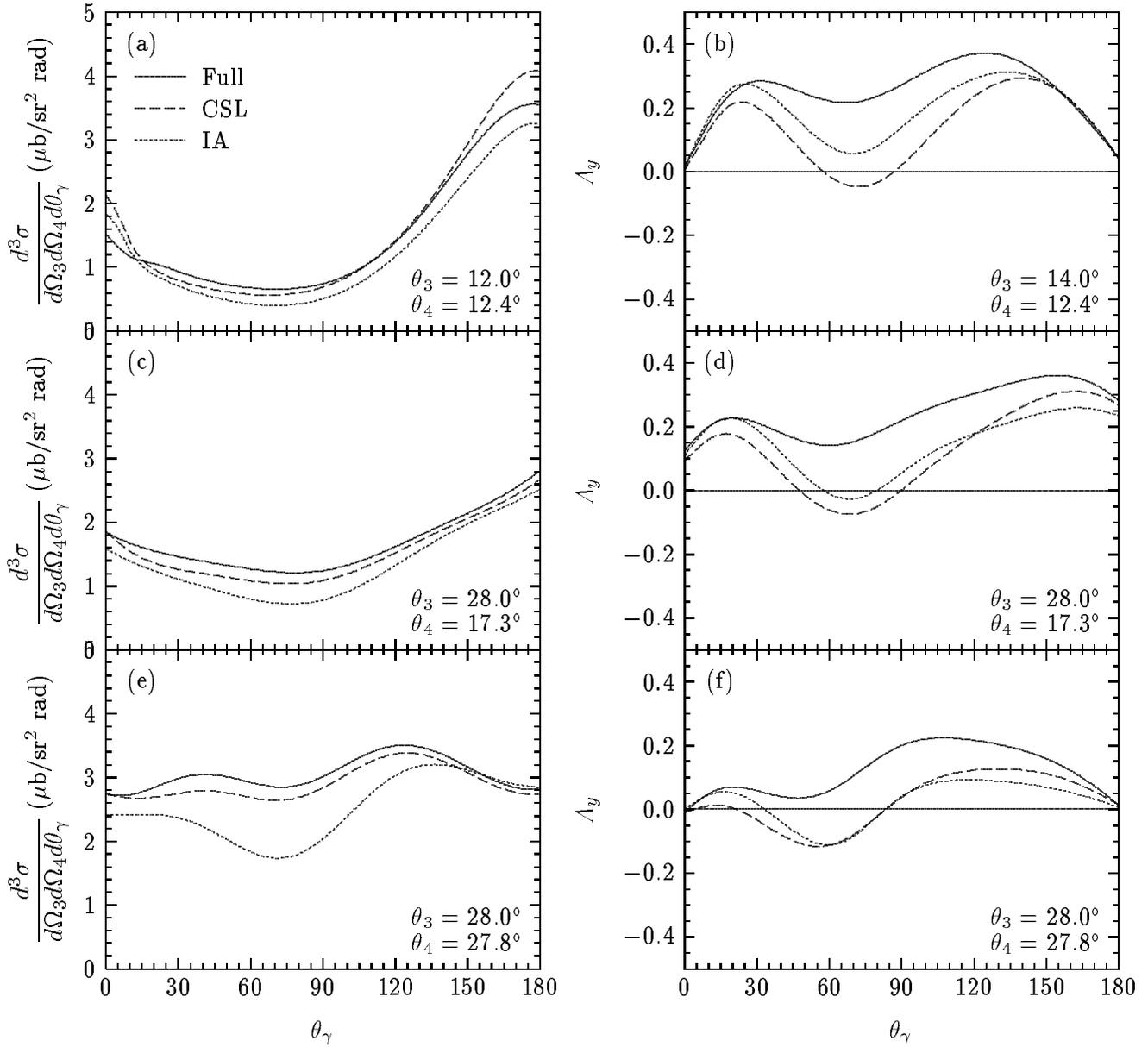}}
\vglue 3 true mm
\caption{ 
Selected results from figs~15-18 using the ps NN$\pi$ coupling
with (Full) and without (IA) the relativistic exchange currents,
now compared with the (CSL) results of ref [1]. The CSL results differ
from the full ones in that the wave function re-orthonormalization, 
meson recoil and N$\bar{\rm N}$-pair currents are neglected entirely
and the N$\Delta\gamma(\pi)$ exchange current is taken in the 
complete static limit. 
The largest numerical difference in the meson-exchange current results
can be traced to the neglect of the N$-\Delta$ mass difference in the
CSL operators. See also sections~\protect\ref{sec3B5} and \protect\ref{sec4D}
for a detailed discussion.
}}
\end{figure}
\begin{multicols}{2}
such, a meaningful interpretation of experiment obviously requires a
completely consistent description of the meson-baryon dynamics defining the
NN-interaction, the exchange currents and the form factors that they contain.

Although the wave function re-normalization and meson-recoil contributions 
are well known to cancel in the static limit for soft photons, this 
cancellation is poorly satisfied in bremsstrahlung  experiments where the
kinematics has been contrived to maximize the photon energy. As such, retaining
these contributions is necessary if the orthonormality of the wave functions
is to be preserved. Although these two-body currents have never before been 
included in bremsstrahlung calculations, they are necessary for a relativistic 
description of the NN$\gamma$ vertex.

The motivation for developing such a relativistic scheme is found in one of 
the oldest outstanding puzzles in nuclear physics. The Lorentz structure of 
the NN$\pi$ Lagrangian ${\cal L}_{{\rm NN}\pi}$ is universally accepted 
to comprise an unknown mixture of ps $(\lambda=1)$ and pv $(\lambda=0)$ 
couplings. This mixture cannot be distinguished by any non-relativistic 
calculation and, for quite different reasons, makes almost no impact on
relativistic calculations for the $N\pi$ scattering lengths. However,
within our relativistic framework we have shown that the existing
pp-bremsstrahlung data indicates $\lambda$ is small. This is surely the most
reliable assessment of the NN$\pi$ Lagrangian available to date. 

We quantified the VP$\gamma$ exchange current contributions and noted their
obvious relationship to the NN-interaction. While the small vector-meson 
couplings in the RuhrPot NN interaction render these effects no larger
than the NN-interaction model differences in impulse approximation, we noted 
the necessity to include these currents in calculations for models using 
large NN$\omega$ and NN$\rho$ couplings (e.g. Bonn~B).
However, the largest isovector exchange current in pp-bremsstrahlung results
from intermediate-state isobar excitation via $\pi$-exchange. We investigated
a series of approximations for this current and found that a relativistic 
description is necessary. Given the practical importance of this result, a 
compact closed form expression was provided and the sources of error in various
approximations were identified.

We demonstrated our earlier assertion \cite{JE95,JE94}
that the N$\Delta\gamma$ wave function re-orthonormalization and meson-recoil 
contributions do {\it not} vanish - even in the static limit for soft photons. 
Their neglect in recent applications \cite{MJ94,JO94} indicates that the
the orthonormality of the initial and final state wave functions is not 
preserved. In a perturbative analysis we showed that the assumed Lorentz 
invariance of the NN$\rightleftharpoons$NN interactions is accurate to
about 3\%, but that a similar assumption \cite{MJ94,JO94} for the 
NN$\rightleftharpoons\Delta$N interaction implies unacceptable errors of 
around 20\%. We noted that a non-perturbative development of our parameter-free 
calculations requires inclusion of the N$\Delta\gamma$ wave function 
re-orthonormalization and meson-recoil exchange currents (for which exact 
expressions were provided) and the application of boost operators. 
The need for more precise experimental data has been stressed.
\acknowledgements
This work is supported by COSY-KFA J\"ulich (41140512),
the Deutsche Forschungsgemeinschaft (Ga 153/11-4) and
BMFT.
\appendix
\section{Non-perturbative N$\Delta\gamma$ amplitudes}\label{appA}
Within the formalism of sec~\ref{sec2B}, a non-perturbative description
of the $\Delta$ isobar contributions requires our partition of the
Hilbert space to modified such that,
\begin{eqnarray}\label{eqA.1} &&
{\cal H}_{\eta_1}= \Bigl\{ \vert \rm NN\rangle \Bigr\}
,\quad 
{\cal H}_{\eta_2}= \Bigl\{ \vert \Delta{\rm N}\rangle \Bigr\}
,\quad 
{\cal H}_{\eta_3}= \Bigl\{ \vert {\rm N}\Delta\rangle \Bigr\}
,
\nonumber \\ &&
{\cal H}_{\eta_4}= \Bigl\{ \vert \Delta\Delta\rangle \Bigr\}
,\quad 
{\cal H}_{\lambda} = \Bigl\{ \vert \hbox{the rest}\rangle  \Bigr\} 
\end{eqnarray}
with projection operators satisfying $\eta=\sum_{i=1}^4\eta_i$ and
$\eta_i\eta_j=\eta_i\delta_{ij}$. Denoting an arbitrary operator 
causing transitions from the $\eta_i$-space to the $\eta_j$-space
as $\eta_j{\cal O}\eta_i={\cal O}_{ji}$, we require (in analogy to 
eq~(\ref{eq2.22})) matrix elements of the form,
\begin{eqnarray}\label{eqA.2}
[{\cal M}_{fi}]_{11} =&&
N \vec{\epsilon} (\vec{k}, \lambda) \sum_{i,j=1}^{4}
\langle \widetilde{\vec{p}_3 \vec{p}_4;\alpha_f}\vert \eta_1
[\openone+t_{1i}^{(-)\dag}G_{1i}]
\nonumber \\ &&
\times [\vec{J}_{\rm eff}]_{ij}
[\openone+G_{j1}t_{j1}^{(+)}]
\eta_1 \vert \widetilde{\vec{p}_1 \vec{p}_2;\alpha_i}\rangle 
\end{eqnarray}
where we have denoted the effective 1+2-body current density as
\begin{eqnarray}\label{eqA.3}
[\vec{J}_{\rm eff}]_{ij} &=&
\eta_i \Bigl[ J + J A + A^\dagger J 
            + A^\dagger J A
\nonumber \\ && 
            - {1\over 2} J A^\dagger A 
            - {1\over 2} A^\dagger A J + \cdots\Bigr] \eta_j
\end{eqnarray}
First consider the amplitudes involving an effective current describing
NN$\rightarrow$NN transitions, which must therefore be
taken between inital- and final-state wave functions constructed from 
the NN$\rightarrow$NN $t$-matrix.
These amplitudes include the dominant NN$\gamma$ impulse currents, as well as 
contributions from the NN$\gamma$ wave function re-orthonormalization and 
meson-recoil currents. (Adopting (\ref{eqA.1}) does not alter the exact 
expressions for these currents given by eq.~(\ref{eq3.8}) and (\ref{eq3.9})). 
However, we now have additional contributions, as shown in Fig.~19, which are
given by eq.~(\ref{eq4.7}) 
with 
\begin{eqnarray}\label{eqA.4}
D^\beta_1 &=& 
{1\over [E_3-E_{\Delta 1k}-\omega_\beta(\vec{q}_2)]
        [E_2-E_4-\omega_\beta(\vec{q}_2)]}
\nonumber \\ &&
-{1 \over 2[E_4-E_2-\omega_\beta(\vec{q}_2)] 
          [E_1-E_3-\omega_\beta(\vec{q}_2)] }
\nonumber \\ &&
-{1\over 2[E_3-E_{\Delta 1k}-\omega_\beta(\vec{q}_2)]
}\nonumber\\&& \qquad \times {1 \over
         [E_1+E_2-E_{\Delta 1k}-E_4-\omega_\beta(\vec{q}_2)]}
\nonumber \\ 
D^\beta_3  &=&
{1\over [E_4-E_2-\omega_\beta(\vec{q}_2)]
        [E_1-E_{\Delta 3k}-\omega_\beta(\vec{q}_2)]}
\nonumber \\ &&
-{1 \over 2[E_3-E_1-\omega_\beta(\vec{q}_2)] 
          [E_2-E_4-\omega_\beta(\vec{q}_2)] }
\nonumber \\ &&
-{1\over 2[E_3+E_4-E_{\Delta 3k}-E_2-\omega_\beta(\vec{q}_2)] 
}\nonumber\\&& \qquad \times {1 \over
         [E_1-E_{\Delta 3k}-\omega_\beta(\vec{q}_2)]}
\end{eqnarray}
where $\beta$=$\vec{\pi}$ or $\vec{\rho}$. Unlike their analogous NN$\gamma$ 
terms, these N$\Delta\gamma$ wave function re-orthonormalization and 
meson-recoil currents do {\it not} vanish in the static limit. Their 
inclusion is required to preserve the orthonormality of the hadronic wave 
functions. 

Next consider the amplitudes involving an effective current describing
$\Delta{\rm N}\rightleftharpoons$NN transitions, which must therefore be 
taken between inital- and final-state wave functions constructed from 
$\Delta{\rm N}\rightleftharpoons$NN and NN$\rightarrow$NN $t$-matrices.
\begin{eqnarray}\label{eqA.5}
[{\cal M}_{fi}]_{11} 
&=& N \vec{\epsilon} (\vec{k}, \lambda) .
\langle \widetilde{\vec{p}_3 \vec{p}_4;\alpha_f}\vert \eta_1
[\openone+t_{11}^{(-)\dag}G_{11}]
\nonumber \\ && \times
[\vec{J}_{\rm eff}]_{12}
[G_{21}t_{21}^{(+)}]
\eta_1 \vert \widetilde{\vec{p}_1 \vec{p}_2;\alpha_i}\rangle 
\nonumber \\  &+&
N \vec{\epsilon} (\vec{k}, \lambda) .
\langle \widetilde{\vec{p}_3 \vec{p}_4;\alpha_f}\vert \eta_1
[t_{12}^{(-)\dag}G_{12}]
\nonumber \\ && \times
[\vec{J}_{\rm eff}]_{21}
[\openone+G_{11}t_{11}^{(+)}]
\eta_1 \vert \widetilde{\vec{p}_1 \vec{p}_2;\alpha_i}\rangle 
\nonumber \\ &+& (1,3 \rightleftharpoons 2,3)
\end{eqnarray}
These are the N$\Delta\gamma$ initial- and final-state interaction amplitudes
and should obviously be specified in a consistent frame.
However, the $t$-matrices are available only in the barycentric frame, yet 
are required for the initial- and final-state interactions of eq.~(\ref{eqA.5})
in barycentric frames which differ by the photon momentum. The procedure
adopted in recent works \cite{MJ94,JO94} is to attach (guessed) off-shell
minimal relativity factors and assume this renders the  initial- and
final-state interaction terms individually Lorentz invariant. Rather than 
adopt this assumption,  we provide an exact specification of the leading-order 
contributions to these amplitudes via eq.~(\ref{eq4.7}) with,
\begin{eqnarray}\label{eqA.6}
D^\beta_1 &=& {1\over 2(E_3+E_4-E_{\Delta 1k}-E_2)}
\nonumber \\ && \times
\Bigl[ {1\over (E_3-E_{\Delta 1k} -\omega_\beta)} 
     - {1\over (E_4-E_2 -\omega_\beta)} \Bigr]
\nonumber \\ 
D^\beta_3 &=& {1\over 2(E_1+E_2-E_{\Delta 3k}-E_4})
\nonumber \\ && \times
\Bigl[ {1\over (E_1-E_{\Delta 3k} -\omega_\beta)} 
     - {1\over (E_2-E_4 -\omega_\beta}) \Bigr]
\end{eqnarray}
\end{multicols}\widetext
%
\appendix
\addtocounter{section}{1}
\LeftLineBreak
\section{Vertex Functions}\label{appB}
\subsection{Interaction energies for strong vertices}
For a meson with momentum $\vec{q}$ and mass $m_\beta$
we define,
\begin{equation}
N_\beta=g_{{\rm NN}\beta}\left\{{\cal E}_i{\cal E}_f\over(2\pi)^3 8\omega_{\beta}E_f E_i\right\}^{1/2},\qquad
\omega_\beta=\sqrt{(\vec{q})^2 + m_\beta^2}
\end{equation}
The interaction energies
$\langle 0\vert  b(\vec{p}_f)\, -\int d^3x \Gamma_{{\rm NN}\beta}\, b^\dagger(\vec{p}_i)\vert 0\rangle $
for the coupling of mesons to the positive frequency nucleon current are,
\begin{mathletters}
\begin{eqnarray}
H_{{\rm NN}\pi}  &=& N_\pi h_{{\rm NN}\pi}\, \vec{\tau}, \qquad
H_{{\rm NN}\rho}  =  N_\rho \epsilon_\mu h^\mu_{{\rm N}{\rm N}\rho}\, \vec{\tau},  \qquad
H_{{\rm NN}\delta}= N_\delta h_{{\rm NN}\delta}\,\vec{\tau} ,
\\
h_{{\rm NN}\pi} &=&
-i F_{{\rm NN}\pi} 
\vec{\sigma}.\left[ {\vec{p}_f\over{\cal E}_f} - {\vec{p}_i\over{\cal E}_i}\right]
\\
h^0_{{\rm N}{\rm N}\rho} &=&
  \Biggl[ F_{{\rm NN}\rho}^{(1)} + \kappa_\rho F_{{\rm NN}\rho}^{(2)} \Bigl[ 1 - {E_f+E_i\over 2m} \Bigr] \Biggr]
+ \Biggl[ F_{{\rm NN}\rho}^{(1)} + \kappa_\rho F_{{\rm NN}\rho}^{(2)} \Bigl[ 1 + {E_f+E_i\over 2m} \Bigr] \Biggr]
  \left[ {\vec{p}_f.\vec{p}_i\over{\cal E}_f{\cal E}_i}   + { i\vec{\sigma}.(\vec{p}_f\times\vec{p}_i) \over{\cal E}_f{\cal E}_i}  \right]
\\
\vec{h}_{{\rm N}{\rm N}\rho} &=&
  \left[ F_{{\rm NN}\rho}^{(1)} + \kappa_\rho F_{{\rm NN}\rho}^{(2)} \right]
  \left[ {\vec{p}_f\over{\cal E}_f} + {\vec{p}_i\over{\cal E}_i} 
         + i\vec{\sigma}\times\Bigl[ {\vec{p}_f\over{\cal E}_f} - {\vec{p}_i\over{\cal E}_i} \Bigr]
  \right]
-
  {\kappa_\rho F_{{\rm NN}\rho}^{(2)}\over 2m} (\vec{p}_f+\vec{p}_i)
  \left[ 1 - {\vec{p}_f.\vec{p}_i \over{\cal E}_f{\cal E}_i}  - {i\vec{\sigma}.(\vec{p}_f\times\vec{p}_i) \over{\cal E}_f{\cal E}_i}  \right]
\\
h_{{\rm NN}\delta} &=&
F_{{\rm NN}\delta} 
\left[1 -  {\vec{p}_f.\vec{p}_i\over{\cal E}_f{\cal E}_i}   + { i\vec{\sigma}.(\vec{p}_f\times\vec{p}_i) \over{\cal E}_f{\cal E}_i}  \right]
\end{eqnarray}
\end{mathletters}
The interaction energies $\langle 0\vert  b(\vec{p}_f)\, -\int d^3x \Gamma_{{\rm NN}\beta}\, d(-\vec{p}_i)\vert 0\rangle $ 
for meson couplings to the pair-creation current are,
\begin{mathletters}
\begin{eqnarray}
H_{\bar{\rm N}{\rm N}\pi} &=& N_\pi h_{\bar{\rm N}{\rm N}\pi} \vec{\tau}, \qquad
H_{\bar{\rm N}{\rm N}\rho} = N_\rho \epsilon_\mu h^\mu_{\bar{\rm N}{\rm N}\rho}\,\vec{\tau},\qquad
H_{\bar{\rm N}{\rm N}\delta}= N_\delta h_{\bar{\rm N}{\rm N}\delta} \,\vec{\tau} 
\\
h_{\bar{\rm N}{\rm N}\pi} &=&
+ i F_{{\rm NN}\pi} 
  \Biggl\{
  \Bigl[ 1 - (1-\lambda){E_i\over m} \Bigr]
 +\Bigl[ 1 + (1-\lambda){E_i\over m} \Bigr]
  \Bigl[ 
  {\vec{p}_f.\vec{p}_i \over {\cal E}_f {\cal E}_i }
  +i\vec{\sigma}.\Bigl( {\vec{p}_f\times\vec{p}_i \over {\cal E}_f {\cal E}_i }\Bigr) 
  \Bigr]
  \Biggr\}
\\
h^0_{\bar{\rm N}{\rm N}\rho} &=&
  \left[ F_{{\rm NN}\rho}^{(1)} + \kappa_\rho F_{{\rm NN}\rho}^{(2)} \Bigl[ 1 + {E_f-E_i\over 2m} \Bigr] \right]
  {\vec{\sigma}.\vec{p}_f\over{\cal E}_f}
-
  \left[ F_{{\rm NN}\rho}^{(1)} + \kappa_\rho F_{{\rm NN}\rho}^{(2)} \Bigl[ 1 - {E_f-E_i\over 2m} \Bigr] \right]
  {\vec{\sigma}.\vec{p}_i\over{\cal E}_i}
\\
\vec{h}_{\bar{\rm N}{\rm N}\rho} &=&
 \left[ F_{{\rm NN}\rho}^{(1)} + \kappa_\rho F_{{\rm NN}\rho}^{(2)}\Bigl[1-{E_i\over m} \Bigr] \right] \vec{\sigma}
+ 
{\kappa_\rho F_{{\rm NN}\rho}^{(2)} \over 2m} (\vec{p}_f+\vec{p}_i) \left[ {\vec{\sigma}.\vec{p}_f\over{\cal E}_f} 
 + {\vec{\sigma}.\vec{p}_i\over{\cal E}_i} \right]
\nonumber \\ &&
+ \left[ F_{{\rm NN}\rho}^{(1)} + \kappa_\rho F_{{\rm NN}\rho}^{(2)}\Bigl[1+{E_i\over m} \Bigr] \right]
  \left[ {i(\vec{p}_f\times\vec{p}_i) \over {\cal E}_f{\cal E}_i } 
       + {\vec{\sigma}(\vec{p}_f.\vec{p}_i)     \over {\cal E}_f{\cal E}_i } 
       - {\vec{p}_f(\vec{\sigma}.\vec{p}_i)     \over {\cal E}_f{\cal E}_i } 
       - {(\vec{\sigma}.\vec{p}_f)\vec{p}_i     \over {\cal E}_f{\cal E}_i } 
  \right]
\\
h_{\bar{\rm N}{\rm N}\delta} &=&
- F_{{\rm NN}\delta}
\left[  {\vec{\sigma}.\vec{p}_f\over{\cal E}_f} + {\vec{\sigma}.\vec{p}_i\over{\cal E}_i}\right]
\end{eqnarray}
\end{mathletters}
The interaction energies 
$\langle 0\vert  d^\dagger(-\vec{p}_f)\, -\int d^x \Gamma_{{\rm NN}\beta}\, b^\dagger(\vec{p}_i)\vert 0\rangle $ 
for meson couplings to the pair-annihilation current are,
\begin{mathletters}
\begin{eqnarray}
H_{{\rm N}\bar{\rm N}\pi} &=& N_\pi h_{{\rm N}\bar{\rm N}\pi}\, \vec{\tau}, \qquad
H_{{\rm N}\bar{\rm N}\rho} = N_\rho \epsilon_\mu h^\mu_{{\rm N}\bar{\rm N}\rho}\,\vec{\tau} \qquad
H_{{\rm N}\bar{\rm N}\delta} = N_\delta h_{{\rm N}\bar{\rm N}\delta} \,\vec{\tau} 
\\
h_{{\rm N}\bar{\rm N}\pi} &=&
- i F_{{\rm NN}\pi} 
  \Biggl\{
  \Bigl[ 1 - (1-\lambda){E_f\over m} \Bigr]
 +\Bigl[ 1 + (1-\lambda){E_f\over m} \Bigr]
  \Bigl[ 
  {\vec{p}_f.\vec{p}_i \over {\cal E}_f {\cal E}_i }
  +i\vec{\sigma}.\Bigl( {\vec{p}_f\times\vec{p}_i \over {\cal E}_f {\cal E}_i }\Bigr) 
  \Bigr]
  \Biggr\}
\\
h^0_{{\rm N}\bar{\rm N}\rho} &=&
  \left[ F_{{\rm NN}\rho}^{(1)} + \kappa_\rho F_{{\rm NN}\rho}^{(2)} \Bigl[ 1 - {E_f-E_i\over 2m} \Bigr] \right]
  {\vec{\sigma}.\vec{p}_i\over{\cal E}_i} 
- \left[ F_{{\rm NN}\rho}^{(1)} + \kappa_\rho F_{{\rm NN}\rho}^{(2)} \Bigl[ 1 + {E_f-E_i\over 2m} \Bigr] \right]
  {\vec{\sigma}.\vec{p}_f\over{\cal E}_f}
\\
\vec{h}_{{\rm N}\bar{\rm N}\rho} &=&
  \left[F_{{\rm NN}\rho}^{(1)} +\kappa_\rho F_{{\rm NN}\rho}^{(2)}\Bigl[1-{E_f\over m}\Bigr]\right]\vec{\sigma}
+ {\kappa_\rho F_{{\rm NN}\rho}^{(2)} \over 2m} (\vec{p}_f+\vec{p}_i) \left[ {\vec{\sigma}.\vec{p}_f\over{\cal E}_f} + {\vec{\sigma}.\vec{p}_i\over{\cal E}_i} \right]
\nonumber \\ &+&
  \left[ F_{{\rm NN}\rho}^{(1)} + \kappa_\rho F_{{\rm NN}\rho}^{(2)}\Bigl[1+{E_f\over m} \Bigr] \right]
  \left[ {i(\vec{p}_f\times\vec{p}_i) \over {\cal E}_f{\cal E}_i } 
       + {\vec{\sigma}(\vec{p}_f.\vec{p}_i)     \over {\cal E}_f{\cal E}_i } 
       - {\vec{p}_f(\vec{\sigma}.\vec{p}_i)     \over {\cal E}_f{\cal E}_i } 
       - {(\vec{\sigma}.\vec{p}_f)\vec{p}_i     \over {\cal E}_f{\cal E}_i } 
  \right]
\\
h_{\bar{\rm N}{\rm N}\delta} &=&
- F_{{\rm NN}\delta} 
\left[  {\vec{\sigma}.\vec{p}_f\over{\cal E}_f} + {\vec{\sigma}.\vec{p}_i\over{\cal E}_i}\right]
\end{eqnarray}
\end{mathletters}
where all strong form factors are evaluated at $Q^2= -q^2 = -(E_f-E_i)^2 + \vec{q}^{\,2}$.

For the ${\rm N}\Delta\beta$ vertices we define ( see eq.~(\ref{eq3.6})),
\begin{equation}
N^\Delta_\pi= 
{i g_{{\rm N}\Delta\pi} F_{{\rm N}\Delta\pi} \over 2m} 
\left[ { {\cal E}_{\Delta} {\cal E} \over (2\pi)^3 8\omega_\pi E_{\Delta} E}\right]^{1\over 2},
\qquad
N^\Delta_\rho= 
{ g_{{\rm NN}\rho} G_{\rm M}^{{\rm N}\Delta\rho} \over 2m} {g_{\Delta{\rm N}\pi}\over g_{{\rm NN}\pi}}
\left[ { {\cal E}_{\Delta} {\cal E} \over (2\pi)^3 8\omega_\rho E_{\Delta} E}\right]^{1\over 2}
\end{equation}
The interaction energies 
$\langle 0\vert  b_{\Delta}(\vec{p}_) -\int d^3x \Gamma_{\Delta{\rm N}\beta} b^\dagger(\vec{p})\vert 0\rangle $
for meson couplings to the ${\rm N}\rightarrow\Delta$ current are,
\begin{mathletters}
\begin{eqnarray}
H_{\Delta{\rm N}\pi} &=& N^\Delta_\pi h_{\Delta{\rm N}\pi} \vec{\tau}_{{\rm N}\Delta},
\qquad
H_{\Delta{\rm N}\rho} = N^\Delta_\rho h^\nu_{\Delta{\rm N}\rho} \epsilon_\nu \vec{\tau}_{{\rm N}\Delta} 
\\
h_{\Delta{\rm N}\pi} &=& 
- \vec{\sigma}_{{\rm N}\Delta} . \left[ \vec{p}_{\Delta} \left( {E\over m_{\Delta}} - 
{(\vec{p}_{\Delta}.\vec{p})\over {\cal E}_\Delta m_\Delta } \right) - \vec{p}\right]
\left[\openone - ({\vec{\sigma}.\vec{p}_{\Delta}\over {\cal E}_{\Delta}})({\vec{\sigma}.\vec{p}\over {\cal E}_{}})\right]
\\
h_{\Delta{\rm N}\rho} &=& + 
\left[ {-\vec{p}_{\Delta}.\vec{\sigma}_{{\rm N}\Delta}\over m_{\Delta}} ; \vec{\sigma}_{{\rm N}\Delta} 
+ \vec{p}_{\Delta} {(\vec{\sigma}_{{\rm N}\Delta}.\vec{p}_{\Delta})\over {\cal E}_\Delta m_\Delta } \right]^\mu
\left[ ({\vec{\sigma}.\vec{p}_{\Delta}\over {\cal E}_{\Delta}}) + ({\vec{\sigma}.\vec{p}\over {\cal E}_{}}) ;
\vec{\sigma}+
{(\vec{\sigma}.\vec{p}_{\Delta})\vec{\sigma}(\vec{\sigma}.\vec{p})\over {\cal E}_{\Delta}{\cal E} }
\right]^\lambda
\nonumber \\ &\times &
\left[ (p_\Delta - p)_\mu \delta_\lambda^\nu - (p_\Delta - p)_\lambda \delta_\mu^\nu)\right] 
\end{eqnarray}
\end{mathletters}
The interaction energies 
$\langle 0\vert  b(\vec{p}) -\int d^3x \Gamma_{\Delta{\rm N}\beta} b_{\Delta}^\dagger(\vec{p}_{\Delta})\vert 0\rangle $
for meson couplings to the $\Delta\rightarrow{\rm N}$ current are,
\begin{mathletters}
\begin{eqnarray}
H_{{\rm N}\Delta\pi} &=&  N^\Delta_\pi h_{{\rm N}\Delta\pi} \vec{\tau}_{{\rm N}\Delta}^\dagger,
\qquad
H_{{\rm N}\Delta\rho} = N^\Delta_\rho h^\nu_{{\rm N}\Delta\rho} \epsilon_\nu \vec{\tau}^\dagger_{{\rm N}\Delta} 
\\
h_{{\rm N}\Delta\pi} &=& + 
\left[\openone - ({\vec{\sigma}.\vec{p}_{\Delta}\over {\cal E}_{\Delta}})({\vec{\sigma}.\vec{p}\over {\cal E}_{}})\right]
\vec{\sigma}_{{\rm N}\Delta}^\dagger . \left[ \vec{p}_{\Delta} \left( {E\over m_{\Delta}} - 
{\vec{p}_{\Delta}.\vec{p})\over {\cal E}_\Delta m_\Delta } \right) - \vec{p}\right]
\\
h^\nu_{{\rm N}\Delta\rho} &=& 
-  \left[ ({\vec{\sigma}.\vec{p}_{\Delta}\over {\cal E}_{\Delta}}) + ({\vec{\sigma}.\vec{p}\over {\cal E}_{}}) ;
       \vec{\sigma}+ {(\vec{\sigma}.\vec{p}_{\Delta})\vec{\sigma}(\vec{\sigma}.\vec{p})\over {\cal E}_{\Delta}{\cal E} }
\right]^\lambda
\left[ {-\vec{\sigma}_{{\rm N}\Delta}^\dagger.\vec{p}_{\Delta} \over m_{\Delta}} ; \vec{\sigma}_{{\rm N}\Delta}^\dagger
+ \vec{p}_{\Delta} {(\vec{\sigma}_{{\rm N}\Delta}^\dagger.\vec{p}_{\Delta})\over {\cal E}_\Delta m_\Delta } \right]^\mu
\nonumber \\ &\times &
\left[(p-p_\Delta)_\mu\delta_\lambda^\nu-(p-p_\Delta)_\lambda\delta_\mu^\nu)\right]
\end{eqnarray}
\end{mathletters}
\subsection{Currents for electromagnetic vertices}
We define,
\begin{equation}
N_\gamma= {e_p\over (2\pi)^3}\left\{{\cal E}_i{\cal E}_f\over 4E_f E_i\right\}^{1/2}
\end{equation}
The photon coupling to the positive frequency nucleon current 
$\langle 0\vert  b(\vec{p}_f)\ J_{\rm eff}(0) b^\dagger(\vec{p}_i)\vert 0\rangle $
is given by,
\begin{mathletters}
\begin{eqnarray}
J^\mu_{{\rm NN}\gamma}  &=&  N_\gamma j^\mu_{{\rm N}{\rm N}\gamma}
\\
j^0_{{\rm N}{\rm N}\gamma} &=&
  \Biggl[ F_{{\rm NN}\gamma}^{(1)} + \kappa F_{{\rm NN}\gamma}^{(2)} \Bigl[ 1 - {E_f+E_i\over 2m} \Bigr] \Biggr]
+ \Biggl[ F_{{\rm NN}\gamma}^{(1)} + \kappa F_{{\rm NN}\gamma}^{(2)} \Bigl[ 1 + {E_f+E_i\over 2m} \Bigr] \Biggr]
  \left[ {\vec{p}_f.\vec{p}_i\over{\cal E}_f{\cal E}_i}   + { i\vec{\sigma}.(\vec{p}_f\times\vec{p}_i) \over{\cal E}_f{\cal E}_i}  \right]
\\
\vec{j}_{{\rm N}{\rm N}\gamma} &=&
  \left[ F_{{\rm NN}\gamma}^{(1)} + \kappa F_{{\rm NN}\gamma}^{(2)} \right]
  \left[ {\vec{p}_f\over{\cal E}_f} + {\vec{p}_i\over{\cal E}_i} 
         + i\vec{\sigma}\times\Bigl[ {\vec{p}_f\over{\cal E}_f} - {\vec{p}_i\over{\cal E}_i} \Bigr]
  \right]
- {\kappa F_{{\rm NN}\gamma}^{(2)}\over 2m} (\vec{p}_f+\vec{p}_i)
  \left[ 1 - {\vec{p}_f.\vec{p}_i \over{\cal E}_f{\cal E}_i}  - {i\vec{\sigma}.(\vec{p}_f\times\vec{p}_i) \over{\cal E}_f{\cal E}_i}  \right]
\end{eqnarray}
\end{mathletters}
The photon coupling to the  pair-creation current 
$\langle 0\vert  b(\vec{p}_f)\, J_{\rm eff}(0) d(-\vec{p}_i)\vert 0\rangle $ 
is given by,
\begin{mathletters}
\begin{eqnarray}
J_{\bar{\rm N}{\rm N}\gamma} &=& N_\gamma \epsilon_\mu j^\mu_{\bar{\rm N}{\rm N}\gamma}\qquad
\\
j^0_{\bar{\rm N}{\rm N}\gamma} &=&
  \left[ F_{{\rm NN}\gamma}^{(1)} + \kappa F_{{\rm NN}\gamma}^{(2)} \Bigl[ 1 + {E_f-E_i\over 2m} \Bigr] \right]
  {\vec{\sigma}.\vec{p}_f\over{\cal E}_f}
-
  \left[ F_{{\rm NN}\gamma}^{(1)} + \kappa F_{{\rm NN}\gamma}^{(2)} \Bigl[ 1 - {E_f-E_i\over 2m} \Bigr] \right]
  {\vec{\sigma}.\vec{p}_i\over{\cal E}_i}
\\
\vec{j}_{\bar{\rm N}{\rm N}\gamma} &=&
 \left[ F_{{\rm NN}\gamma}^{(1)} + \kappa F_{{\rm NN}\gamma}^{(2)}\Bigl[1-{E_i\over m} \Bigr] \right] \vec{\sigma}
+ 
{\kappa F_{{\rm NN}\gamma}^{(2)} \over 2m} (\vec{p}_f+\vec{p}_i) \left[ {\vec{\sigma}.\vec{p}_f\over{\cal E}_f} 
 + {\vec{\sigma}.\vec{p}_i\over{\cal E}_i} \right]
\nonumber \\ &&
+ \left[ F_{{\rm NN}\gamma}^{(1)} + \kappa F_{{\rm NN}\gamma}^{(2)}\Bigl[1+{E_i\over m} \Bigr] \right]
  \left[ {i(\vec{p}_f\times\vec{p}_i) \over {\cal E}_f{\cal E}_i } 
       + {\vec{\sigma}(\vec{p}_f.\vec{p}_i)     \over {\cal E}_f{\cal E}_i } 
       - {\vec{p}_f(\vec{\sigma}.\vec{p}_i)     \over {\cal E}_f{\cal E}_i } 
       - {(\vec{\sigma}.\vec{p}_f)\vec{p}_i     \over {\cal E}_f{\cal E}_i } 
  \right]
\end{eqnarray}
\end{mathletters}
The photon coupling to the  pair-annihilation current 
$\langle 0\vert  d^\dagger(-\vec{p}_f)\, J_{\rm eff}(0) b^\dagger(\vec{p}_i)\vert 0\rangle $ 
is given by,
\begin{mathletters}
\begin{eqnarray}
J_{{\rm N}\bar{\rm N}\gamma} &=& N_\gamma \epsilon_\mu j^\mu_{{\rm N}\bar{\rm N}\gamma}
\\
j^0_{{\rm N}\bar{\rm N}\gamma} &=&
  \left[ F_{{\rm NN}\gamma}^{(1)} + \kappa F_{{\rm NN}\gamma}^{(2)} \Bigl[ 1 - {E_f-E_i\over 2m} \Bigr] \right]
  {\vec{\sigma}.\vec{p}_i\over{\cal E}_i} 
- \left[ F_{{\rm NN}\gamma}^{(1)} + \kappa F_{{\rm NN}\gamma}^{(2)} \Bigl[ 1 + {E_f-E_i\over 2m} \Bigr] \right]
  {\vec{\sigma}.\vec{p}_f\over{\cal E}_f}
\\
\vec{j}_{{\rm N}\bar{\rm N}\gamma} &=&
  \left[F_{{\rm NN}\gamma}^{(1)} +\kappa F_{{\rm NN}\gamma}^{(2)}\Bigl[1-{E_f\over m}\Bigr]\right]\vec{\sigma}
+ {\kappa F_{{\rm NN}\gamma}^{(2)} \over 2m} (\vec{p}_f+\vec{p}_i) \left[ {\vec{\sigma}.\vec{p}_f\over{\cal E}_f} + {\vec{\sigma}.\vec{p}_i\over{\cal E}_i} \right]
\nonumber \\ &+&
  \left[ F_{{\rm NN}\gamma}^{(1)} + \kappa F_{{\rm NN}\gamma}^{(2)}\Bigl[1+{E_f\over m} \Bigr] \right]
  \left[ {i(\vec{p}_f\times\vec{p}_i) \over {\cal E}_f{\cal E}_i } 
       + {\vec{\sigma}(\vec{p}_f.\vec{p}_i)     \over {\cal E}_f{\cal E}_i } 
       - {\vec{p}_f(\vec{\sigma}.\vec{p}_i)     \over {\cal E}_f{\cal E}_i } 
       - {(\vec{\sigma}.\vec{p}_f)\vec{p}_i     \over {\cal E}_f{\cal E}_i } 
  \right]
\end{eqnarray}
\end{mathletters}
where the electromagnetic form form factors are defined as
\begin{mathletters}
\begin{eqnarray}
&&F_{{\rm NN}\gamma}^{(1)} = {1\over 2}F_{{\rm NN}\gamma}^{(1);is} + {1\over 2}F_{{\rm NN}\gamma}^{(1);iv}\tau^0
\qquad
\kappa_{{\rm N}} F_{{\rm NN}\gamma}^{(2)} = {1\over 2}\kappa^{is}F_{{\rm NN}\gamma}^{(2);iv}
                             + {1\over 2}\kappa^{iv}F_{{\rm NN}\gamma}^{(2);iv}\tau^0
\\ \Rightarrow 
{\rm protons:} \hskip .5 cm && F_{{\rm NN}\gamma}^{(1)}(k^2=0) + \kappa_{{\rm N}} F_{{\rm NN}\gamma}^{(2)}(k^2=0) 
                           = (1) + ({\kappa^{is}+\kappa^{iv}\over 2}) 
                           = \mu_p 
\\ \Rightarrow 
{\rm neutrons:}\hskip .5 cm && F_{{\rm NN}\gamma}^{(1)}(k^2=0) + \kappa_{{\rm N}} F_{{\rm NN}\gamma}^{(2)}(k^2=0) 
                           = (0) + ({\kappa^{is}-\kappa^{iv}\over 2}) 
                           = \mu_n 
\end{eqnarray}
\end{mathletters}

For the ${\rm N}\Delta\gamma$ vertices we define ( see eq.~(\ref{eq3.6})),
\begin{equation}
N^\Delta_\gamma= 
{e_p \over (2\pi)^3 2m} {G_M^{iv}\over 2}{g_{\Delta{\rm N}\pi}\over g_{{\rm NN}\pi}}
\left[ { {\cal E}_{\Delta} {\cal E} \over 4 E_{\Delta} E}\right]^{1\over 2};
\qquad
G_M^{iv}(0)= 1 + \kappa^{iv} = 4.706
\end{equation}
The photon coupling to the 
${\rm N}\rightarrow\Delta$ current
$\langle 0\vert  b_{\Delta}(\vec{p}_) J_{\rm eff}(0) b^\dagger(\vec{p})\vert 0\rangle $
is given by,
\begin{mathletters}
\begin{eqnarray}
J_{\Delta{\rm N}\gamma} &=& N^\Delta_\gamma j^\nu_{\Delta{\rm N}\gamma} \epsilon_\nu ({\tau}_{{\rm N}\Delta})^0
\\
j^\nu_{\Delta{\rm N}\gamma} &=& + 
\left[ {-\vec{\sigma}_{{\rm N}\Delta}.\vec{p}_{\Delta} \over m_{\Delta}} ; \vec{\sigma}_{{\rm N}\Delta} 
+ \vec{p}_{\Delta} {(\vec{\sigma}_{{\rm N}\Delta}.\vec{p}_{\Delta})\over {\cal E}_\Delta m_\Delta } \right]^\mu
\left[ ({\vec{\sigma}.\vec{p}_{\Delta}\over {\cal E}_{\Delta}}) + ({\vec{\sigma}.\vec{p}\over {\cal E}_{}}) ;
\vec{\sigma}+
{(\vec{\sigma}.\vec{p}_{\Delta})\vec{\sigma}(\vec{\sigma}.\vec{p})\over {\cal E}_{\Delta}{\cal E} }
\right]^\lambda
\nonumber \\ &\times &
\left[ (p_\Delta - p)_\mu \delta_\lambda^\nu - (p_\Delta - p)_\lambda \delta_\mu^\nu)\right] 
\end{eqnarray}
\end{mathletters}
The photon coupling to the $\Delta\rightarrow{\rm N}$ current
$\langle 0\vert  b(\vec{p}) J_{\rm eff}(0) b_{\Delta}^\dagger(\vec{p}_{\Delta})\vert 0\rangle $
is given by,
\begin{mathletters}
\begin{eqnarray}
J_{{\rm N}\Delta\gamma} &=& N^\Delta_\gamma j^\nu_{{\rm N}\Delta\gamma} \epsilon_\nu ({\tau}^\dagger_{{\rm N}\Delta})^0
\\
j^\nu_{{\rm N}\Delta\gamma} &=& 
-  \left[ ({\vec{\sigma}.\vec{p}_{\Delta}\over {\cal E}_{\Delta}}) + ({\vec{\sigma}.\vec{p}\over {\cal E}_{}}) ;
       \vec{\sigma}+ {(\vec{\sigma}.\vec{p}_{\Delta})\vec{\sigma}(\vec{\sigma}.\vec{p})\over {\cal E}_{\Delta}{\cal E} }
\right]^\lambda
\left[ {-\vec{p}_{\Delta}.\vec{\sigma}_{{\rm N}\Delta}^\dagger\over m_{\Delta}} ; \vec{\sigma}_{{\rm N}\Delta}^\dagger
+ \vec{p}_{\Delta} {(\vec{\sigma}_{{\rm N}\Delta}^\dagger.\vec{p}_{\Delta})\over {\cal E}_\Delta m_\Delta } \right]^\mu
\nonumber \\ &\times &
\left[(p-p_\Delta)_\mu\delta_\lambda^\nu-(p-p_\Delta)_\lambda\delta_\mu^\nu)\right]
\end{eqnarray}
\end{mathletters}

The photon coupling to the vector-meson-decay current is given by,
\begin{equation}
J_{{\rm V}{\rm P}\gamma}^\mu(0)(q_{{\rm P}},q_{{\rm V}})
 =  
{- e_p g_{{\rm V}{\rm P}\gamma} F_{{\rm V}{\rm P}\gamma}\over m_{{\rm V}}}
{1\over (2\pi)^3}
{1 \over  \sqrt{4 \omega_{{\rm V}}\omega_{{\rm P}}}}
\epsilon^{\mu \xi  a b } \epsilon_\xi(\vec{q}_{{\rm V}},\lambda_{{\rm V}}) (q_{{\rm P}})_a (q_{{\rm V}})_b
\end{equation}
where $F_{{\rm V}{\rm P}\gamma}(k^2=0)$=1, $q_{{\rm V}}$ and $q_{{\rm P}}$ are the 4-momenta delivered to 
the nucleons by the vector and pseudo-scalar mesons respectively.
\section{The Momentum-Space NN$\gamma$ Rescattering Calculation}
\label{appC}
The NN$\gamma$ rescattering amplitudes in bremsstrahlung were first calculated by
Brown \cite{VB69} in $r$-space. For $p$-space calculations, a recipe has 
already been given in ref~\cite{VH91}, although the residue terms are valid 
only for soft photons and the dominant final-state interaction is scaled by 
(guessed) off-shell minimal relativity factors and cast into the I-frame with 
a barycentric momentum $k$. This choice of frame does not minimize the effect 
of the neglected boost operators and the inclusion of minimal relativity 
factors disturbs the convergence properties of the p-space integral.

Casting the entire rescattering amplitude into the A-frame of eq~(\ref{eq3.32})
requires a somewhat different numerical procedure. We start from 
eq.~(\ref{eq3.43}) and simplify our notation by defining,
\begin{eqnarray}
{\cal M}^{\rm R}_{fi} 
=&& K \int \! \int H(\hat{p}) d\theta d\phi  \qquad\qquad
H(\hat{p}) =  \int {F(\vec{p}_{}) \over G_i(\vec{p}_{}) G_f(\vec{p}_{})} dp
\end{eqnarray}
with,
\begin{eqnarray}
 F(\vec{p}_{}) &&= p^2 \sin\theta 
         [E(\vec{p}_{3}+{1\over 4}\vec{k}) + E(\vec{p}_{}+{1\over 4}\vec{k})]
         [E(\vec{p}_{1}-{1\over 4}\vec{k}) + E(\vec{p}_{}-{1\over 4}\vec{k})] 
\nonumber \\ && \times  
\sum_{M_S M_S'} 
\langle \widetilde{\vec{p}_{3}+{1\over 4}\vec{k}};S_f M_{S_f}; T_f M_T\vert   [1+i\chi(-{\vec{k}\over 2})] 
   t^{(-)\dag}
[1-i\chi(-{\vec{k}\over 2})]\vert \vec{p} + {1\over 4}\vec{k};S_f M_S';T_f M_T\rangle  
\nonumber \\ && \times  
\langle S_f M_S';T_f M_T\vert  
J_{{\rm NN}\gamma}[1](-\vec{p}_{}+{\vec{k}\over 2},-\vec{p}_{}-{\vec{k}\over 2})  
\vert S_i M_S;T_i M_T\rangle 
\nonumber \\ && \times  
\langle +\vec{p}_{} - {1\over 4}\vec{k};S_i M_S; T_i M_T\vert   [1+i\chi(+{\vec{k}\over 2})]
   G_i t^{(+)}
[1-i\chi(+{\vec{k}\over 2})] \vert \widetilde{\vec{p}_{1} - {1\over 4}\vec{k}}; S_i M_{S_i}; T_i M_T\rangle  
\nonumber \\ 
G_i(\vec{p}_{}) &&= (\vec{p}_{1}-{1\over 4}\vec{k})^2 - (\vec{p}_{}-{1\over 4}\vec{k})^2 + i\eta_i 
              = \Delta_i^2 - (p - {1\over 4}|\vec{k}|\cos\vartheta)^2 + i\eta_i   
\nonumber \\
G_f(\vec{p}_{}) &&= (\vec{p}_{3}+{1\over 4}\vec{k})^2 - (\vec{p}_{}+{1\over 4}\vec{k})^2 + i\eta_f 
                   = \Delta_f^2 - (p + {1\over 4}|\vec{k}|\cos\vartheta)^2 +i\eta_f   
\nonumber \\
&& \Delta_i^2 = \left({1\over 4}|\vec{k}|\cos\vartheta\right)^2 -\vec{p}_{1}.\vec{p}_{2} \qquad\qquad
\Delta_f^2 = \left({1\over 4}|\vec{k}|\cos\vartheta\right)^2 -\vec{p}_{3}.\vec{p}_{4}
\nonumber \\
K &&=  {N\over 2} (-1)^{(S_f+S_i+T_f+T_i)} 
\end{eqnarray}
with $\cos\vartheta = \hat{p}.\hat{k} = 
\sin\theta_p\sin\theta_k \cos(\phi_p - \phi_k) + \cos\theta_p\cos\theta_k$.
Within the (maximally symmetric) A-frame the effects of the boost operators
$\chi(+{1\over 2}\vec{k})$ and $\chi(-{1\over 2}\vec{k})$ 
can be expected to substantially cancel and will therefore be neglected 
in the present numerical applications. Their inclusion would change nothing 
that we will discuss in this appendix.

For $\Delta_i, \Delta_f < 0$, there are no poles on the real 
axis, but it is easy to see that there are singularies in the Greens' 
functions at $\vec{p}$ = $\vec{p}_1$, $\vec{p}_2$, $\vec{p}_3$, $\vec{p}_4$ 
as well as
$(\vec{p}_1-\vec{p}).(\vec{p}_2-\vec{p})$ and
$(\vec{p}_3-\vec{p}).(\vec{p}_4-\vec{p})$, the latter two requiring
particular attention when
$\cos\vartheta_i=\pm4\sqrt{\vec{p}_{1}.\vec{p}_{2}}/ |\vec{k}|$ or
$\cos\vartheta_f=\pm4\sqrt{\vec{p}_{3}.\vec{p}_{4}}/|\vec{k}|$ 
cause $\Delta_i=0$ or $\Delta_f=0$, so that $G_i^{-1}$ and/or $G_f^{-1}$
each have a 2$^{nd}$-order pole (on the real axis).
We adopt the utilatarian attitude of noting that for experiments 
below the $\pi$-production threshold, the second-order poles only occur at 
$\theta_1^{\rm Lab},\theta_2^{\rm Lab} < 6^{\circ}$, and even then, 
only for $\theta_{\gamma}^{\rm Lab} < 1^{\circ}$. Since no data exists in 
this region, we simply defer a treatment of second order singularities
and confine our attention to the kinematics containing the simple and 
separable poles,
\begin{eqnarray}
p &&=p_i^{(\pm)} = +{1\over 4}|\vec{k}|\cos\vartheta\pm\Delta_i \qquad\qquad
\hskip 1 true mm
\hbox{if~~} \Delta_i^2 \ge 0 \hbox{~~then~~} 
\Delta_i= |p_i^{(\pm)}-{1\over 4}|\vec{k}| \cos\vartheta | \nonumber \\
p &&=p_f^{(\pm)} = -{1\over 4}|\vec{k}|\cos\vartheta\pm\Delta_f \qquad\qquad
\hbox{if~~} \Delta_f^2 \ge 0 \hbox{~~then~~} 
\Delta_f= |p_f^{(\pm)}+{1\over 4}|\vec{k}| \cos\vartheta | 
\end{eqnarray}
and evaluate $\Delta_i^2$ and $\Delta_f^2$ to determine if poles exist 
on the positive real $p-$axis.
Defining such poles to be vectors so that
$\hat{p}_{i}^{\pm}$=$\hat{p}_{f}^{\pm}$=$\hat{p}$, we obtain,
\begin{eqnarray}
H(\hat{p}) 
=&& {\rm V.P.}\int dp {F(\vec{p}_{}) \over 
\Bigl[ \Delta_i^2 - (p - {1\over 4}|\vec{k}|\cos\vartheta)^2 \Bigr]
\Bigl[ \Delta_f^2 - (p + {1\over 4}|\vec{k}|\cos\vartheta)^2  \Bigr]  } - Z
\end{eqnarray}
where
\begin{eqnarray}
Z =&& {i\pi F(\vec{p}_{i}^{\;(+)}) \over 2 \Delta_i G_f(p_{i}^{(+)})}
  +  {i\pi F(\vec{p}_{i}^{\;(-)}) \over 2 \Delta_i G_f(p_{i}^{(-)})} 
  +  {i\pi F(\vec{p}_{f}^{\;(+)}) \over 2 \Delta_f G_i(p_{f}^{(+)})} 
  +  {i\pi F(\vec{p}_{f}^{\;(-)}) \over 2 \Delta_f G_i(p_{f}^{(-)})} \nonumber \\
\end{eqnarray}
remains well defined.
This formally completes the specification of the integral. 
However, for practical purposes, we need to add special forms of zero 
to smooth the divergences near the poles. With a simple generalization
of
\begin{equation}
{\rm V.P.} \int_{a_1}^{a_2} {1\over k^2 - p^2} dp = 
{1 \over 2k} \ln\vert {a_2+k \over a_2-k}\;.\;{k-a_1 \over k+a_1}\vert  
\end{equation}
we obtain,
\begin{eqnarray}
H =&&  H_{p_{i}^{(+)}} + H_{p_{i}^{(-)}} + H_{p_{f}^{(+)}} + H_{p_{f}^{(-)}} \nonumber \\
\end{eqnarray}
where
\begin{mathletters}
\begin{eqnarray}
H_{p_{i}^{(+)}} =&& \int_{a1}^{a2} 
{1 \over G_i(\vec{p}_{}) } 
\Bigg[ {F(\vec{p}_{}) \over G_f(\vec{p}_{}) }  - 
       {F(\vec{p}_{i}^{\;(+)}) \over G_f(\vec{p}_{i}^{\;(+)}) } \Biggr] dp
+ {1\over 2\Delta_i} { F(\vec{p}_{i}^{\;(+)}) \over G_f(p_{i}^{(+)})}
  \left[\ln\vert {(a2-p_i^{(-)}) \over (a2-p_i^{(+)})}
	      {(a1-p_i^{(+)}) \over (a1-p_i^{(-)})}\vert  -i\pi \right]
\nonumber \\ &&
\end{eqnarray}
\begin{eqnarray}
H_{p_{i}^{(-)}} =&& \int_{b1}^{b2} 
{1 \over G_i(\vec{p}_{}) }
\Biggl[ {F(\vec{p}_{}) \over G_f(\vec{p}_{}) } -
        {F(\vec{p}_{i}^{\;(-)}) \over G_f(p_{i}^{(-)})} \Biggr] dp
+ {1 \over 2\Delta_i} { F(\vec{p}_{i}^{\;(-)}) \over G_f(p_{i}^{(-)})}
  \left[\ln\vert  {(b2-p_i^{(-)}) \over (b2-p_i^{(+)})}
         {(b1-p_i^{(+)}) \over (b1-p_i^{(-)})}\vert  - i\pi\right]
\nonumber \\ &&
\end{eqnarray}
\begin{eqnarray}
H_{p_{f}^{(+)}} =&& \int_{c1}^{c2} 
{ 1 \over G_f(\vec{p}_{}) }
\Biggl[ {F(\vec{p}_{}) \over G_i(\vec{p}_{}) } -
        {F(\vec{p}_{f}^{\;(+)}) \over G_i(\vec{p}_{f}^{\;(+)}) } \Biggr] dp
+ {1 \over 2\Delta_f} { F(\vec{p}_{f}^{\;(+)}) \over G_i(p_{f}^{(+)})}
  \left[\ln\vert  {(c2-p_f^{(-)}) \over (c2-p_f^{(+)})}
         {(c1-p_f^{(+)}) \over (c1-p_f^{(-)})}\vert  -i\pi \right]
\nonumber \\ &&
\end{eqnarray}
\begin{eqnarray}
H_{p_{f}^{(-)}} =&& \int_{d1}^{d2} 
{ 1 \over G_f(\vec{p}_{}) } 
\Biggl[ {F(\vec{p}_{}) \over G_i(\vec{p}_{}) } -
        {F(\vec{p}_{f}^{\;(-)}) \over G_i(p_{f}^{(-)})} \Biggr] dp
+ {1\over 2\Delta_f} { F(\vec{p}_{f}^{\;(-)}) \over G_i(p_{f}^{(-)})}
  \left[\ln\vert  {(d2-p_f^{(-)}) \over (d2-p_f^{(+)})}
                 {(d1-p_f^{(+)}) \over (d1-p_f^{(-)})}\vert  -i\pi \right]
\nonumber \\ &&
\end{eqnarray}
\end{mathletters}
where  the domains [a1:a2] + [b1:b2] + [c1:c2] + [d1:d2] span [0:$\infty$]
and contain the poles $p_{i}^{(+)}$, $p_{i}^{(-)}$, $p_{f}^{(+)}$ and 
$p_{f}^{(-)}$ respectively. In the event that $p_\alpha^{(\pm)}$ does not 
exist in [0:$\infty$], then our expressions require $F(p_\alpha^{(\pm)})=0$.
\begin{multicols}{2}
%

%
%
%
%
%
%
%
%
\end{multicols}
\end{document}